\documentclass[11pt,a4paper,oneside]{report}

\usepackage[utf8]{inputenc}
\usepackage[T1]{fontenc}
\usepackage[english]{babel}

\usepackage{amsmath,amssymb}
\usepackage{graphicx}
\usepackage{caption}
\usepackage{subcaption}
\usepackage{tabularx}
\usepackage{tabulary}
\usepackage{adjustbox}
\usepackage{siunitx}
\usepackage{bbm}
\usepackage{comment}
\usepackage{float}
\usepackage{url}
\usepackage[
    colorlinks=true,
    linkcolor=blue,
    citecolor=blue,
    urlcolor=blue
]{hyperref}

\numberwithin{equation}{chapter}

\begin{document}


\begin{titlepage}

\centering

\vspace*{2cm}

{\Huge\bfseries
Development of an RPC-based gaseous photodetector
with picosecond resolution
\par}

\vspace{2cm}

{\Large Simone Garnero\par}

\vfill

{\large
Master Thesis\\
International Master on Advanced methods in Particle Physics
\par}

\vspace{0.5cm}

{\large
September 2025
\par}

\vspace{2cm}

\begin{flushleft}
Supervisor: Prof.\ Kodai Matsuoka\\
Co-supervisor: Dr.\ Diego Tonelli\\
Reviewer: Dr.\ Jens Weingarten
\end{flushleft}

\vfill

\end{titlepage}


\pagenumbering{roman}

\tableofcontents

\clearpage

\thispagestyle{plain}

\section*{Abstract}

This experimental particle-physics thesis reports the latest developments on the GasPM, a novel gaseous photodetector aimed at suppressing beam-induced backgrounds in the electromagnetic calorimeter for a potential upgrade of the Belle~II experiment. The GasPM technology is based on combining a photocathode with a resistive-plate chamber offering high efficiency, excellent time resolution, and cost-effective scalability. A further advantage is that, combined with a radiator, the GasPM offers precise Cherenkov-based charged-particle identification.

As part of a project launched in 2017, this work aims at addressing the degradation in time resolution observed in a previous beam test over what was achieved earlier  with laser light. 
I focus specifically on ultraviolet-photon emission during excitation and de-excitation of the gas molecules, which leads to a secondary signal that in turn spoils time resolution (photon feedback). 
I design and execute an improved beam test that, along with several configuration changes, newly introduces single-vs-multiple electron discrimination and high-frequency signal readout.
In addition, I probe through a cosmic-ray test the quantum efficiency of a new LaB$_6$ photocathode resistant to damage from ions drifting backwards, for use in future beam tests.

The principal results are the development of an algorithm to efficiently suppress photon feedback; a preliminary calibration of a novel digitiser; the achievement of discrimination between single- and multiple-electron events; and an early qualification of a LaB$_6$ photocathode. These results are being prepared for showing at the 7th International Workshop on New Photon Detectors organized in Bologna in December 2025 and pave the way for an upcoming beam test of an improved GasPM prototype.

\clearpage


\pagenumbering{arabic}

\chapter*{Introduction}
\label{chap:Introduction}
\addcontentsline{toc}{chapter}{Introduction}

The Standard Model of elementary particles and their interactions (SM) is the currently accepted theory of particle physics. It is recognized as the ultimate success of the reductionist paradigm for describing microphysics at its most fundamental level. By means of about twenty parameters, the Standard Model describes accurately thousands of measurements involving processes mediated by the electromagnetic, weak, and strong interactions that span more than ten orders of magnitude in energy.

However, theoretical considerations and potential experimental inconsistencies support the general belief that the Standard Model might be an \textit{effective theory} --- a theory valid at the energies probed thus far, that is incorporated in a yet-unknown and more general theory encompassing higher energies. Completing the Standard Model is the principal goal of today's particle physics.

\textit{Direct} approaches, which broadly consist in searching for decay products of non-SM particles produced on mass-shell in high-energy collisions, have been traditionally fruitful. However, reach is limited by the collision energy of accelerators and by the large investments needed to further it. Complementary approaches consist in comparing with predictions precise measurements of lower-energy processes in which virtual non-SM particles could contribute. The reach of such \textit{indirect} approaches is not constrained by collision energy, but rather by the precision attainable, both in observations and predictions.
Precision requires large samples of data, and therefore high collision intensities, along with detectors capable to cope with the associated experimental challenges.

Belle~II is an international collaboration of about 700 physicists aiming to indirectly test the Standard Model by studying billions of decays of mesons containing the $b$ and $c$ quarks, as well as $\tau$ leptons, produced in high-intensity electron-positron collisions. The core physics program focuses on decays into photons or neutrinos, as these probe key aspects of the weak interactions that are difficult to study by the competitor experiment LHCb. This makes the electromagnetic calorimeter and its performance essential.

However, calorimeter performance is degraded by significant and persistent beam-induced backgrounds due to the unprecedentedly harsh collision environment. The extreme intensities needed to pursue the Belle~II program require steering high-current beams to collide within a luminous region of just a few tens of nanometres. This highly constrained collision environment generates significant backgrounds from particles that do not originate from the collision itself, but result from beam interactions with residual gas molecules in the enclosure or with the collider infrastructure. 
The performance of the Belle~II calorimeter suffers serious limitations due to beam-background photons, which degrade the purity of reconstructed decay signals and spoil the resolution of physics quantities, such as the $\pi^{0}$ mass.
Suppressing or mitigating beam-backgrounds and their effects is therefore essential to enable the full exploitation of the Belle~II program.

A promising approach relies on exploiting the distinctive time distribution of beam-backgrounds.
Because beam-background photons are preferentially out of time with respect to the collision, an efficient photodetector with a time resolution of about 20~ps, installed between the collision point and the inner surface of the calorimeter, would enable selection of collision photons while strongly rejecting background photons. However, such a photodetector would need to cover the entire inner surface of the calorimeter, which would make costs prohibitive with current technology.

Since 2017, the group I joined for my thesis is developing a promising proposal based upon the gaseous photomultiplier, a photosensitive gaseous detector based on coupling a photocathode with a resistive-plate chamber. Photons hit the photocathode, which releases photoelectrons. These are subsequently accelerated by an electric field in the gas and multiplied. The induced current signal is read out by the anode and can be measured with excellent time resolution and high efficiency. In addition, combined with a radiator, the GasPM offers precise Cherenkov-based charged-particle identification. This technology is based on affordable components offering opportunities for large coverage at reasonable costs.  

The objective of my thesis is to advance the development of this new detector. My work addresses the major causes of the time-resolution degradation observed in a 2023 beam test with respect to that achieved in an earlier laser test. I prepare and execute an improved beam test, and analyse its data, focusing on the understanding and mitigation of one of the main causes of the degradation: secondary signals generated by UV photons impinging on the photocathode (photon feedback).
In addition, I develop a selection for single-electron events, to ensure the time resolution from beam tests is unbiased.
Finally, I investigate the feasibility of a future beam test based on a new LaB$_6$ photocathode  aiming at reducing  damage from ions that drift backwards.

This thesis is structured as follows. Chapter~\ref{chap:phys} outlines succinctly the flavour sector of the Standard Model and its potential in indirectly searching for non-SM physics; Chapter~\ref{chap:detector} introduces briefly the Belle II experiment and introduces beam-induced backgrounds; Chapter~\ref{chap:GasPM} describes the GasPM architecture and properties, along with previous performance achievements; Chapter~\ref{chap:NALU} presents a novel digitiser first deployed in this work to address photon feedback, and its calibration; Chapter~\ref{chp:beamtest} describes the
beam test I conducted and the analyses to study single-electron events selection and photon feedback;  Chapter~\ref{chap:cosmic} describes work done on the new photocathode. The final Chapter~\ref{chap:conclusion} presents the summary and conclusions.
\chapter{Flavour physics to overcome the Standard Model}\label{chap:phys}
\chaptermark{Flavour Physics}

\section{The Standard Model of particle physics}
The Standard Model (SM) is an effective quantum field theory that describes all fundamental interactions in nature without gravity~\cite{SM-weinberg,SM-glashow,Salam:1968rm,Englert:1964et,Higgs:1964pj}.

The quantum-field-theory framework results from the unification of quantum mechanics with special relativity and offers the most fundamental description of nature known to date.

A field is a set of values, associated to certain physical properties, assigned to every point in space and time.
Quantum fields are fields that pervade the whole spacetime and obey the rules of quantum mechanics.
If a quantum field is modified by an appropriate perturbation, the resulting oscillatory states, called field excitations, carry more energy than the resting state and are called `particles'. For instance, the electron is the massive excitation of the electron field. The quantized nature of the description implies that only certain perturbations that satisfy precise energetic conditions are capable of generating field excitations. It is not possible, for example, to generate a wave in the electron field that corresponds to half an electron.

Quantum fields interact with each other. The Standard Model is the theory that describes their dynamics at energy scales relevant for the subnuclear world.
Particles and their interactions are described in a Lagrangian formalism, in which every combination of fields and interaction operators that is not forbidden by the symmetries of the dynamics is, in principle, included. Local gauge symmetry, i.e., the invariance of the Lagrangian under space-time-dependent transformations applied to the phases of fields, is the key overarching concept. Interaction terms appear in the free-field Lagrangian after requiring it to be invariant under local gauge symmetries. 
\enlargethispage{\baselineskip}
The Standard Model is based on the symmetry group $${SU(3)}_{C}\otimes{SU(2)}_{L}\otimes{U(1)}_{Y},$$ where ${SU(3)}_{C}$ is the standard unitary group that describes the strong interactions (quantum chromodynamics, QCD), and $C$ stands for the colour charge; ${SU(2)}_{L}\otimes{U(1)}_{Y}$ is the product of groups that describe the combination of the weak and electromagnetic interactions, where ${SU(2)}_{L}$ is the standard unitary group of weak isospin doublets ($L$ standing for \textit{left}\footnote{Only particles with \textit{left} chirality are influenced by the weak interaction.}) and ${U(1)}_{Y}$ stands for the unitary group of \textit{hypercharge} $Y$.

Spin-$1$ particles called \textit{gauge bosons} mediate the interactions. 
Strong interactions are mediated by eight massless particles corresponding to the ${SU(3)}_{C}$ generators, called \textit{gluons}: they carry a charge that can be of three kinds, called \textit{colour}. Weak interactions are mediated by two charged massive bosons, $W^{\pm}$, and a neutral massive boson, $Z^0$. Electromagnetic interactions occur between particles carrying electric charge and are mediated by a neutral massless boson, the photon $\gamma$. The physical electroweak bosons ($W^{\pm}$, $Z^0$, $\gamma$) arise from the following linear combinations of ${SU(2)}_{L}\otimes{U(1)}_{Y}$ generators: $$ W^{\pm}=\frac{1}{\sqrt{2}}(W_1 \mp iW_2)\ \ \mathrm{and}\ \ 
\begin{pmatrix}
\gamma \\ Z^0
\end{pmatrix}
=
\begin{pmatrix}
\cos\theta_W & \sin\theta_W \\ -\sin\theta_W & \cos\theta_W
\end{pmatrix}
\begin{pmatrix}
B \\ W_3
\end{pmatrix},$$
where ${\theta}_{W}$ is a free parameter, called \textit{Weinberg angle}. The $W^{\pm}$ mass depends on the $Z$ mass via ${\theta}_{W}$.
Particles acquire mass via the interaction with the Higgs field, which is mediated by a spin-$0$ particle, the Higgs boson.

Matter particles correspond to excitations of spin-$\frac{1}{2}$ fields and are called \textit{fermions}. Their masses are free parameters of the theory. Each fermion is also associated with an antiparticle that has the same mass and opposite internal quantum numbers. Fermions are further classified into two classes, quarks, which are the fundamental constituents of nuclear matter, and leptons, each organized in three weak-isospin doublets.

Quark doublets are each composed of an up-type quark, with charge $\frac{2}{3}e$, and a down-type quark, with charge $-\frac{1}{3}e$, 
\begin{equation*}
\begin{pmatrix}
u \\ d
\end{pmatrix}
\begin{pmatrix}
c \\ s
\end{pmatrix}
\begin{pmatrix}
t \\ b
\end{pmatrix}\mbox{ .}
\end{equation*}
They couple with both the strong and electroweak interactions. Each quark has colour and a flavour quantum number, which comes in six varieties and is conserved in the electromagnetic and strong interactions, but not in the weak interactions. Due to colour confinement free quarks are not observable. They are only observed in their colourless bound states, which include mesons, typically composed of a quark and an anti-quark, and baryons, composed of three quarks. Baryons are assigned a quantum number, called baryon number, found to be conserved even if no symmetry of the Lagrangian implies that.
Lepton doublets are each composed by an almost massless neutral neutrino and a massive particle with electric charge $-e$;
\begin{equation*}
\begin{pmatrix}
\nu_e \\ e
\end{pmatrix}
\begin{pmatrix}
\nu_{\mu} \\ \mu
\end{pmatrix}
\begin{pmatrix}
\nu_{\tau} \\ \tau
\end{pmatrix}\mbox{ .}
\end{equation*}
They couple only with the electroweak interaction. Each lepton has a lepton-family quantum number; their sum in a process, called global lepton number, is found to be conserved in all interactions, although no symmetry of the dynamics prescribes that; individual lepton numbers are not conserved in neutrino oscillations.

\begin{figure}[htb]
	\centering
	\includegraphics[width=\textwidth]{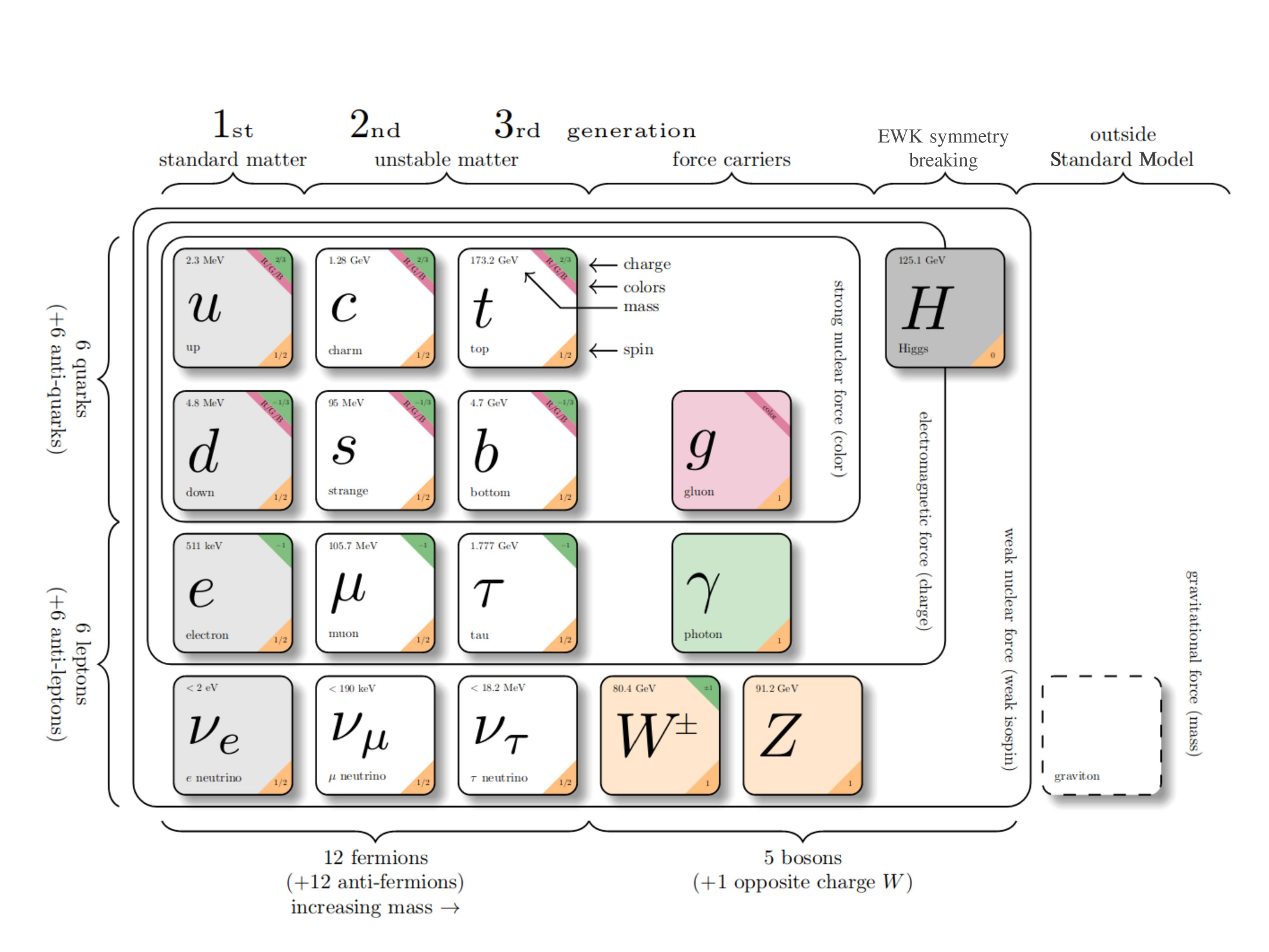}
	\caption{Scheme of particles and interactions in the Standard Model \cite{Manfredi}.}
	\label{fig: SM}
\end{figure}

\noindent Figure \ref{fig: SM}  shows a scheme of the Standard Model particles and their interactions.

\noindent In addition to gauge symmetry, discrete symmetries are important in constraining the dynamics. Parity ($P$) is a transformation that inverts all spatial coordinates; charge conjugation ($C$) is the exchange of every particle with its own antiparticle; and time reversal ($T$) inverts the time axis. The product of these three discrete symmetry transformations is found to be conserved in all interactions, as prescribed by foundational axioms of field theory~\cite{1957AnPhy...2....1L,1957AnPhy...2....1L2}, but the symmetries are not conserved individually. Parity symmetry is maximally violated in the weak interactions, while the combined ${\textit CP}$ symmetry is violated in the weak interactions at the $0.1\%$ level. In principle, the strong interaction too could violate ${\textit CP}$ symmetry, but no experimental evidence of that has ever been observed. The existence of as-yet unobserved particles, called axions, has been postulated to account for that~\cite{adams2023axion}.

\section{Where do we stand?}

The Standard Model was completed in the 1970's and has been successfully tested since, in thousands of measurements whose fractional precisions reach one part per trillion~\cite{ParticleDataGroup:2024cfk}. However, observations and theoretical considerations suggest that the Standard Model is likely to be an effective theory, valid at the $\si{\electronvolt}$--$\si{\tera\electronvolt}$ energies probed so far, that should be completed by a more general full theory valid over a broader range of high energies. Open questions that support this interpretation include the lack of an explanation for a dynamical origin for the observed asymmetry between matter and antimatter in the universe, the strikingly large differences observed between fermion masses, the possible instability of the Higgs vacuum, the conceptual and technical difficulties in achieving a description of gravity consistent with quantum mechanics, or the postulated large amounts of non-interacting matter (dark matter), introduced to justify cosmological observations.

Extending the Standard Model to higher energy-scales is the main goal of today's particle physics, in an attempt at addressing these and other open issues. Current strategies to extend the Standard Model can broadly be classified into two synergic approaches. 

The energy-frontier, \textit{direct} approach aims at using high-energy collisions to produce on-shell particles (that is, particles satisfying the energy-momentum conservation at production) not included in the Standard Model, and detect directly their decay products, thus gaining direct evidence of their existence.\footnote{\textit{Mass shell} is jargon for mass hyperboloid, which identifies the hyperboloid in energy–momentum space describing the solutions to the mass-energy equivalence equation $E^{2}=(pc)^{2}+m^2c^{4}$. A particle \textit{on-mass-shell} satisfies this relation.} Historically this offered striking experimental evidence of new phenomena, when energetically accessible, but its reach is limited by the maximum energy available at colliders. 

The intensity-frontier, \textit{indirect} approach broadly consists in searching for significant differences between precise measurements and equally precise SM predictions in lower-energy processes sensitive to non-SM contributions. A semi-intuitive, although simplified conceptual representation of the subtending idea is that exchanges of virtual (off-mass-shell) particles of arbitrary high mass, including those not described in the Standard Model, occur in the transition, thus altering the amplitudes in an observable manner. The presence of virtual particles, which may imply a temporary non-conservation of energy if interpreted classically, is allowed by Heisenberg's uncertainty principle $\Delta E \Delta t > \hbar/2$. Experimental evidence is typically harder to establish, but the reach is not bounded by the maximum collision energy reachable by experiments. A large portion of the effort in this approach is centered on the weak-interactions of quarks (so called `flavour physics').

\section{Flavour physics in the Standard Model}\label{sec:flav}

Although technically flavour physics includes also lepton interactions, I restrict the scope by referring solely to the quark interactions here.

The role of flavour in shaping the Standard Model has been central since the early days of particle physics. However, its prominence in determining the theory can perhaps be tracked down to the early 1960's with the apparent inconsistency between weak coupling constants measured in muon decay, neutron decay, and strange-particle decays. Such inconsistency was first addressed by Gell-Mann and Levy~\cite{GellMann-Levy} and then Cabibbo~\cite{cabibbo}, who postulated differing mass (\textit{d}) and weak (\textit{d$^\prime$}) eigenstates for down-type quarks. This was achieved by introducing a mixing angle ($\theta_C$) between the \textit{s} quark and \textit{d} quark, the only two down-type quarks known at the time. While Cabibbo's theory addressed efficiently the difference of weak coupling constants, it also predicted a rate for the $K^0_{L} \rightarrow \mu^+\mu^-$ and other kaon decays inconsistent with the experimental exclusion limits at the time. Glashow, Iliopoulos, and Maiani addressed the conundrum by postulating the existence of a fourth quark (\textit{c}) of $\SI{2}{\giga\electronvolt}/c^2$ mass, whose contribution in the $K^0_{L} \rightarrow \mu^+\mu^-$ decay amplitude would cancel the \textit{u} quark contribution, suppressing the branching fraction down to values consistent with experimental limits \cite{GIM}. The charm quark was then discovered four years after the prediction, showing the compelling power of the indirect approach. In addition, in 1973, when only three quarks were known, Kobayashi and Maskawa generalized Cabibbo's theory from a four-quark model to a six-quark model to accommodate the phenomenon of ${\textit CP}$ violation observed in 1964 \cite{km}. They introduced a complex unitary matrix to describe the relations between mass (unprimed) and weak (primed) interaction eigenstates of quarks as seen by $W^\pm$ bosons. This is known as the Cabibbo-Kobayashi-Maskawa (CKM) quark-mixing matrix or $V_{\textrm {CKM}}$,

\begin{equation*}
\scalebox{0.88}{$ 
\begin{pmatrix}
d' \\
s' \\
b'
\end{pmatrix} =
\begin{pmatrix}
V_{ud} & V_{us} & V_{ub}\\
V_{cd} & V_{cs} & V_{cb}\\
V_{td} & V_{ts} & V_{tb}
\end{pmatrix}
\begin{pmatrix}
d \\
s \\
b
\end{pmatrix} = 
\left[
\left(
\begin{array}{ccc}
1-\lambda^2/2 & \lambda        & A\lambda^3(\rho - i \eta) \\
-\lambda      & 1-\lambda^2/2  & A \lambda^2 \\
A\lambda^3 (1- \rho -i \eta)   &-A \lambda^2 & 1 
\end{array}
\right)
+ \mathcal{O}(\lambda^4)
\right]
\begin{pmatrix}
d \\
s \\
b
\end{pmatrix}\mbox{ .}
$}
\end{equation*}

where
\begin{equation*}
\scalebox{0.95}{$
\lambda =\frac{V_{us}}{\sqrt{V_{ud}^2 + V_{us}^2}}, \quad A\lambda^2 = \lambda \frac{V_{cb}}{V_{us}}, \quad {\textrm{and}} \quad A \lambda^3 (\rho + i\eta) = V_{ub}^*\mbox{ .}
$}
\end{equation*}

The parameter $\lambda$ expresses the mixing between the first and second quark generations, $A$ and $\rho$ are real parameters, and $\eta$ is a complex phase that introduces ${\textit CP}$ violation.
Each $V_{ij}$ matrix element encapsulates the weak-interaction coupling between an up-type $i$ and down-type $j$ quarks.
It is a $N\times N$ CKM matrix with $(N-1)^2$ free parameters, where $N$ is the number of quarks families. If $N=2$, the only free parameter is the Cabibbo angle $\theta_C \approx 13^\circ$, whereas if $N=3$, the free parameters are three Euler angles ($\theta_{12}$, $\theta_{13}$, and $\theta_{23}$) and a complex phase ($\delta$), which allows for ${\textit CP}$-violating couplings~\cite{PDG}. The matrix is most conveniently written in the so-called \textit{Wolfenstein parametrization}, an expansion in the small parameter $\lambda = \sin\theta_C \approx 0.23$ that makes explicit the observed hierarchy between its elements~\cite{wolfenstein}.

\begin{figure}[h!]
	\centering
	\includegraphics[width=0.7\textwidth]{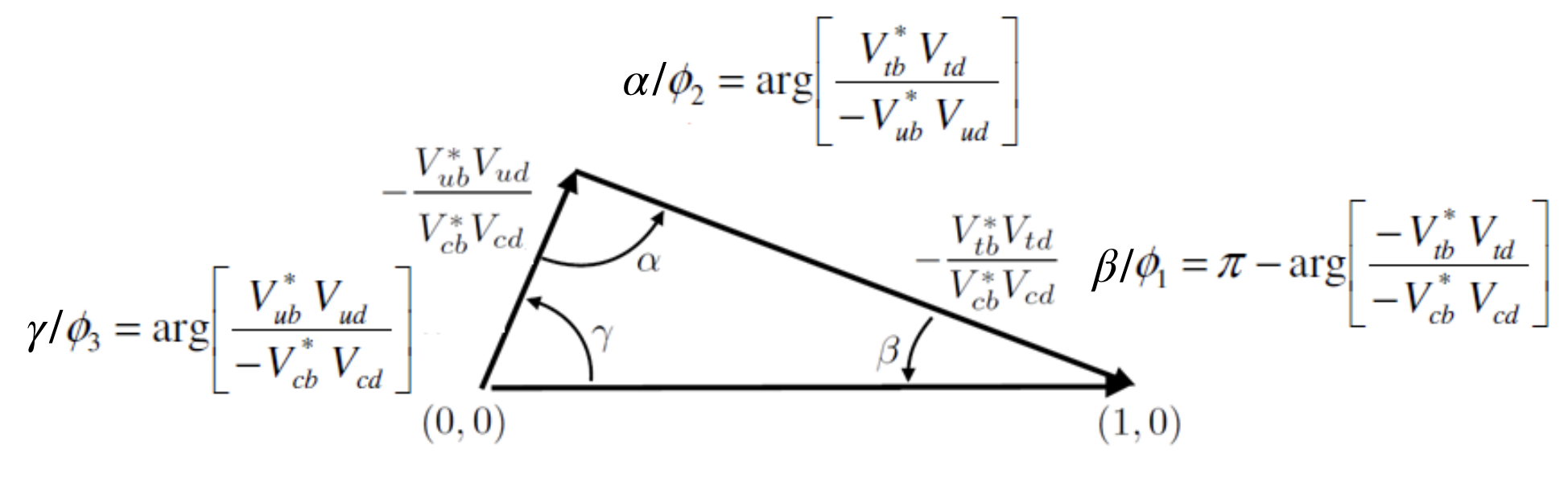}
	\caption{Graphical representation of the Unitarity Triangle \cite{Manfredi}.}
	\label{fig: unitriangle}
\end{figure}

The unitarity condition $V_{\textrm CKM} V_{\textrm CKM}^{\dagger} = \mathbbm{1}$ yields nine relations,
\begin{align*}
	|V_{ud}|^2 + |V_{cd}|^2 + |V_{td}|^2 &= 1 &V^*_{us}V_{ud} + V^*_{cs}V_{cd}+ V^*_{ts}V_{td} &= 0 &V_{ud}V^*_{cd} + V_{us}V^*_{cs} + V_{ub}V^*_{cb} &= 0\mbox{ ,}\\
	|V_{us}|^2 + |V_{cs}|^2 + |V_{ts}|^2 &= 1 &V^*_{ub}V_{ud} + V^*_{cb}V_{cd}+ V^*_{tb}V_{td} &= 0 &V_{ud}V^*_{td} + V_{us}V^*_{ts} + V_{ub}V^*_{tb} &= 0\mbox{ ,}\\
	|V_{ub}|^2 + |V_{cb}|^2 + |V_{tb}|^2 &= 1 &V^*_{ub}V_{us} + V^*_{cb}V_{cs}+ V^*_{tb}V_{ts} &= 0 &V_{cd}V^*_{td} + V_{cs}V^*_{ts} + V_{cb}V^*_{tb} &= 0\mbox{ ,}
\end{align*}

\noindent which are sums of three complex numbers each. The six equations summing to zero prompt a convenient geometric representation in the complex plane in terms of so-called \textit{unitarity triangles}. A ${\textit CP}$ conserving theory would yield null-area triangles or, equivalently, a vanishing Jarlskog invariant $J=\Im(V_{us}V_{cb}V_{ub}^*V_{cs}^*)$~\cite{Jarlskog:1985ht,Jarlskog:1985cw}. All elements of the second equation in the second row have similar magnitudes, yielding a notable  triangle referred to as `the Unitarity Triangle', shown in Fig.~\ref{fig: unitriangle}. Conventionally, side sizes are normalized to the length of the base, and the three angles are labelled $\alpha$ or $\phi_2$, $\beta$ or $\phi_1$, and $\gamma$ or $\phi_3$.

The flavour-mixing phenomenon, which involves flavoured neutral mesons $|M \rangle$, enriches significantly the {\textit CP} violation phenomenology. Flavour quantum numbers are conserved in strong interactions and thus flavour eigenstates are eigenstates of strong interactions. Weak interactions do not conserve flavour, allowing $|M\rangle$ to undergo a transition into $|\overline{M}\rangle$ (or vice versa), which changes flavour by two units. Because the full Hamiltonian contains strong and weak interactions, its eigenstates (which are the particles we observe, with definite masses and lifetimes) are linear superpositions of flavour eigenstates $|M\rangle$ and $|\overline{M}\rangle$.

\begin{figure}[htb]
	\centering
	\includegraphics[width=0.65\textwidth]{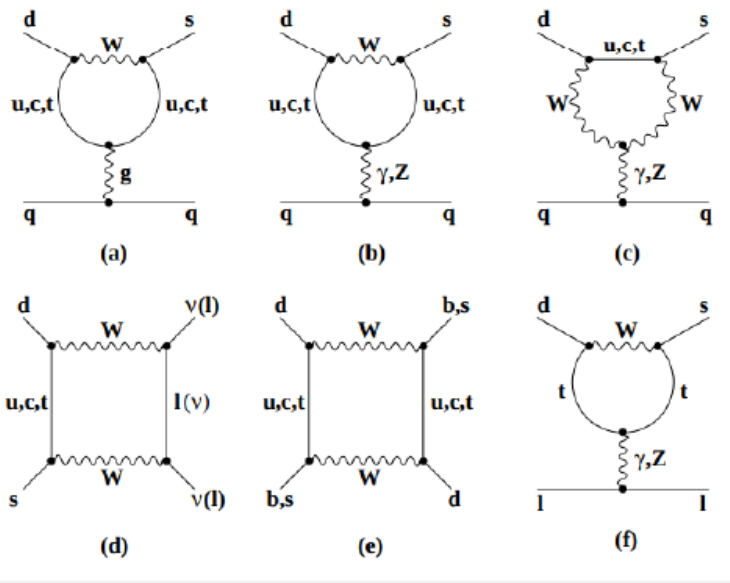}
	\caption{Examples of leading FCNC diagrams \cite{Manfredi}.}
	\label{fig:FCNC}
\end{figure}

\section{Flavour physics to overcome the Standard Model}
Many physicists find the current understanding of flavour dynamics unsatisfactory. The observed hierarchies between quark masses and couplings seem too regular to be accidental and the abundance of free parameters (six quark masses and four couplings) suggests the possibility of a deeper, more fundamental theory possibly based on a reduced set of parameters. In addition, while the CKM mechanism offers a framework to include ${\textit CP}$ violation in the Standard Model, it does not really enlighten the origin for such a singular phenomenon.

But even in the absence of a deeper understanding of the origin of {\textit CP} violation, naturalness arguments indicate that most generic extensions of the Standard Model would involve additional sources of {\textit CP} violation. These and other considerations support the notion that a more detailed and complete study of the phenomenology of quarks dynamics may reveal useful information to guide searches for SM extensions.

The abundance and diversity of experimentally accessible processes to measure redundantly a reduced set of parameters makes indirect searches in the flavour sector  a powerful option for exploring non-SM dynamics. In fact, even if no deviations from the Standard Model will be found, the resulting stringent constraints on SM extensions are expected to remain useful in informing future searches. 

The two classes of flavour-physics processes most promising for probing contributions of non-SM particles are \textit{flavour-changing-neutral-currents} and ${\textit CP}$-\textit{violating} processes. 

Flavour-changing neutral currents (FCNC) are processes in which quark flavour changes in the transition, but quark electric charge does not. The processes are suppressed in the Standard Model, because they occur only through higher-order amplitudes involving the internal exchange of $W^\pm$ bosons (`loop amplitudes'), as shown in Fig.~\ref{fig:FCNC}. 
Such amplitudes are naturally sensitive to non-SM contributions, since any particle with proper quantum numbers and nearly arbitrary mass can replace the SM-quark closed-line in these diagrams thus altering the rate. Hence, FCNC are powerful in signaling contributions from non-SM particles if rate enhancements, or suppressions, with respect to Standard Model expectations are observed.
In addition to rate alterations, the phenomenon of ${\textit CP}$ violation offer additional avenues to uncover or characterize possible non-SM contributions. Alterations of the ${\textit CP}$-violating phases with respect to those predicted by the SM are generically expected in a broad class of SM-extensions. Observing experimental evidence of those phases offers further opportunities to explore the dynamics, even if total rates are unaffected.

\chapter{The Belle II experiment at the SuperKEKB collider}\label{chap:detector}

\section{The SuperKEKB collider}\label{sec:SuperKEKB}

SuperKEKB is an electron-positron ($e^+e^-$) energy-asymmetric collider, designed to produce more than 600 $B\overline{B}$ pairs per second ($B^0\overline{B}^0$ and $B^+B^-$ in approximately equal proportions) via decays of $\Upsilon(4{\textrm S})$ mesons produced at threshold~\cite{acc_design}. Such colliders are called `$B$-factories', and were proposed in the late 1980's for the dedicated exploration of ${\textit CP}$ violation in $B$ mesons. The main goal of $B$-factories is to produce low-background quantum-correlated $B\overline{B}$ pairs at high rates and with sufficient boost to study their time evolution.

\begin{figure}[h]
	\centering
	\includegraphics[width=0.6\textwidth]{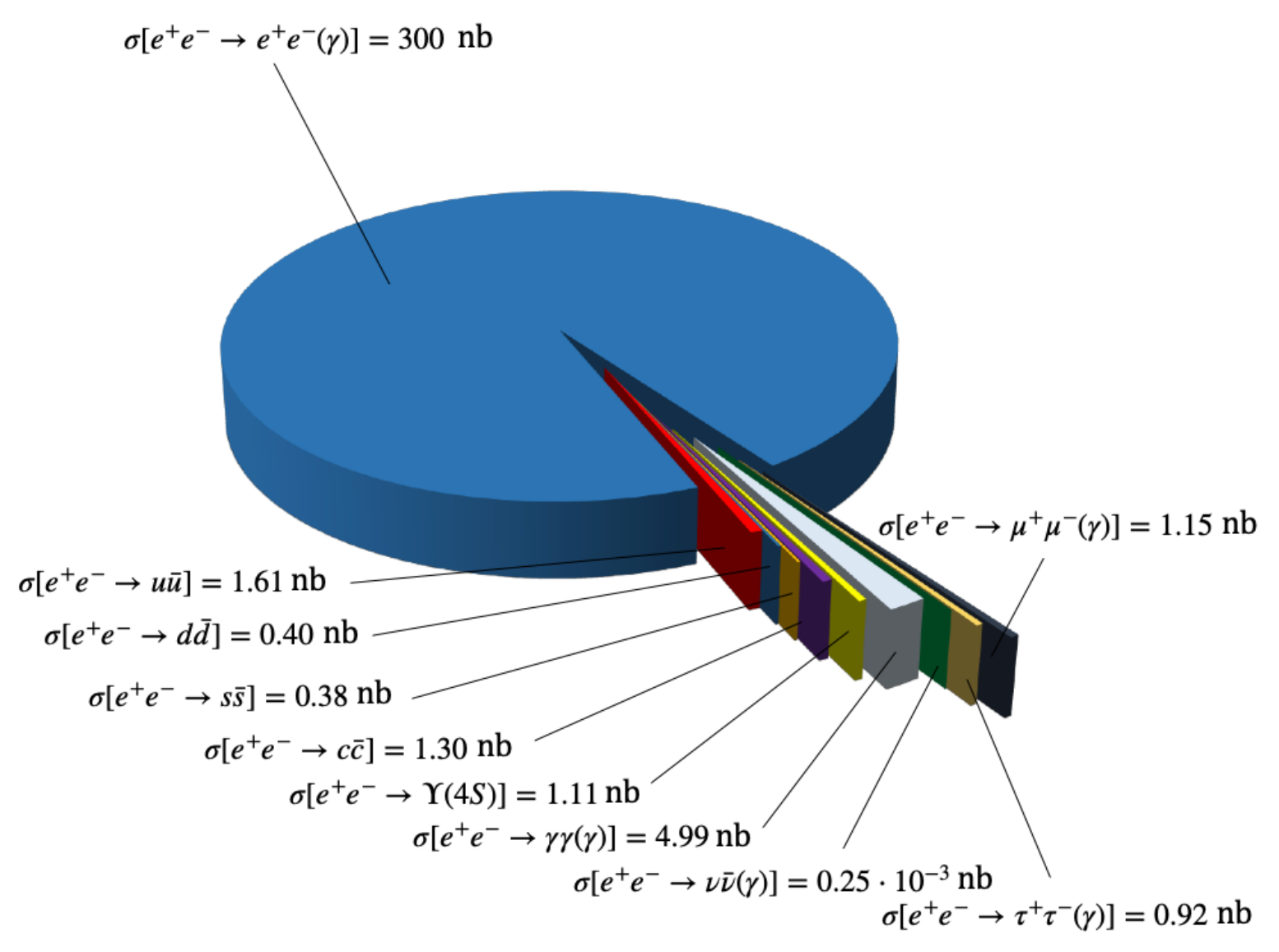}
	\caption{Pie chart of the cross sections for the main processes produced in $e^+e^-$ collision at the $\Upsilon(4{\textrm S})$ centre-of-mass energy \cite{Manfredi}.}
	\label{fig:Cross_Sections_Upsilon}
\end{figure}

\begin{figure}[h]
	\centering
	\includegraphics[width=11cm]{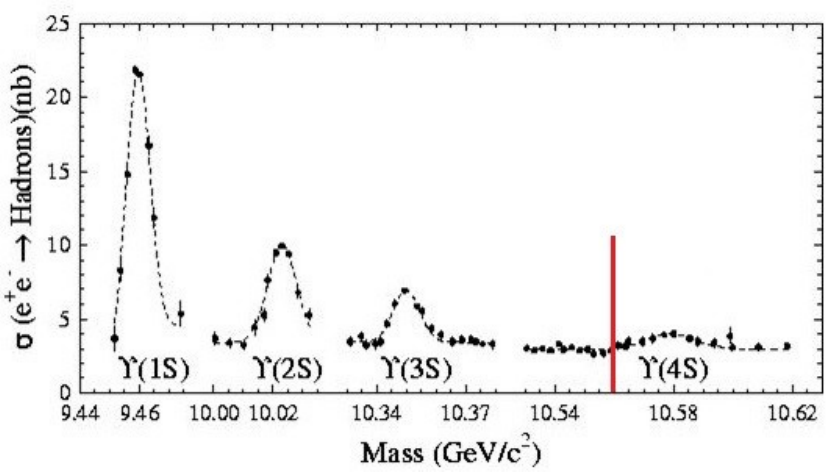}
	\caption{Hadron production cross section from $e^+e^-$ collisions as a function of the final-state mass. The vertical red line indicates the $B\overline{B}$ production threshold~\cite{10.1093/ptep/pts083}.}
	\label{fig: upsilon}
\end{figure}

Intense beams of electrons and positrons are brought to collision at the energy corresponding to the $\Upsilon(4{\textrm S})$ meson mass, $\SI{10.58}{GeV}$, which is just above the $B\overline{B}$ production kinematic threshold. The great majority of collisions yield electromagnetic processes as $e^+e^-\to e^+e^-$, $e^+e^-\to \mu^+\mu^-$, $e^+e^-\to \tau^+\tau^-$, $e^+e^-\to \gamma \gamma$ etc. (see Fig.~\ref{fig:Cross_Sections_Upsilon}); the rest are collisions that produce hadrons. 
Figure~\ref{fig: upsilon} shows the hadron-production cross-section as a function of the final-state mass. The various peaks are radial excitations of the $\Upsilon$ meson. They overlap the nearly uniform background at about $\SI{4}{\nano\barn}$ from so-called continuum of lighter-quark pair-production from the process $e^+e^-\rightarrow q\overline{q}$, where $q$ identifies a $u$, $d$, $c$, or $s$ quark. These are useful for charm physics, some selected topics in hadron physics, and as control channels. 
The rest are $\Upsilon(4{\textrm S})$ events, which decay to $B\overline{B}$ pairs more than $96\%$ of the time. At-threshold production implies little available energy to produce additional particles, resulting in low-background. 
In addition, colliding beams of point-like particles imply that the collision energy is precisely known, which sets stringent constraints on the collision's kinematic properties, thus offering means of further background suppression. Since bottom mesons are produced in a strong-interaction decay, flavour is conserved, and the null net bottom content of the initial state implies production of a flavourless $B\overline{B}$ pair. Even though $B^0$ and $\overline{B}^0$ undergo flavour oscillations before decaying, their time-evolution is quantum-correlated in such a way that no $B^0B^0$ or $\overline{B}^0\overline{B}^0$ pairs are present at any time before either decays. Angular-momentum conservation implies that the decay of the $\Upsilon(4S)$, which is spin-1, in the two spin-0 bottom mesons yields total angular momentum $J=1$. Because the simultaneous presence of two identical particles in an antisymmetric state would violate Bose statistics, the system evolves coherently as an oscillating $B^0\overline{B}^0$ particle-antiparticle pair until either one decays. This allows efficient identification of the bottom (or antibottom) content of one meson at the time of decay of the other, if the latter decays in a final state accessible only by either bottom or antibottom states. This important capability is called `flavour tagging' and allows measurements of flavour-dependent decay rates, as needed in many determinations of ${\textit CP}$-violating quantities.

\begin{figure}[htb]
 \centering
 \includegraphics[width=0.6\textwidth]{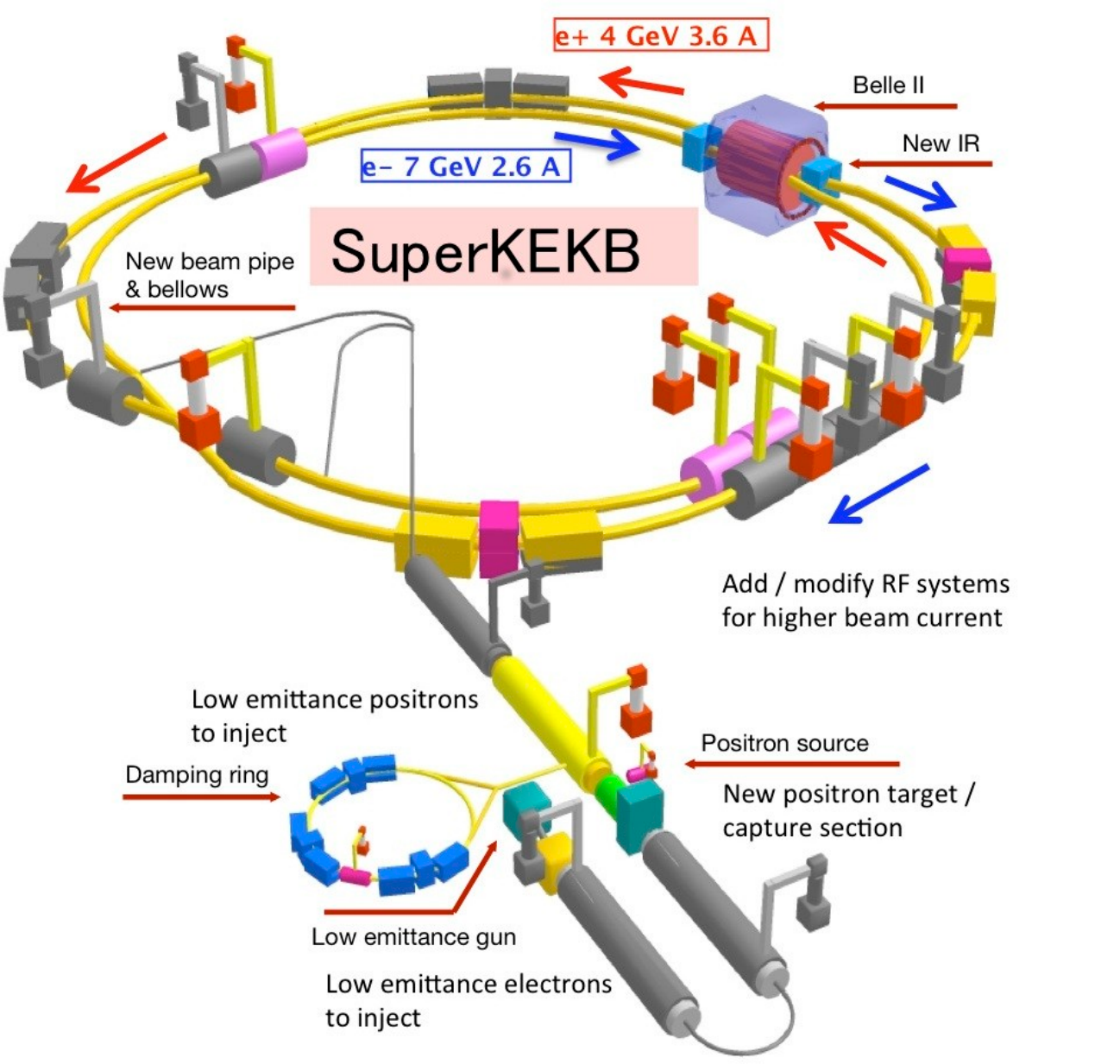}
 \caption{Illustration of the SuperKEKB collider \cite{Eidelman}.}
 \label{fig:kekb}
\end{figure}

Because the $\Upsilon(4{\textrm S})$ mesons are produced at threshold, in an energy-symmetric collider they would be nearly at rest in the laboratory. The resulting $B$ mesons too would be produced with low momentum (about $\SI{10}{MeV}$/$c$) in the laboratory, because of the $\SI{21}{\mega\electronvolt}/c^2$ difference between the $\Upsilon(4{\textrm S})$ mass and the $B\overline{B}$ pair mass. With such low momenta the $B$ would only travel approximately $\SI{1}{\mu m}$ before decaying, rendering the $10\ \mu$m typical spatial resolution of vertex detectors insufficient to separate their decay vertices and study the decay-time evolution. Asymmetric beam energies are used to circumvent this limitation. By boosting the collision centre-of-mass along the beam in the laboratory frame, $B$-decay vertex become resolvable with current vertex detectors~\cite{Oddone}. SuperKEKB (Fig.~\ref{fig:kekb}) implements a 7-on-4 GeV energy-asymmetric double-ring design, which achieves a vertex displacement of about $\SI{130}{\mu m}$. 

\begin{figure}[tb]
	\centering
	\includegraphics[width=0.9\textwidth]{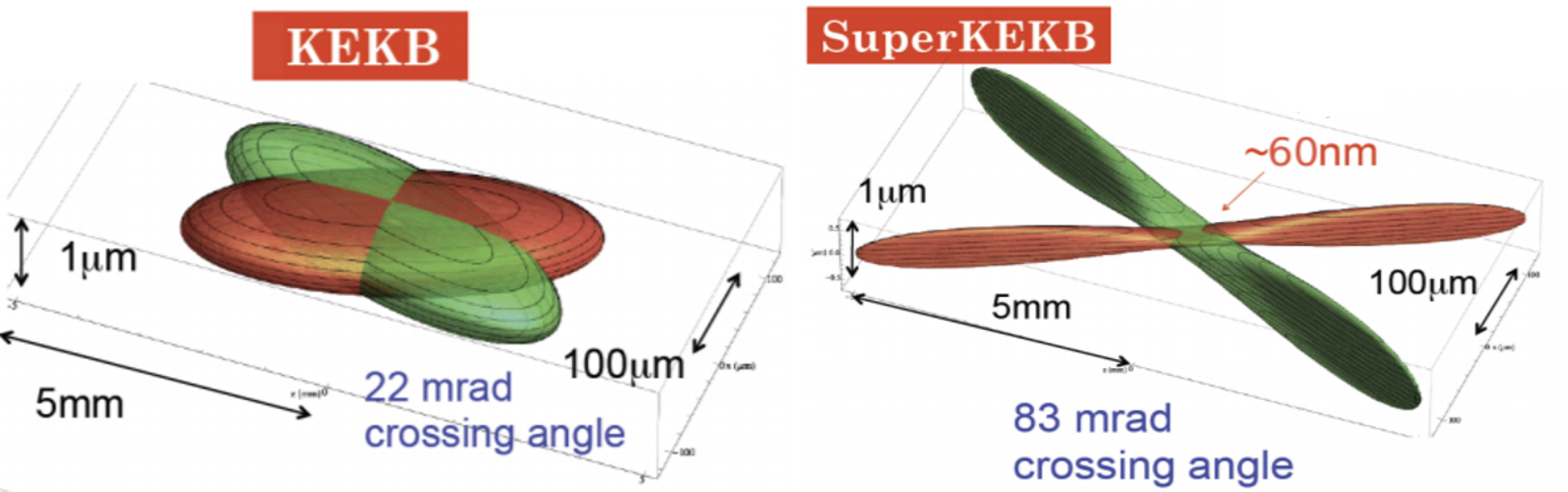}
	\caption{Two-dimensional sketch of the nano-beam mechanism implemented in SuperKEKB (right) compared with the previous KEKB collision scheme (left) \cite{10.1093/ptep/pts083}.}
	\label{fig: NanoBeam}
\end{figure}

Electrons are produced in a thermionic gun with a barium-impregnated tungsten cathode, then accelerated to 7 GeV with a linear accelerator (linac) and injected in the high-energy ring. Positrons are produced by colliding electrons on a tungsten target, then isolated by a magnetic field, accelerated to 4 GeV with the linac and injected in the low-energy ring.
The electrons and positrons continuously collide at a single interaction point, surrounded by the Belle~II detector. To achieve high collision intensities, a nano-beam, large crossing-angle collision scheme is implemented \cite{nanobeam}. This is an innovative configuration based on keeping small horizontal and vertical emittance, which is a measure of the spread and size of the particle beam in position and momentum, phase-space and large crossing angle, as shown in Fig.~\ref{fig: NanoBeam}. Such configuration is obtained with low emittance beams, in addition to a sophisticated final-focus superconducting-quadrupole-magnet system, that include also corrector coils and compensation solenoids installed at each longitudinal end of the interaction region. Conceptually the nano-beam scheme mimics a collision, in a spatial region 20 times smaller than in conventional approaches, of many short micro-bunches. This provides an order-of-magnitude increase in intensity with respect to previous conventional schemes. The penalties are significant challenges in achieving the design performance and operating steadily, and higher beam-induced backgrounds.

The performance of the SuperKEKB collider is mainly characterized in terms of the instantaneous luminosity $\mathcal{L}$, which is a measure of collision intensity, 
\begin{center}
    $\mathcal{L}=\frac{\gamma_{\pm}}{2er_e}\left(1+\frac{\sigma^*_y}{\sigma^*_x}\right)\frac{I_{\pm}\xi_{y\pm}}{\beta^*_{y\pm}}\cdot\frac{R_{\mathcal{L}}}{R_{\xi_y}}$, \\
\end{center}
where $\gamma$ is the relativistic Lorentz factor, $e$ is the absolute value of the electron charge, $r_e$ is the classical radius of the electron, $\sigma^*_x$ and $\sigma^*_y$ are the bunch widths at the interaction point (IP) in the plane orthogonal to the beam direction (transverse plane), $I$ is the current of the beam, $\beta^*_{y}$ is the vertical betatron function at the IP~\cite{betatronFunctionParameterization}, $\xi_y$ is the vertical beam-beam parameter, $R_{\mathcal{L}}$ and $R_{\xi_y}$ are the luminosity-reduction factors and the vertical beam-beam parameter, respectively, due to non-vanishing crossing angle~\cite{verduAndresLuminosity}. The ratio of these reduction factors is close to unity. Design values for the other parameters are reported in table~\ref{tab:SKBparams}.
The rate of any given process

\begin{center}
rate [events s$^{-1}$] = $\mathcal{L}$ [cm$^{-2}$ s$^{-1}$] $\times$ $\sigma$ [cm$^2$], \\
\end{center}
is the product of its cross-section and $\mathcal{L}$. The integral of instantaneous luminosity over time $T$, called integrated luminosity,

\begin{center}
$\mathcal{L}_{\textrm int}$ = $\int^{T}_{0}\mathcal{L}(t^\prime)dt^\prime$ \\
\end{center}
is a measure of the number of produced events of interest $N =\mathcal{L}_{\textrm int}$$\sigma$.

\begin{table}[!htb]
    \centering
    \begin{tabular}{l c c}
    \hline\hline 
          & Design & Achieved (as of mid 2025) \\
          \hline
        Energy [GeV] & 4.0/7.0  & 4.0/7.0\\
        $\xi_y$ & 0.090/0.088  & 0.0407/0.0279\\
        $\beta^*_{y}$ [mm] & 0.27/0.41  & 1.0/1.0\\
        $I$ [A] & 3.6/2.62  & 1.321/1.099\\
          \hline\hline
    \end{tabular}
    \caption{Design and achieved values for SuperKEKB fundamental parameters for high/low energy rings.}
    \label{tab:SKBparams}
\end{table}

Physics data-taking started in March 2019, and Belle II has integrated $531 \, {\textrm{fb} }^{-1}$ 
of luminosity at the time of this writing. In 2024, SuperKEKB achieved the instantaneous-luminosity world record, 5.1$\times$10$^{34}$ cm$^{-2}$s$^{-1}$. In spite of these achievements, a number of technological and scientific challenges have significantly reduced SuperKEKB performance compared to design. Issues associated with unexplained sudden total beam losses, beam injection, collimation, and short beams lifetime which also causes high beam backgrounds, limited the capability to deliver the expected samples of data in its first six years. Consolidation, improvement and development work is ongoing to overcome these difficulties.

\section{The Belle~II detector}

Belle~II (Fig.~\ref{fig: belle2}) is a large-solid-angle, multipurpose magnetic spectrometer surrounded by a calorimeter and particle-identification systems, installed around the SuperKEKB interaction point. It is designed to determine energy, momentum, and identity of a broad range of particles produced in 10.58 GeV $e^+e^-$ collisions. Belle~II is approximately a cylinder of about 7 m in length and 7 m in diameter that comprises several subsystems, each dedicated to one or few specific aspects of event reconstruction. From the interaction point outward, a particle would traverse the beam pipe, a two-layer silicon-pixel vertex-detector, a four-layer silicon-strip vertex-detector, a central wire drift-chamber, a time-of-propagation central Cherenkov counter or an aerogel threshold forward Cherenkov counter, an array of CsI(Tl) crystals, a superconducting solenoidal magnet, and multiple layers of resistive plate counters. 

The principal experimental strengths are hermetic coverage, which allows for reconstruction of final states involving neutrinos; efficient and precise reconstruction of charged-particle trajectories (tracks), which provide accurately reconstructed decay-vertices and good momentum resolution; high-purity charged-particle identification and neutral-particle reconstruction. A detailed description of Belle~II and its performance is in Ref.~\cite{TDR}. In the following, I focus on the electromagnetic calorimeter, which is the subdetector more relevant for the purpose of the study reported in this thesis.

\subsection{Charged-particle reconstruction}
 
At Belle~II, reconstruction of charged particles and measurement of their momenta and charges is achieved through an integrated system consisting of six layers of silicon sensors and a drift chamber, surrounding the beam pipe and immersed in a 1.5~T axial magnetic field maintained in a cylindrical volume 3.4~m in diameter and 4.4~m in length. The field is oriented along the longitudinal direction and is provided by an aluminium-stabilized superconducting solenoid made of NbTi/Cu alloy. 
The solenoid surrounds all the subdetectors except for the outer muon detectors. An iron yoke serves as the return path of the magnetic flux. 
Charged particles are reconstructed with  greater than 94\% efficiency in the polar range from 17$^\circ$ to 150$^\circ$, achieving a transverse momentum resolution of 0.5\% p$_T$ (GeV/c) and a typical impact-parameter resolution of $15\ \si{\mu m}$.

\subsection{Charged-particle identification}
Belle~II combines measurements of time-of-propagation, Cherenkov radiation, and ionization-energy loss in the silicon and drift chamber to identify charged particles.
In particular, it is capable of separating kaons from pions with an identification efficiency of 94\% and a misidentification rate of less than 4\%~\cite{Sandilya:2016rpm}.

\subsection{Electromagnetic calorimeter}    \label{sec:ECL}

The electromagnetic calorimeter (ECL) is a highly segmented array of 8736 cesium-iodide crystals doped with thallium (CsI(Tl)) that measures the energy of photons and electrons~\cite{Aulchenko:2015}. High energy photons and electrons entering the crystals initiate an electromagnetic shower through bremsstrahlung and electron-positron pair production. The energy is mostly converted to photons, which are collected by the photodiodes. In contrast to hadrons, which pass through the calorimeter with minimal energy loss, most photons and electrons dissipate their entire energy.

\begin{figure}[!htb]
 \centering
 \includegraphics[width=1.0\textwidth]{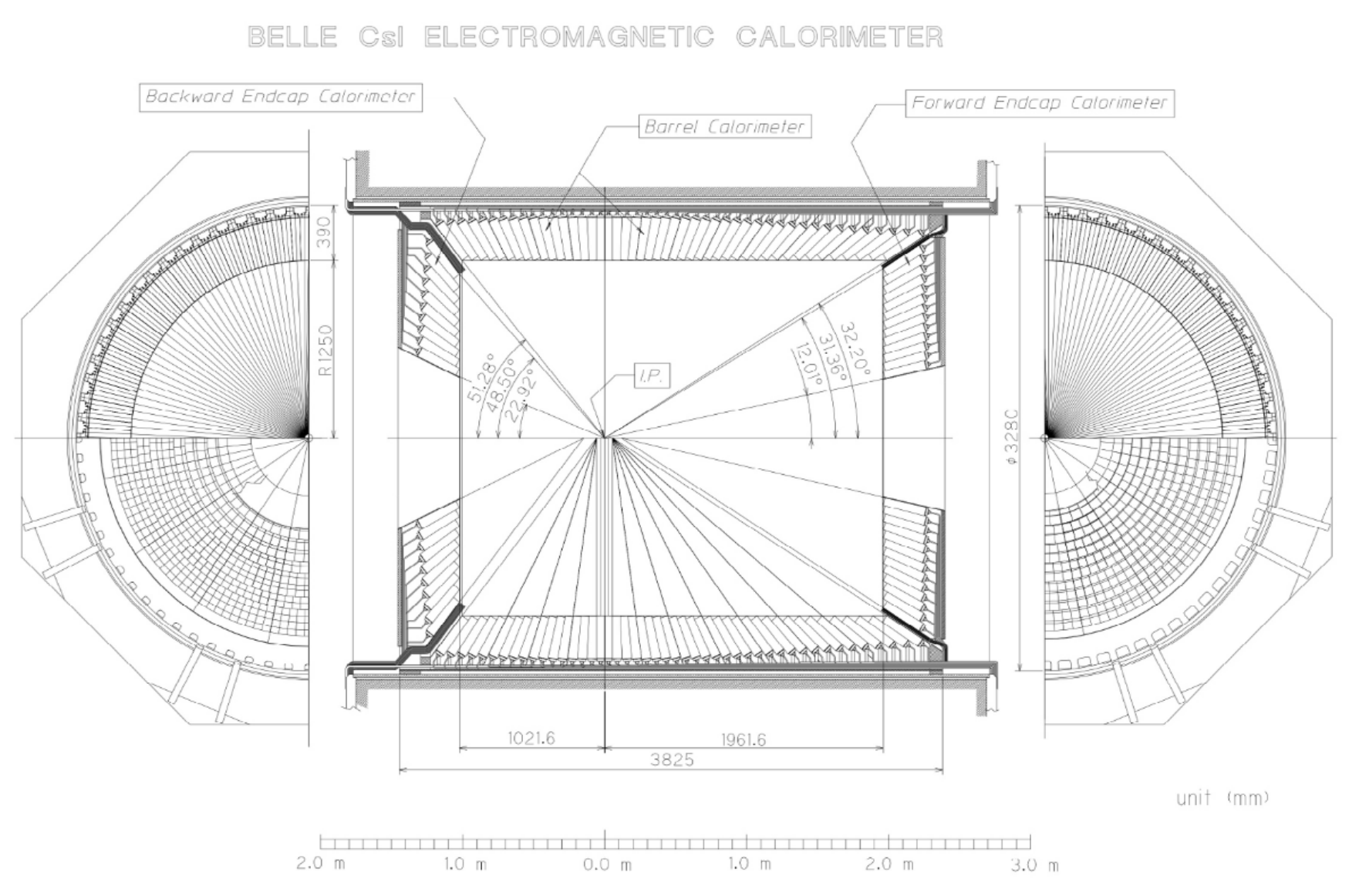}
 \caption{ECL layout \cite{adachi2018detectors}.}
 \label{fig:ecl}
\end{figure}

\begin{figure}[htb]
 \centering
 \includegraphics[width=0.4\textwidth]{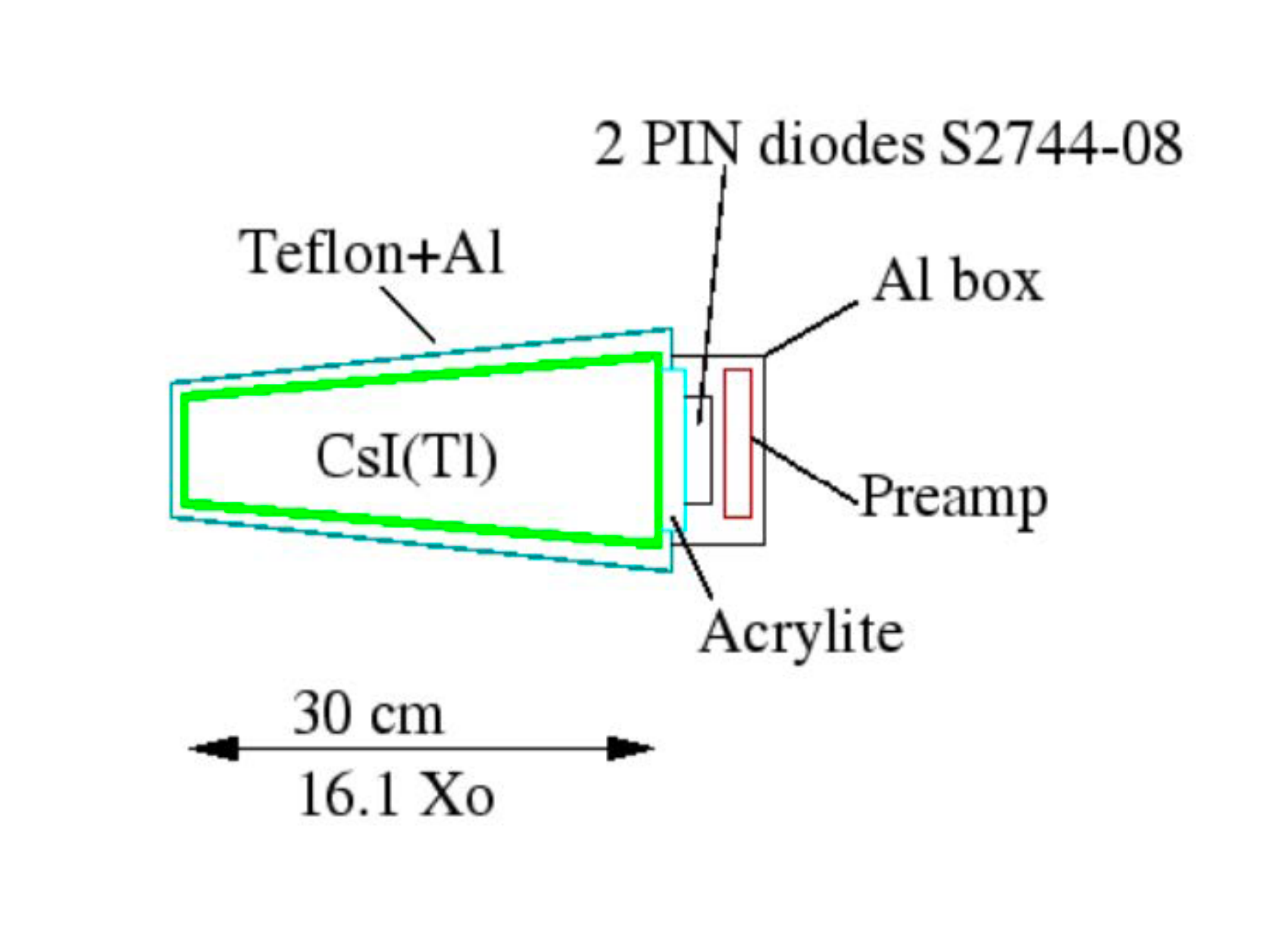}
 \caption{Schematic design of a CsI(Tl) crystal with attached readout electronic circuits \cite{milesi}.}
 \label{fig:eclF1}
\end{figure}

The ECL consists of three polar compartments (Fig.~\ref{fig:ecl}): the barrel, the forward endcap, and the backward endcap section. The barrel section is 3.0 m long with 1.25 m of inner radius; the endcaps are located at longitudinal distances $z$ = +2.0 m (forward) and $-$1.0 m (backward) from the interaction point. A typical crystal in the barrel section has a 55$\times$55 mm$^2$ active surface on the front face and 65$\times$65 mm$^2$ on the rear face; the dimensions of the crystals in the endcap sections vary from 44.5 to 70.8 mm and from 54 to 82 mm for front and rear faces, respectively. A scheme of an ECL crystal is shown in Fig.~\ref{fig:eclF1}. The 30-cm crystal length, corresponding to 16.1$X_0$, reduces the fluctuations of shower leakages out of the outermost end of the crystals, which spoils energy resolution. The crystals are designed in such a way that a photon impinging the centre of the crystal would deposit 80$\%$ of its energy in the crystal on average.


\begin{figure}[h]
	\centering
	\includegraphics[width=1.33\textwidth, angle=90]{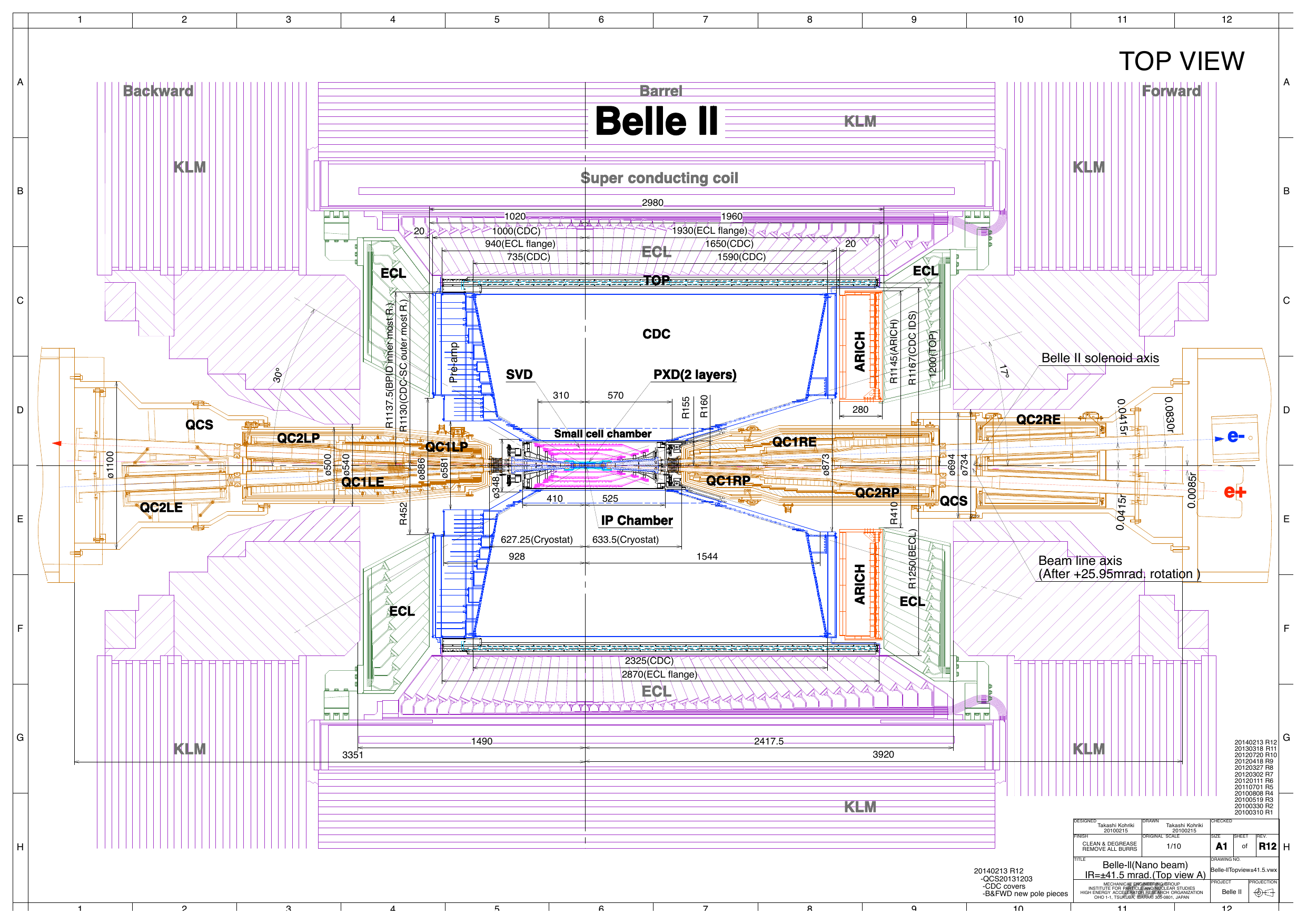}
	\caption{Top view of Belle II, the beam pipe at IP and final-focus magnets \cite{adachi2018detectors}.}
	\label{fig: belle2}
\end{figure}

\clearpage

The principal axes of the crystals do not point exactly to the nominal interaction point, but they are inclined to prevent photons from escaping through gaps by about 1.3$^\circ$ in the $\theta$ and $\phi$ directions in the barrel section, and by about 1.5$^\circ$ and about 4$^\circ$ in the $\theta$ direction in the forward and backward sections. 
Thallium inside the CsI, shifts the energy of the excitation light into the visible spectrum. The light is detected by a independent pair of silicon PIN photodiodes~\cite{Aulchenko:2015} whose signals are fed to charge-sensitive preamplifiers installed at the outer end of each crystal. 
The fraction of photons that do not leave a detectable signal in the calorimeter is only 0.2\%. The typical energy resolution is $\sigma_E/E \approx 3\%/\sqrt{E}$.
The ECL also uses Bhabha scattering to measure luminosity. Because the Bhabha cross section is predicted accurately in QED, a precise inference of luminosity is achieved from the observed rate of Bhabha events in an instrumented volume of known acceptance.

\section{Beam-induced backgrounds}     \label{sec:beam_back}
\begin{figure}[H]
	\centering
	\includegraphics[width=\textwidth]{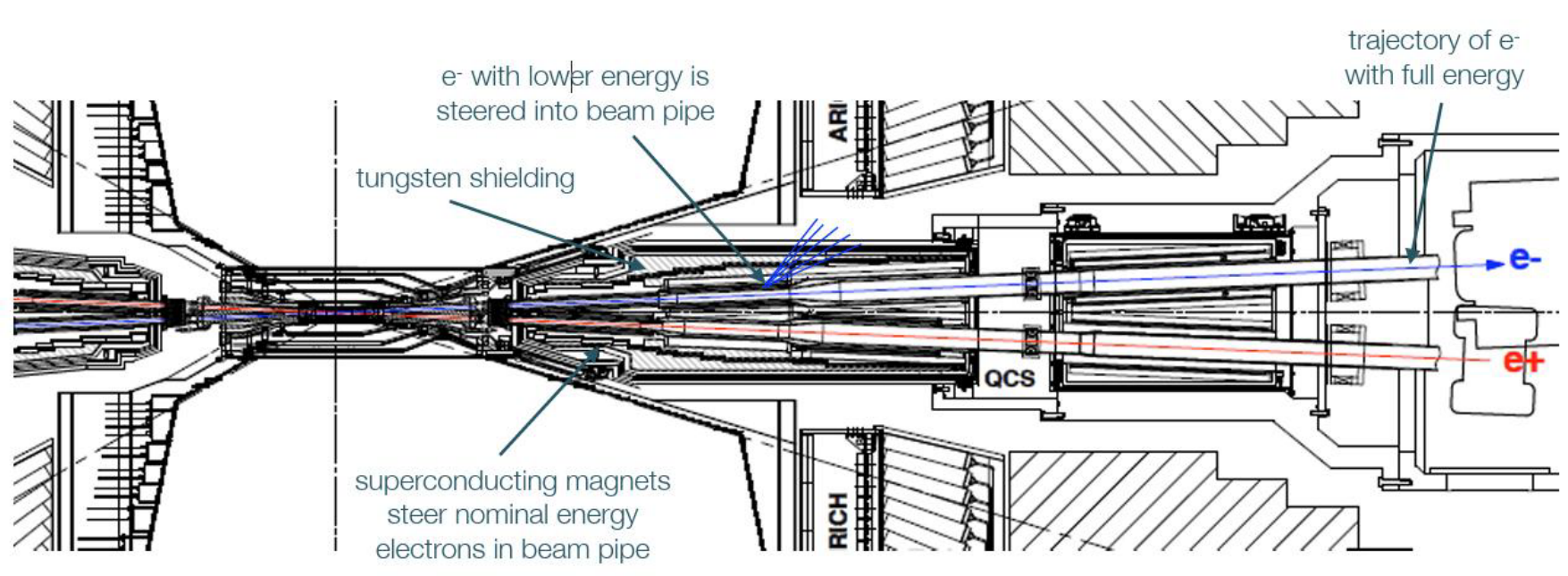}
	\caption{Sketch of an off-momentum electron steered into the beam pipe by the           final focussing magnets, producing background through interactions with the surrounding material~\cite{Hearty}.}
	\label{fig:beambackground}
\end{figure}

The calorimeter is central in the Belle II physics program, as its hermeticity and capability to precisely reconstruct photons and electrons enable unique key measurements.  
However, its proximity to the beams, large acceptance, and technology make it sensitive to beam-background photons.  
These are energy deposits in the calorimeter crystals that do not come from the $e^+e^-$ collision.  
They are associated with the harsh and tightly constrained collision environment required by the nano-beam scheme. They can originate from single-beam processes such as interactions with residual gas in the vacuum, accelerator infrastructure, or interactions between neighbouring electrons in the same beam. They are mostly photons of 1-2 MeV energy, but they can reach sufficient energies to be reconstructed as calorimeter clusters~\cite{Liptak:2021tog}.
Figure \ref{fig:beambackground} shows a simplified example of an electron beam that has fluctuated to have slightly less energy than nominal and is steered towards the beam pipe by the focusing magnets close to the interaction point. This makes the beam to interact with the infrastructure material and generates background.

Beam-background photons complicate calorimeter clustering and bias the reconstructed energy of collision products. This degrades the reconstruction of photons or neutrinos, preventing Belle~II from reaching its full physics potential. Furthermore, beam-backgrounds are highly correlated with the instantaneous luminosity. It is therefore expected that they will increase further as SuperKEKB will keep approaching its design intensity goals.
Mitigating these backgrounds and their effects is essential.

\begin{figure}[htp]
	\centering
	\includegraphics[width=0.5\textwidth]{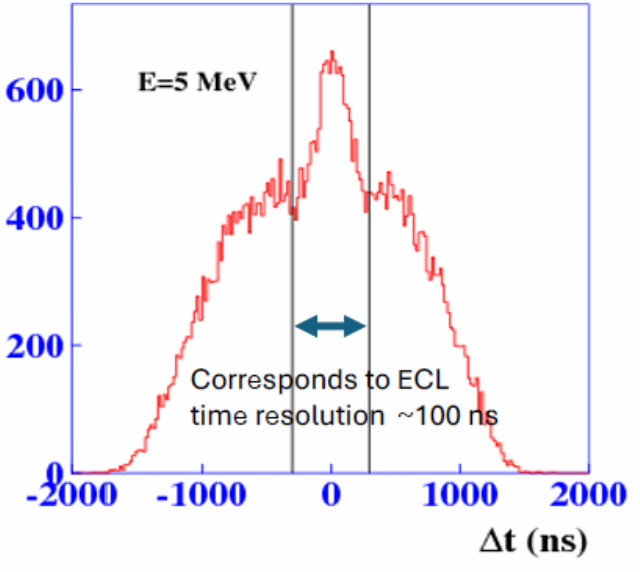}
	\caption{Calorimeter-cluster time distribution for 5 MeV energy deposition~\cite{TDR}.}
	\label{fig:ECLcluster}
\end{figure}

Because the collision time is precisely known from the SuperKEKB radio-frequency, one promising strategy to cope with beam-backgrounds is to exploit detection-time to recognise photons not originated from the collision, which are preferentially out-of-time. Figure~\ref{fig:ECLcluster} shows the time-of-flight distribution of reconstructed photons. The cluster of events around 0.0 ns is in time with the collision and has roughly 100 ns width, which is the current calorimeter timing resolution~\cite{TDR}. The larger, broader component of photons detected with larger time differences with respect to the collision are beam-background photons. The current calorimeter time resolution already helps suppressing 80\% of them, but the 20\% that is still accepted is a significant nuisance for reconstruction of rare signals. Improving the calorimeter time resolution would allow for better event reconstruction removing an even larger fraction of beam-backgrounds.

\section{An upgrade proposal}
Detector upgrades tentatively planned for 2030 to mitigate the impact of beam backgrounds are discussed within the collaboration.  The group I joined for my thesis proposes an upgrade of the calorimeter based on improved time information.

The main requirements of such an upgrade are high time resolution, for precise timing determination, along with high efficiency and large area, so as to cover the maximum possible inner surface of the calorimeter (Fig.~\ref{fig: belle2}). Because the inner calorimeter surface is about 30 m$^2$, the chosen technology must be cost-effective.

Many photodetectors with the desired time resolution and efficiency are available, but their significant cost and difficulties in scalability make them hardly suitable for our purpose.
The best single-photon time resolution achieved with silicon devices is 7.8~ps (FWHM), obtained with a small (20~$\mu$m diameter) avalanche diode integrated with advanced CMOS circuitry~\cite{instruments2040019}. However, such designs are hardly scalable to large surfaces. For solid-state photodetectors, the large active area increases capacitance, leading to higher noise and therefore a degradation of time resolution. Vacuum-based photomultipliers suffer from transit-time spread, which worsens with size. Microchannel plate photomultipliers (MCP-PMT) currently provide the best compromise, with time resolutions of about 30~ps and planar geometries suitable for tiling~\cite{matsuoka2019performance,Conneely_2015}. However, their large-scale deployment is limited by significant cost. Superconducting nanowire single-photon detectors~\cite{10.1063/1.1388868} push the time resolution even further, down to 2.6~ps~\cite{Korzh:2018oqv}, but their scalability remains prohibitive. 
Photosensitive gaseous detectors offer a promising opportunity as they can be scaled to large areas at relatively low cost. Devices using CsI photocathodes have already been widely deployed, and recent developments, such as the PICOSEC-Micromegas detector, demonstrated a time resolution of 44 ps for single photons~\cite{Sohl_2020}.
Another possible option, which is being  developed, are large-area picosecond photodetectors, a cost-efficient solution based on the MCP-PMT technology~\cite{Shin:2022ybc}.

Our approach is based on a novel gaseous photodetector, the GasPM, featuring a photocathode coupled with the avalanche electron-multiplication principle of resistive plate chambers.
It aims at measuring photon timing with a resolution of 20 ps or better at 90\% efficiency. This would allow identification of off-time photons from beam-backgrounds and would reduce background calorimeter clusters to 10\%, according to simulations.

In addition, the GasPM can also be used as a Cherenkov detector to perform charged-particle identification by measuring time-of-flight with high resolution.

\chapter{The gaseous photomultiplier}  \label{chap:GasPM}

This chapter introduces the GasPM, the novel gaseous photodetector concept which is studied and developed in this work. Design details and performance are discussed.

\section{Design and technology}
\begin{figure}[htb]
	\centering
	\includegraphics[width=0.5\textwidth]{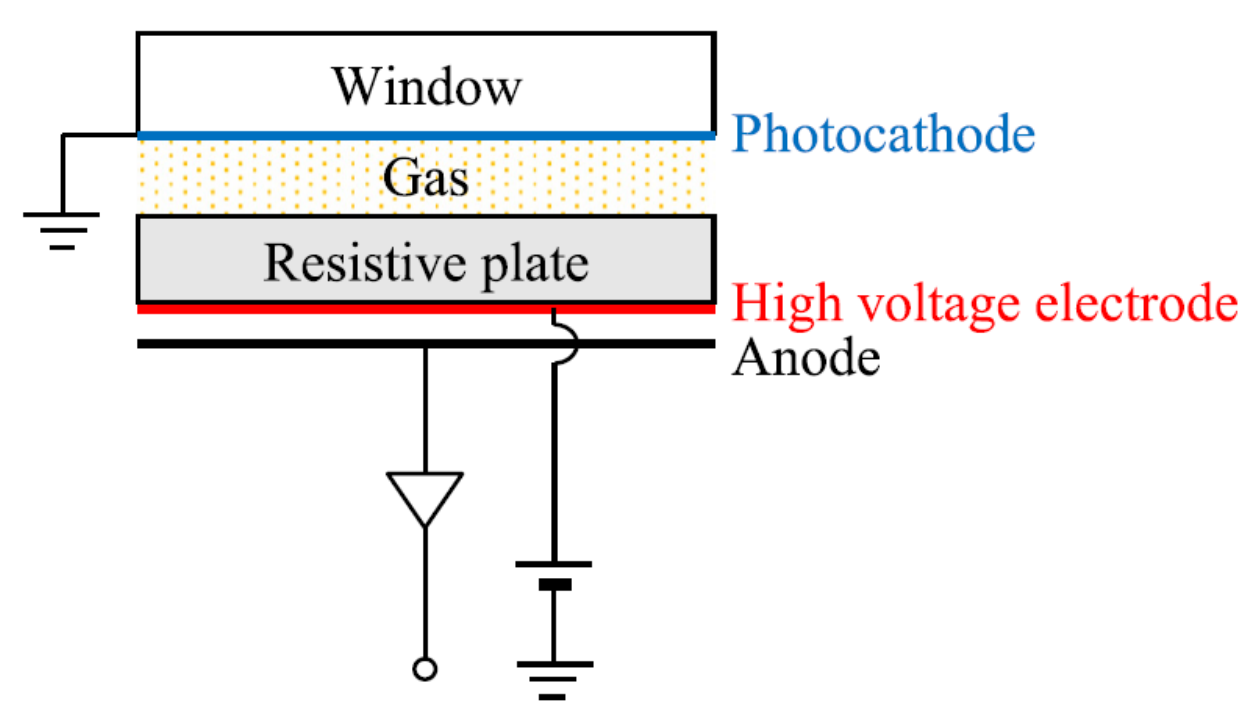}
	\caption{Schematic design of the GasPM \cite{MATSUOKA2023168378}.}
	\label{fig:schematicGasPM}
\end{figure}

GasPM~\cite{MATSUOKA2023168378} is a novel concept for detecting photons with high efficiency, excellent time resolution, and affordable costs, based on pairing a photocathode with a resistive-plate chamber.
Figure~\ref{fig:schematicGasPM} illustrates the design. The detector features a planar photocathode deposited on the inner surface of a transparent light‐input window, facing a resistive plate placed in parallel. Photocathode and plate are separated by a thin gas volume. Behind the resistive plate lies a low‐resistivity high-voltage electrode, while an adjacent conductive anode pad collects the induced signals. A uniform electric field is generated between the photocathode and the anode, which is held at ground potential. The resistive plate is biased to the same potential as the HV electrode. 

When incident photons of energies of $\mathcal{O}(MeV)$, coming through the input window, strike the photocathode, electrons are emitted through photoelectric effect for a fraction of them; electrons are subsequently accelerated by the electric field and gain enough energy between collisions to ionise gas atoms, producing further electrons and positive ions. The newly released electrons are also accelerated, creating a chain reaction (avalanche) that typically yields a factor of $10^6$ multiplication. A resulting mirror current is induced on the anode pad. This current cannot be instantaneously supplied by the high-voltage source, leading to a local drop of the electric potential across the plate in the region of the avalanche. Consequently, the local electric field is reduced, quenching the avalanche before it develops into a discharge. The principle of electron multiplication is analogous to that of resistive-plate chambers (RPCs)~\cite{Santonico1981}.

The GasPM has an additional advantage: a slight modification allows usage as a Cherenkov-based time-of-flight detector, in which the entrance window is repurposed as a Cherenkov radiator.
However, when operated as a Cherenkov detector, the GasPM records signals not only from Cherenkov photons, but also from ionisation, as the charged particles that produce the Cherenkov photons inevitably traverse the detector, which therefore also functions as a conventional RPC. 

\begin{figure}[tb]
	\centering
	\includegraphics[width=\textwidth]{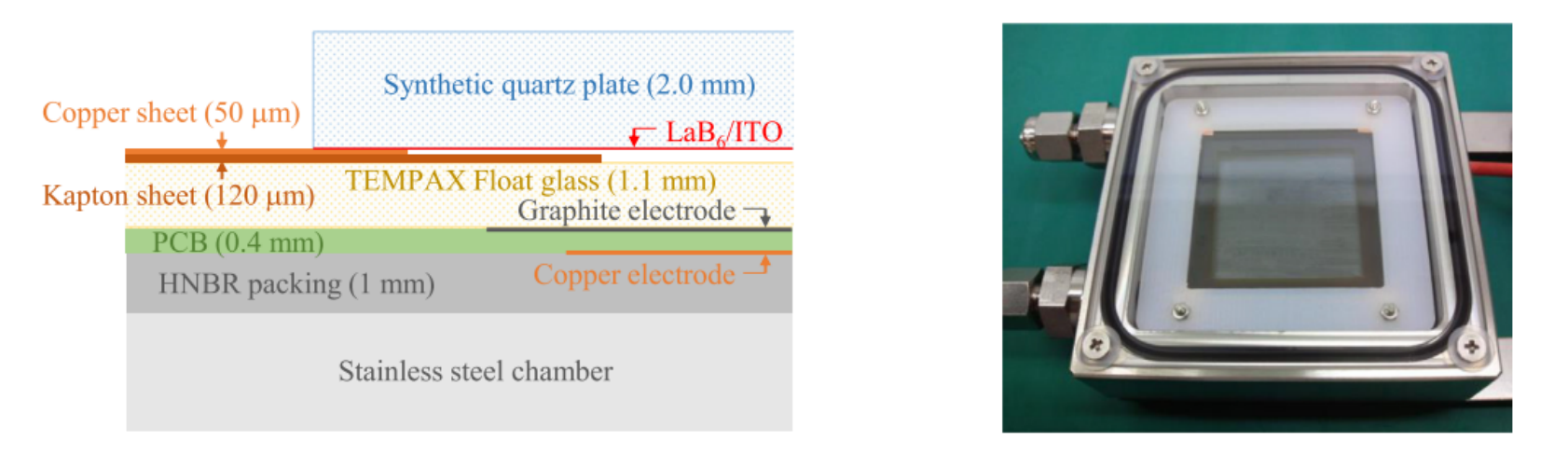}
	\caption{(Left) Cross-sectional schematic of the GasPM prototype inside the stainless steel chamber. Only the region around the edge of the photocathode is shown. The thickness of each component is indicated in parentheses. (Right) Photograph of the GasPM prototype from above~\cite{MATSUOKA2023168378}}
	\label{fig:prototype_scheme}
\end{figure}

Figure~\ref{fig:prototype_scheme} shows the layout of a typical GasPM prototype. The device features several layers stacked within a stainless steel gas-tight chamber. 
The top side of the chamber is sealed with a 2-mm-thick TEMPAX float glass window. TEMPAX float glass is less transparent than materials like quartz, but it offers a convenient balance between performance and affordability. Beneath the window lies the photocathode, supported by a synthetic quartz plate. The initially adopted photocathode was made of lanthanum hexaboride (LaB$_6$), which can be easily handled in air. In a later beam test, the LaB$_6$ photocathode was replaced with a cesium iodide (CsI) one. This provides adequate quantum efficiency in the vacuum ultraviolet (VUV), at wavelengths below 200 nm region, as adequate for Cherenkov photons, and exhibits good tolerance to the gas mixture~\cite{FRANCKE20041}.
A thin copper sheet connects the photocathode to ground, while a Kapton foil underneath provides electrical insulation. Together, the Kapton and copper layers also act as spacers, enclosing the gas gap between the photocathode and the underlying TEMPAX glass plate. The gap thickness can be adjusted depending on experimental requirements and is filled with a mixture of tetrafluoroethane (C$_2$H$_2$F$_4$, R134a) and sulphur hexafluoride (SF$_6$), which are both standard choices for RPCs.
Below the gap lies a 1.1-mm-thick TEMPAX float glass plate. On its upper face, a graphite electrode with $\mathcal{O}(1)~\mathrm{k\Omega/sq}$ sheet resistivity is kept at high voltage, while its lower side faces a printed-circuit board carrying a copper signal anode. At the bottom of the stack, a hydrogenated-nitrile-butadiene rubber layer seals the high-voltage and signal feedthroughs.
The electrode surfaces are $36.0\times36.0~\mathrm{mm^2}$ for the photocathode, $34.0\times34.0~\mathrm{mm^2}$ for the high-voltage electrode, and $31.2\times31.2~\mathrm{mm^2}$ for the signal anode. The active area, defined by the Kapton spacer, is $30.0\times30.0~\mathrm{mm^2}$ and corresponds to a single channel.
The GasPM operates in avalanche (proportional) mode. The SF$_6$ quenching gas suppresses uncontrolled and self-sustaining discharges. 

A slightly modified prototype is used as Cherenkov detector.
We apply negative high voltage directly to the photocathode, removing the dedicated high-voltage electrode. The resulting electric field preserves the same geometry as in the standard design, although both the resistive plate and the anode are kept at ground potential. 
Furthermore, the resistive-plate material is replaced with soda glass, to improve the rate tolerance by reducing the resistivity of the plate. The 2.4-mm-thick quartz window serves as a radiator.

\section{Previous results}       \label{sec:prev_res}
The GasPM project began in 2017 at the university of Nagoya and KEK. I joined in March 2025, when several preliminary studies and tests had already been carried out. 

\begin{figure}[tb]
    \centering
    \includegraphics[width=0.495\textwidth]{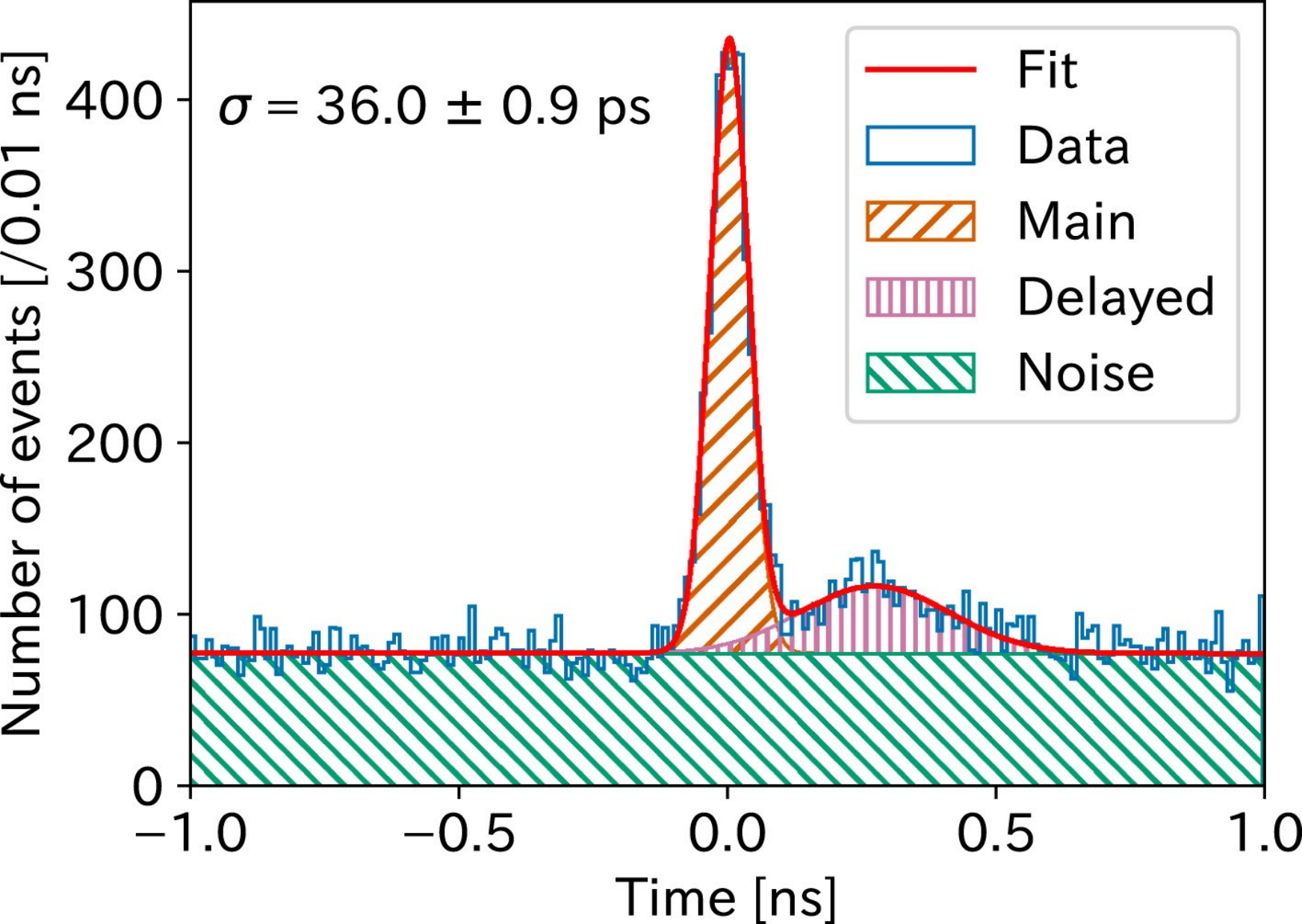}
    \hfill
    \includegraphics[width=0.495\textwidth]{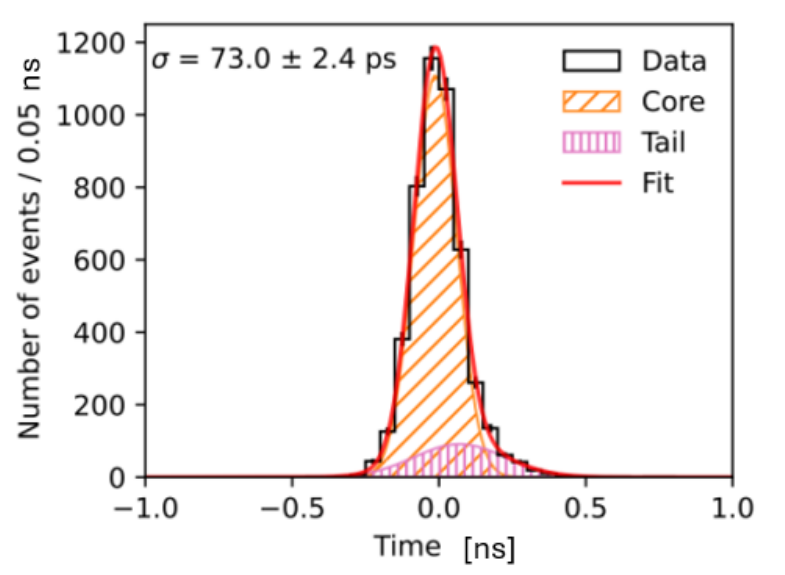}
    \captionsetup{width=0.9\linewidth} 
    \caption{Time distribution of GasPM signals from laser test (left) and 2023 beam test (right) with  fit projections overlaid~\cite{Okubo-GasPM}.}
    \label{fig:laser_res}
\end{figure}

An important early achievement was the measurement of the single-photon detection performance using a picosecond-pulse diode laser with 375~nm centre wavelength~\cite{MATSUOKA2023168378}. The prototype had an LaB$_6$ photocathode and featured 3~kV bias voltage over a 170~$\mu$m gas gap, corresponding to an electric field of 176~kV/cm. The photocathode quantum efficiency, defined as the number of emitted photoelectrons over the number of incident photons, was extremely low ($<10^{-5}$). While such inefficiency is in general a limitation, it was suitable for the single-photon demonstration, as detected signals could be reliably  attributed to individual single avalanches. The gas mixture consisted of 90\% R134a and 10\% SF$_6$.
The time resolution was measured to be $36.0 \pm 0.9$~ps, which improved to $25.0 \pm 1.1$~ps after accounting for the laser-pulse width and the time resolution of the readout system~\cite{MATSUOKA2023168378} (Fig.~\ref{fig:laser_res}(left)).

A modified prototype, prepared as a Cherenkov detector, underwent its first beam test in 2023. A CsI photocathode ensured sufficient quantum efficiency. The electric field was reduced to 140~kV/cm, corresponding to an applied voltage of 2.8~kV over an increased gas gap of 200~$\mu$m. The gas proportion was 50\% R134a and 50\% SF$_6$ to prevent discharges.
However, the resulting time resolution was $73.0 \pm 2.4$~ps, as shown in Fig.~\ref{fig:laser_res}(right), which improved to $62.3 \pm 4.8$~ps when considering only large pulse-height events. 
A significant part of this thesis focuses on investigating the causes of this time-resolution degradation and identifying strategies to improve it.

\section{Performance drivers and limitations}          \label{sec:GasPM_issues}
Knowledge accumulated in previous tests shows that the principal GasPM performance drivers are

\begin{description}
    \item[Input window/radiator] material transparency and thickness modify the number of transmitted (or generated Cherenkov-) photons.
    \item[Photocathode material]
    different materials feature different properties that affect the GasPM performance. Particularly relevant are quantum efficiency, its dependence on the photon wavelength, and resistance to exposure to air or ions that drift backwards in the gap.
    \item[Electric field intensity]
    The intrinsic GasPM time resolution is ~$\sigma_t = 1.28/[(\alpha - \eta)v],$ where $\alpha$ and $\eta$ are the first Townsend and attachment coefficients, respectively, and $v$ is the electron drift velocity~\cite{RIEGLER2003144}.
    Therefore, higher electric field improves $\sigma_t$ as it increases $(\alpha - \eta)$ and $v$. In addition it increases the collection efficiency $\varepsilon = 1 - \eta/\alpha$, which is the probability that a photoelectron leads to an avalanche. Moreover, it increases the gain, which improves signal-to-noise ratio.
    \item[Gas mixture]
    related to the field intensity,  the gas-mixture choice determines the drift properties and allows controlling the avalanche development and the probability of uncontrolled discharges.
\end{description}

\begin{figure}[tb]
    \centering
    \begin{minipage}[t]{0.45\textwidth}
        \centering
        \includegraphics[width=0.90\textwidth]{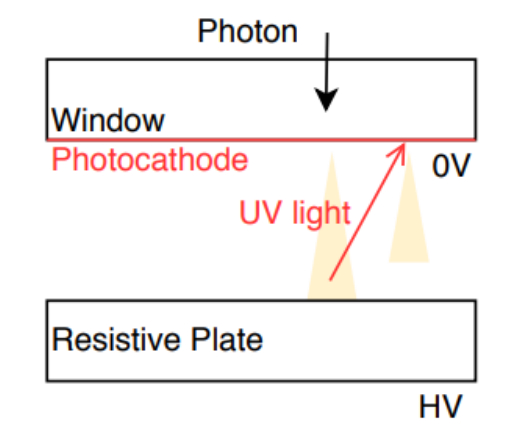}
        \captionsetup{width=\linewidth}
        \caption{Sketch of the photon feedback phenomenon~\cite{Okubo-GasPM}.}
        \label{fig:PF_scheme}
    \end{minipage}%
    \hspace{0.05\textwidth}  
    \begin{minipage}[t]{0.45\textwidth}
        \centering
        \includegraphics[width=\textwidth]{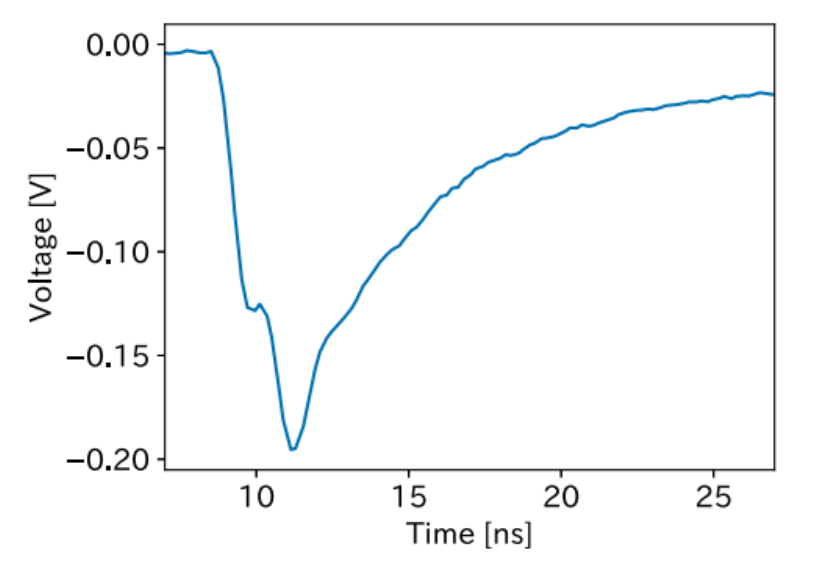}
        \captionsetup{width=\linewidth}
        \caption{Example of secondary photon feedback signal from a laser test~\cite{Okubo-GasPM}.}
        \label{fig:PF_signal}
    \end{minipage}
\end{figure}

In addition, we observed the following limitations:
\begin{description}
    \item[Photon feedback] UV photons emitted during gas excitation and de-excitation may impinge on the photocathode, and cause the emission of additional photoelectrons, triggering secondary electron avalanches in the gas gap (Fig.~\ref{fig:PF_scheme}). These secondary avalanches confuse and degrade the timing information. Signature of photon feedback is a secondary signal peak, as shown in Fig.~\ref{fig:PF_signal}. Here, the secondary peak is clearly distinctive, but often the two peaks overlap, particularly when the gas gap is reduced, making it challenging to identify the phenomenon and thus spoiling time resolution.   
    \item[Ion feedback] avalanche ions in the gas gap drift backwards and hit the photocathode, damaging it over time. Ion feedback does not affect time resolution, but it worsens the efficiency and may represent a serious issue for long-term use in experiments. The 2023 beam test showed that the CsI photocathode is vulnerable to ion feedback and even with low-rate cosmic rays it can be damaged within few weeks.
    \item[Streamer formation] large discharges happen when the electric field is high enough that primary avalanches are not limited to a local gap volume, but extend over the whole gap. Normally, ionisation electrons are accelerated by the field and initiate the primary avalanche.
    However, as the avalanche develops, it can reach the Raether limit of approximately $10^8$ electrons~\cite{peskov2009research}. At this point, the charge of the large number of positive ions accumulated in the avalanche front locally distorts and enhances the electric field. The enhanced field promotes secondary ionisation ahead of the avalanche and enables the release of additional electrons through photoionisation: ultraviolet photons emitted during molecular de-excitation travel through the gas and free electrons at distance from the avalanche.
    The combined action of the distorted electric field and secondary ionisation generates a streamer channel, a highly conductive plasma filament between the electrodes. The streamer propagate rapidly across the gap, creating a self-sustained discharge that blinds readout from the details of the primary signal and may damage the detector.
\end{description}

\section{Thesis goals}
The main goal of this work is to make progress in addressing some of the above limitations and optimise performance through the preparation, execution, and data analysis of an improved beam test and an additional cosmic-ray test.
The principal modifications and improvements introduced in this work are the following:

\begin{itemize}
    \item we upgrade the acquisition system by introducing a new digitiser with a high sampling frequency, which provides better discrimination of secondary signals caused by photon feedback from the primary signals;
    \item we increase the electric field in the GasPM from \(140 \ \mathrm{kV/cm}\) to \(187 \ \mathrm{kV/cm}\) by reducing the gas gap from \(200 \ \mu\mathrm{m}\) to \(150 \ \mu\mathrm{m}\), which enhances the detector gain and single-photon time resolution;
    \item we include a multi-pixel photon counter array in the beam-test setup, which enables us to reliably select single-electron events for an unbiased time-resolution measurement;
    \item finally, I use cosmic rays to explore and prepare use, in future beam tests, of the LaB$_6$ photocathode, which is expected to be more resistant to ion feedback.
\end{itemize}


\chapter{Readout digitiser characterisation and calibration}         \label{chap:NALU}

This chapter discusses the preliminary tests and calibration of a newly deployed fast-sampling prototype digitiser intended for use in the beam test to discriminate secondary signals due to photon feedback from primary signals.

\section{Architecture}
A major objective of the test beam is to discriminate against secondary signals due to photon feedback. The close proximity in time between primary and photon-feedback signals calls for fine-grained signal sampling. To this purpose, we employ a new digitiser.
A digitiser is an electronic device that converts analogue signals into digital data for increased resilience against environmental disturbances and downstream processing. It takes as input a continuous electrical signal, the GasPM output in our case, and samples it at regular time intervals, converting each sampling into a digital number.

\begin{figure}[tb]
  \begin{center}
    \includegraphics[width=0.55\textwidth]{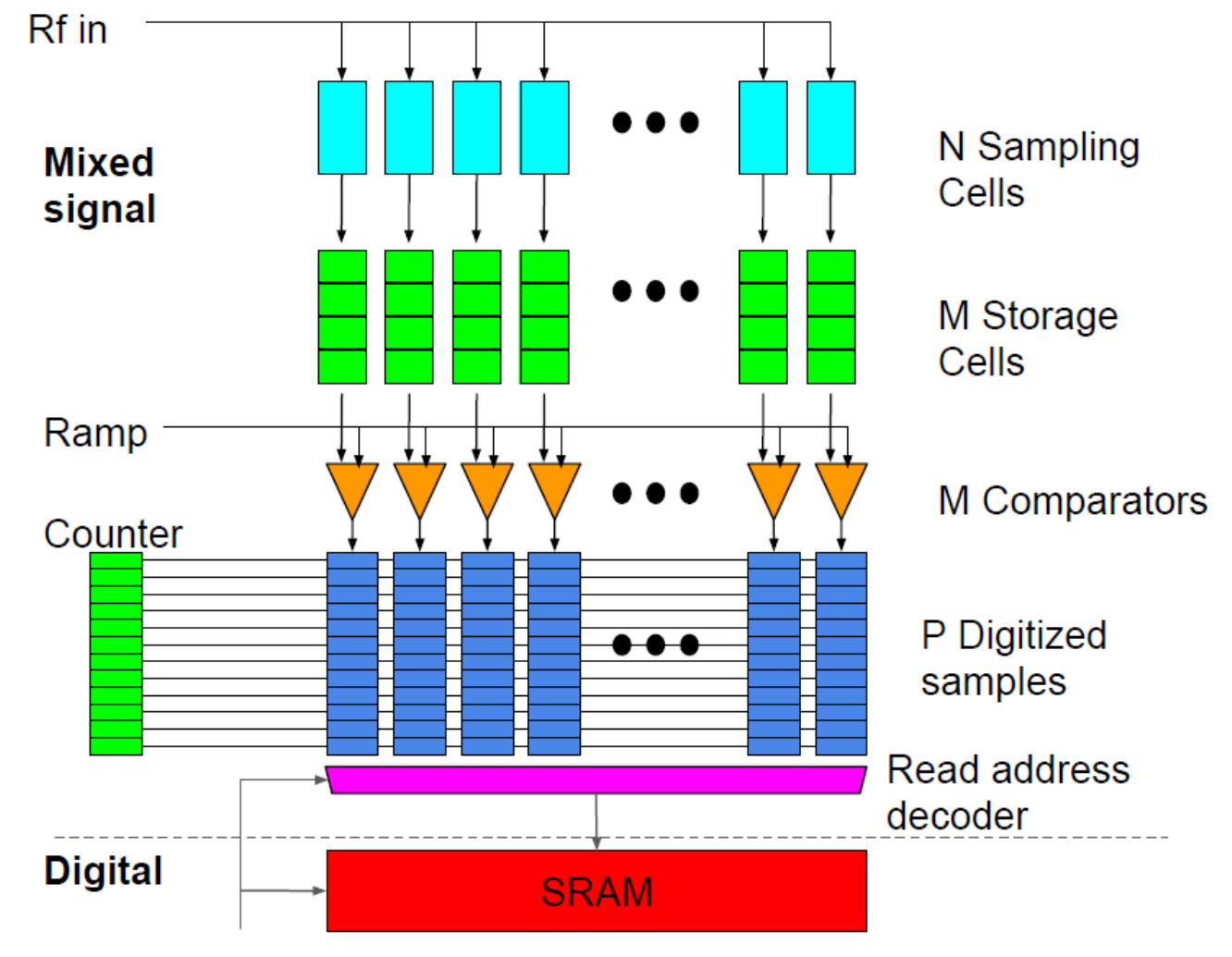}
  \end{center}
  \caption{DSA-C10-8+ waveform digitiser architecture. For this model N=128, M=32768, P=64~\cite{NALU}.}
\label{fig:NALU_architecture}
\end{figure}

\begin{figure}[tb]
  \begin{center}
    \includegraphics[width=0.55\textwidth]{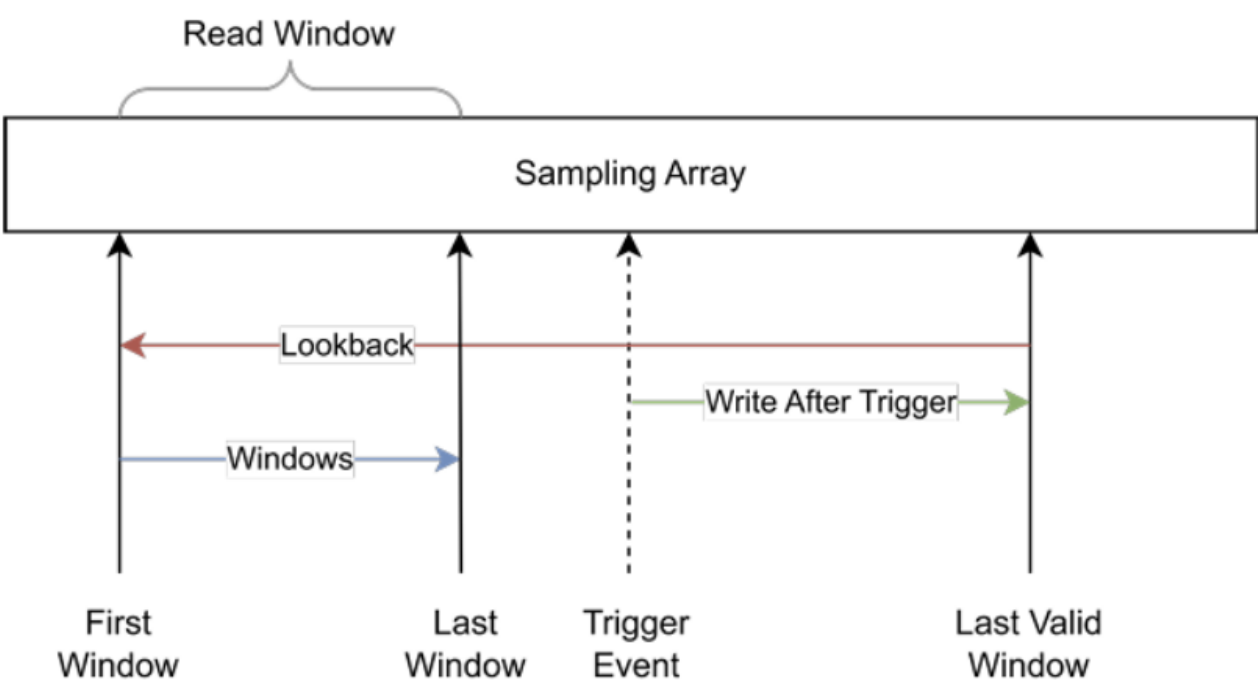}
  \end{center}
  \caption{Scheme of the read-out logic and advanced read windows controls.}
\label{fig:sample_process}
\end{figure}

The DSA-C10-8+ digitiser is an application-specific integrated circuit designed and produced by Nalu Scientific~\cite{NALU}. It is based on the third version of the ARRDVARC chip, that has variable-rate readout and is intended for fast timing and low dead-time measurements. The digitiser has two chips, 4 channels each. Each channel is digitised with a 12-bit ADC, with up to 10 GSPS sampling frequency and negative polarity.
It has a 20 dB input amplifier leading to an effective range of $\pm 100$ mV.
The bandwidth is 1.8 GHz.
The architecture of each channel is shown in Fig.~\ref{fig:NALU_architecture}.
A sampling array of 128 acquisition cells operates in round-robin configuration. Every half period, 64 analogue samples are transferred into a secondary storage array of buffers. The secondary array is divided into 512 physical windows, each containing 64 buffers. The total depth of this storage array is therefore 32768 buffers, of which only 32640 are available because 2 windows are reserved for internal use. Analogue values are stored in these capacitors until the acquisition loop restarts, and the buffer is overwritten. The array is therefore always full during acquisition, even if only a subset of windows is read out.   
Read-out is controlled by  proprietary software and  occurs window by window. It starts when either a self-trigger (signal) or an external trigger (trig\_in) is sent, and the desired window is flagged. Three configurable parameters, allow to (i) continue filling a \textit{write after trigger} number of windows, (ii) look backward for a \textit{lookback} number of windows, (iii) read a \textit{windows} number of windows starting from that position, which means \textit{windows} $\times$ 64 number of samples.
This logic is schematised in Fig. \ref{fig:sample_process}.

The fast-sampling digitiser in our use is a prototype still under development by the vendor.
Before it can be reliably used for data acquisition, it necessitates a thorough characterisation and calibration, both in voltage and time. Due to logistical constraints, beam-test data are acquired using the uncalibrated device; calibration is subsequently performed and applied offline.

\section{Calibration}     \label{sec:calib_proc}
The calibration purpose is to ensure that the output signal reflects faithfully the original input shape and timing. The process is therefore divided into two sequential steps: voltage and time calibration.

The voltage-calibration goal is to make sure that the digitised output is unbiased. One checks that the input is not altered during digitisation and sufficient linearity exists between analogue input and digital output by studying output-vs-input graphs. 

The time-calibration goal is to precisely infer the sampling time of each cell.
The 10 GSPS acquisition frequency implies that the interval between two neighbouring samplings is 100 ps in average. However, each sample is collected by physically different acquisition cells in the looping array. Small differences among the low-level electronic components involved in the sampling, routing, or temperature gradients, can impair precision, yielding slightly irregular sampling intervals.
Since our DSA-C10-8+ is still a development prototype, the documentation is not particularly thorough and we study its properties by empiric testing.

The lower bound of the bandwidth is unknown, and I therefore test it.
I first use a power supply to inject a constant 100 mV signal; in addition, I use a waveform generator to feed in a square wave with 98\% duty cycle, so that the output is essentially constant for almost the entire acquisition window. The output shows no signal under constant input. Only noise is digitised. A similar outcome occurs with square waves: the digitiser only registers the rising and falling transients, showing no activity during the uniform parts. This suggests the presence of a capacitor in the digitiser input circuitry that suppresses the direct current and establishes that a DC based voltage calibration, as suggested in Ref.~\cite{SRitt} for instance, is not an option. 
This complicates considerably the calibration, which cannot rely on testing the digitiser with sequences of constant voltages, observe the output, and build linearity graphs. We therefore devise an alternative approach.
Before converging to it we explore several attempts, occasionally misled by misunderstandings with the board manufacturer on the system architecture.
While most of these attempts are unsuccessful, they prove useful for gaining a better understanding of the device and for establishing procedures later incorporated into the final method.

\begin{figure}[tb]
    \centering
    \begin{minipage}[t]{0.45\textwidth}
        \centering
        \includegraphics[width=\linewidth]{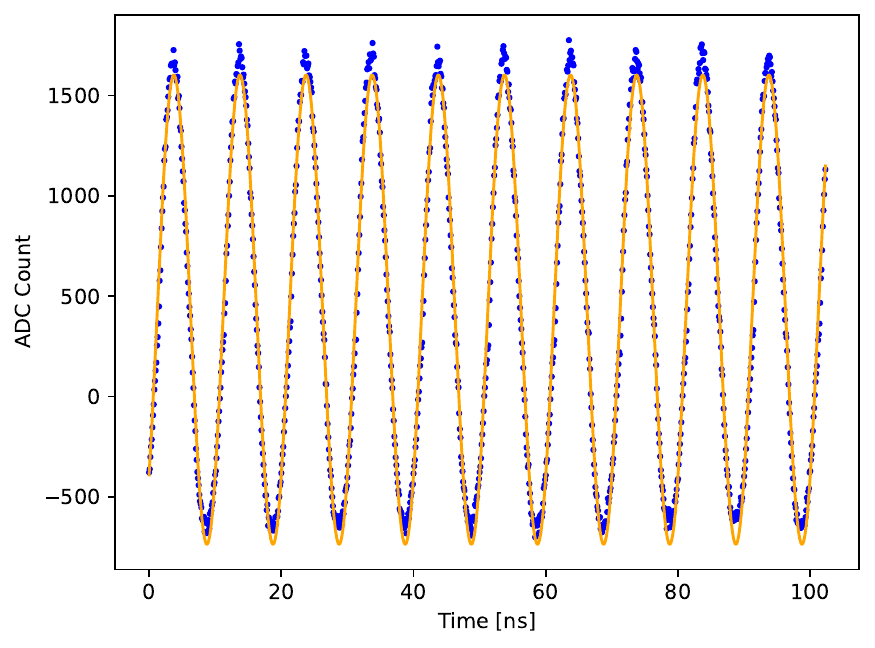}
        \captionsetup{width=\linewidth}
        \caption{Sampled 250 mV$_{\text{pp}}$ sine wave (blue) with sine fit overlaid (orange).}
        \label{fig:sine_fit}
    \end{minipage}%
    \hspace{0.05\textwidth}
    \begin{minipage}[t]{0.443\textwidth}
        \centering
        \includegraphics[width=\linewidth]{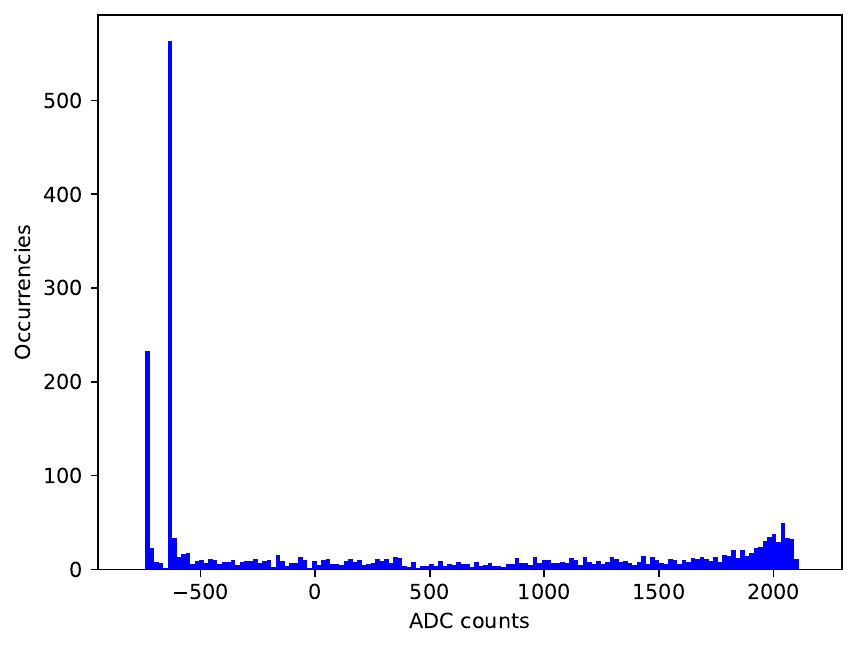}
        \captionsetup{width=\linewidth}
        \caption{ADC distribution for one cell from a 350 mV$_{\text{pp}}$ sine wave. }
        \label{fig:ADC_distr}
    \end{minipage}
\end{figure}

\subsection{Voltage}
The digitiser has a limited dynamic range, whereas the beam-test signal range spans over a factor of 7, from roughly 30 mV to over 200 mV. Finding the optimal input amplification or attenuation to remain within the board dynamic range is challenging, as excessive attenuation leads to a loss in efficiency due to noise. Because the digitiser will work outside the nominal dynamic range, I consistently extend the calibration over a broader range.

To calibrate voltage, I acquire sine waves using the maximum and minimum amplitudes as calibration references. This suboptimal solution brings two complications. One needs large samples, as the events collected at the extrema are only a small fraction of the total samplings. 
A deeper conceptual problem is that a calibration based on sine waves does not allow to decouple voltage from time; the time-dependence of the input voltage correlates the two calibrations, which spoils the results. 

I collect sine waves of various amplitudes generated by an Agilent Technologies 81150A waveform generator, capable of frequencies up to 240 MHz~\cite{Keysight}.
I trigger the digitiser using a clock generator, which produces precise pulses at configurable, equidistant time intervals and is not correlated to the internal clock and the waveform. The trigger-clock NIM signal is converted in TTL through a level adapter to comply with the digitiser specifics.
Despite the board sampling rate of 10 GSPS, data transfer is limited by the USB connection, which slows the acquisition considerably. The trigger rate is therefore set to 10 Hz, and each acquisition takes approximately 1.5 hours.
Because we have 32640 buffers, I acquire roughly 50000 events per waveform, for a total of 1550 samples per cell. 
Acquisition is done with 100 MHz frequency and seven different peak-to-peak amplitudes from 100 mV$_{\text{pp}}$ to 400 mV$_{\text{pp}}$. Figure~\ref{fig:sine_fit} shows a 250 mV$_{\text{pp}}$ sine wave. The blue points are the digitiser output whereas the orange line is the input. Output distortion is visible, with high peaks at varying voltages and clusters at low voltages.

\begin{figure}[tb]
  \begin{center}
    \includegraphics[width=0.7\textwidth]{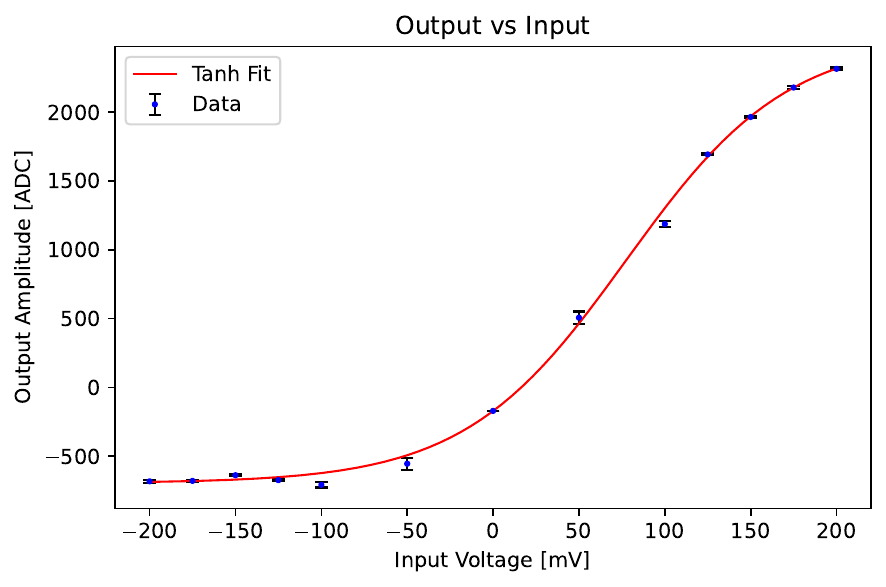}
  \end{center}
  \caption{Digitiser output vs input plot for buffer 0 out of 32640.}
\label{fig:INvsOUT}
\end{figure}

Using a digitiser parameter to map, for each event, the relevant trigger window, I associate each sample to the cell that stores it. For each cell I obtain the ADC distribution over the entire acquisition time. As a typical example, Fig. \ref{fig:ADC_distr} shows the distribution for a 350 mV input. For a linear and unbiased response, we would expect the distribution to be bimodal and symmetric, with two similar peaks corresponding to the minimum and maximum amplitudes. The observed distribution, however, is multimodal and shows a strong asymmetrical clustering with two narrow peaks at negative ADC values, and a much broader positive ADC-count peak. The lowest ADC-count peak corresponds to samplings of the maxima of the positive half-period, which are randomly clipped down, probably as an effect of saturation since the phenomenon increases with amplitude.

These imperfections do not prevent a calibration. I infer the minimum amplitude of the output from the ADC distribution by fitting the main negative ADC peak with a Gaussian distribution. I assume the mean to be a good estimator of the input voltage (in ADC units), and the width an estimator of the uncertainty. I then define the maximum positive amplitude as the ADC value corresponding to the first bin that exceeds an empiric threshold. This approach is preferred for the observed broader peaks, as Gaussian fitting would not reliably yield the actual maximum amplitude. 
I assume the uncertainty to be the same as for negative peaks, since it is approximately constant over all cells.
To sample the pedestal, which is the output corresponding to 0.0 V input, I use pulses and take the first time gate before any pulse to compute the mean collected amplitude.
This improves upon a pure noise acquisition based on no inputs, because the capacitor charge biases the pedestal if the circuit is closed.

Repeating the procedure for various amplitudes allows constructing the output vs input graph for each of the 32640 storage cells (Fig. \ref{fig:INvsOUT}).
An empiric function that fits the data adequately is
\begin{equation} \label{tanh}
f(x) = A \tanh\left(Fx + B\right) + C,  
\end{equation} 

where the variable $x$ is the input voltage, $F = 100$ MHz is the known frequency, and $A$, $B$ and $C$ are the fitting parameters representing amplitude, phase, and offset, respectively. The output saturates fully for negative amplitudes already at $-100$ mV, while saturation begins from about 150 mV on positive amplitudes. We observe a nearly linear response within the restricted range $-50$ mV to $+100$ mV, which is slightly narrower than the nominal range, too short for our purpose.

\begin{figure}[tb]
  \begin{center}
    \includegraphics[width=0.7\textwidth]{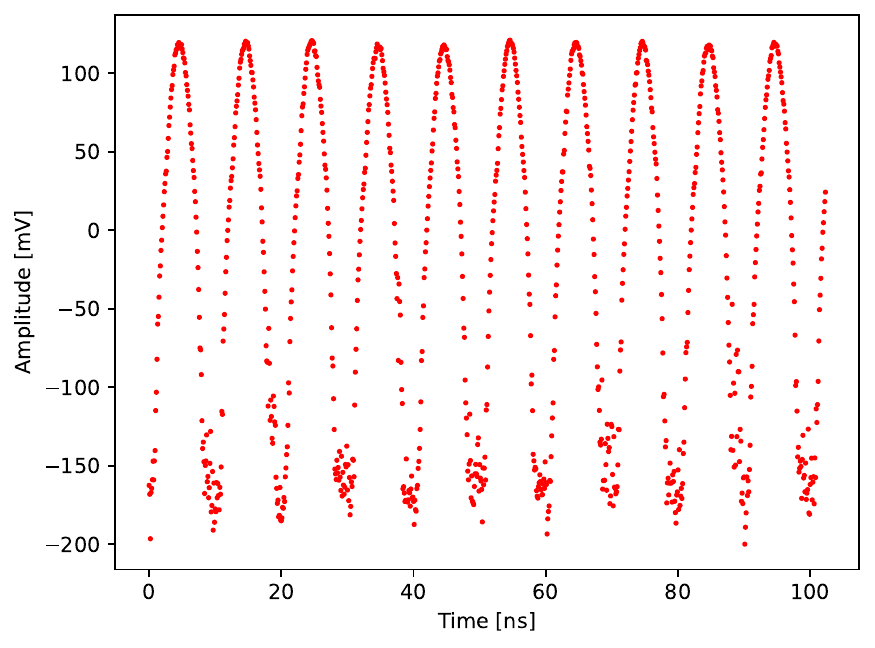}
  \end{center}
  \caption{250 mV$_{\text{pp}}$ amplitude event after voltage calibration.}
\label{fig:voltage_applied}
\end{figure}

I then convert ADC counts into voltage by inverting Eq.~\eqref{tanh} and applying this correction to all datasets after properly associating the correction parameters to the appropriate cell.
Figure~\ref{fig:voltage_applied} shows the results from an example acquisition of the 250 mV$_{\text{pp}}$ dataset. The positive side looks adequate, reasonably smooth and approximately at the right voltage. The negative amplitudes, though, are severely distorted, as expected by the shape of the correction function, which clips to a nearly constant saturation value, resulting in large differences in input to correspond to negligible changes in output. 

However, since the GasPM signal has positive polarity as a result of the inversion of the digitiser, we ignore the negative region and consider the voltage calibration acceptable. We then proceed to the time calibration.

\subsection{Time}
The time calibration is relatively straightforward in principle. The goal is to determine the actual sampling time of each acquisition cell, which ideally should occur every 0.1 ns. At this stage, only the 128 acquisition cells are relevant.
For each channel, I acquire sine waves with frequencies 10, 50, 100, 150, 175, 200, and 240 MHz, while keeping the amplitude at 150 mV$_{\text{pp}}$. I use a signal splitter, which allows to feed the same waveform from a single generator output into four channels simultaneously to increase efficiency. Since frequency rather than amplitude is relevant here, this approach is acceptable. However, the generator has to output a signal four times higher than the desired value to ensure the correct amplitude at each input. 

\begin{figure}[tb]
  \begin{center}
    \includegraphics[width=0.7\textwidth]{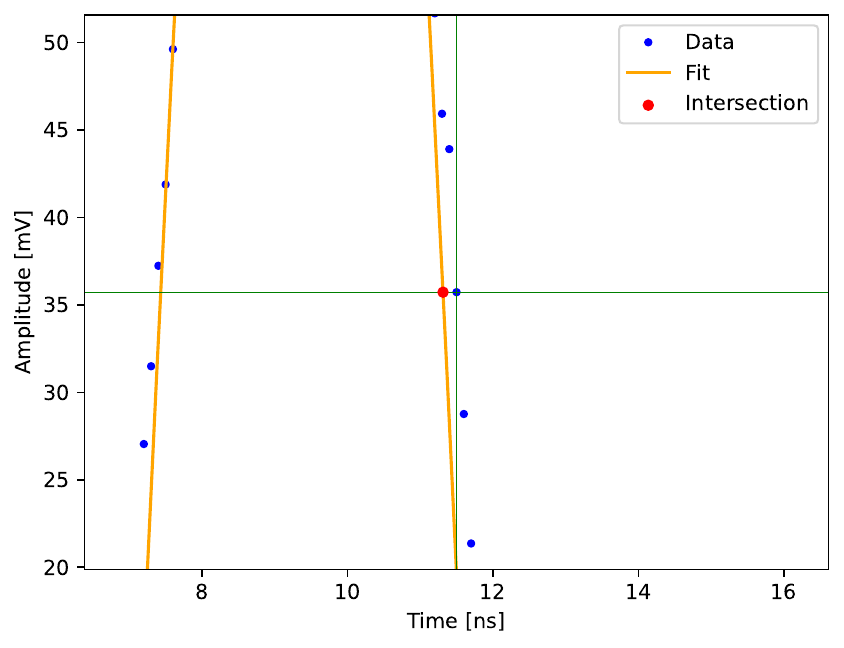}
  \end{center}
  \caption{Graphical representation of the time difference between the sampled data points and the fitting waveform used in the time calibration. The red dot indicates the intersection of the horizontal line determined by the data-point voltage with the sine wave.}
\label{fig:time_distance}
\end{figure}

\begin{figure}[tb]
    \centering
    \begin{minipage}[t]{0.45\textwidth}
        \centering
        \includegraphics[width=\textwidth]{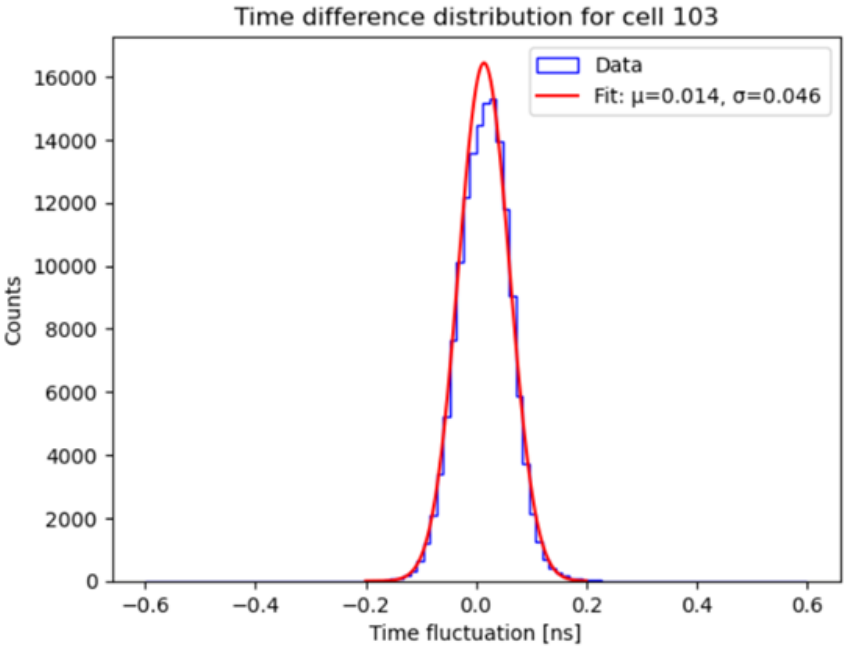}
        \captionsetup{width=\linewidth} 
        \caption{Time fluctuation of the samples acquired by a single cell (number 103) with respect to the sine fit. A Gaussian fit is overlaid.}
        \label{fig:corr_distrib}
    \end{minipage}\hfill
    \begin{minipage}[t]{0.45\textwidth}
        \centering
        \includegraphics[width=\textwidth]{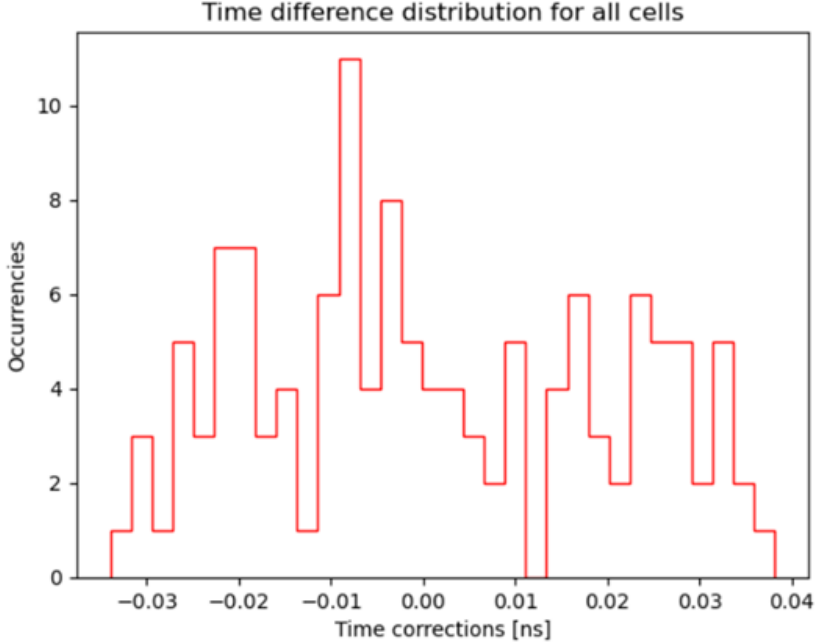}
        \captionsetup{width=\linewidth} 
        \caption{Distribution of the 128 final time corrections.}
        \label{fig:128corr_factors}
    \end{minipage}
\end{figure}

Once the voltage calibration is applied and the output signal is reconverted in volts, I fit the calibrated data with a sine function, assumed to be a good approximation of the input signal. For each sample, I calculate the time difference between the recorded data point and the corresponding position on the fitting curve, as represented in Fig.~\ref{fig:time_distance}. I then assign this time correction to the appropriate acquisition cell.
After repeating for all recorded events, a distribution of time corrections is available for each cell. Figure~\ref{fig:corr_distrib} shows one example. The shift of the mean from zero is the timing calibration factor. Figure~\ref{fig:128corr_factors} shows the distribution of the resulting 128 correction values, which  span the range $-0.03$ ns to $0.04$ ns.

\section{Time resolution}
To validate the calibration, I measure the digitiser time resolution by comparing the rise-time of a pulse signal split into two channels. 

Time calibration is expected to improve the resolution because it reduces time jitter.  In an uncalibrated digitiser, the signal rise time would depend on the sampling cells, making it challenging to extract consistently the correct time. 
Time calibration reduces this dependence, leading to an improved time resolution.
The voltage calibration does not affect time resolution, since the rise-time is extracted by constant-fraction discrimination, unless the linearity correction significantly depends on the single cells.

\begin{figure}[tb]
    \centering
    \begin{minipage}[t]{0.45\textwidth}
        \centering
        \includegraphics[width=\textwidth]{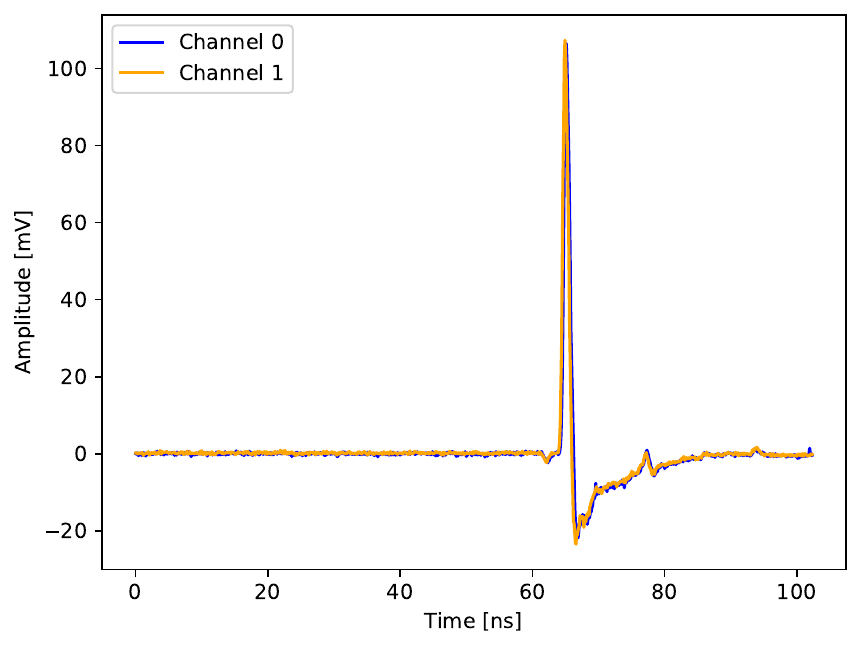}
        \captionsetup{width=\linewidth} 
        \caption{Split pulse signal collected by channel 0 and 1 simultaneously, after both voltage and time calibration.}
        \label{fig:pulse}
    \end{minipage}\hfill
    \begin{minipage}[t]{0.45\textwidth}
        \centering
        \includegraphics[width=\textwidth]{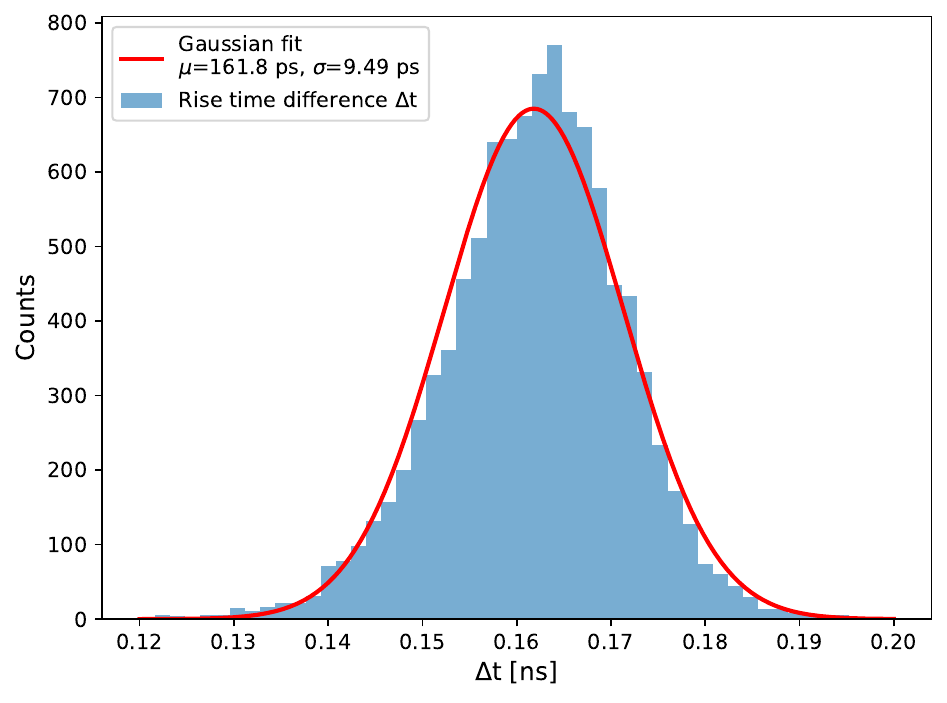}
        \captionsetup{width=\linewidth} 
        \caption{Pulse rise-time distribution between channel 0 and 1 of the DSA-C10-8+ digitiser.}
        \label{fig:time_resolution}
    \end{minipage}
\end{figure}

The test consists on generating a fast 110 mV pulse with a Tektronix PSPL2600C pulse generator, which features a rise time of 800 ps~\cite{tektronix}. Each pulse is then split into two identical and synchronous pulses using a BNC T-connector and delivered to two independent digitiser input channels. An example of such a pulse, after both voltage and time correction, is displayed in Fig.~\ref{fig:pulse} for channels 0 and 1. The signal timing is extracted by fitting the rising edge with a 20th-order polynomial and determining the time at which the signal reaches 50\% of its final amplitude.
The time difference between the two channels is then computed. This procedure is repeated for every event. Finally, a distribution of the time differences is constructed and fit with a Gaussian model (Fig.~\ref{fig:time_resolution}). The standard deviation approximates adequately the digitiser time resolution. I perform the split-pulse test under three conditions: without calibration, after voltage calibration, and after both time and voltage calibration. The data reported in the following are from channels 0 and 1 of the digitiser but the calibration is applied to all channels.

The results are unsatisfactory, although not fully unexpected. The time resolution without calibration is good, namely $6.99 \pm 0.05$ ps. After applying the voltage calibration, the resolution is $7.01 \pm 0.05$ ps, which is also satisfactory and shows that the dependence of the voltage linearity calibration on the single cells is not significant.
The possible issue arises with the time calibration. This worsens the resolution by 30\% to $9.49 \pm 0.07$ ps. Although this value is still sufficient for studying photon feedback, a resolution degradation upon calibration shows that we do not have yet a sufficient control of the digitiser, unless we assume a common systematic bias of the sampling interval that affects most cells coherently. I therefore accept this calibration for our early-stage specific study, even if it is not optimal. For future GasPM time resolution studies we will need either to refine the procedure or to consider using the manufacturer calibration.
\chapter{Beam test} \label{chp:beamtest}

In this chapter I discuss the design, execution, and results of the beam test of the GasPM prototype that I conducted. 
Since the GasPM is designed to be deployed in a Belle II upgrade, or other high energy physics experiments, testing it with particle beams aims at characterising its performance  and limitations as a Cherenkov detector in an environment as similar as possible to its possible future applications. 
A brief introduction to general concepts and ultimate goals is provided before addressing the specific issues investigated in this work.

\section{Concepts} \label{sec:concepts} 

A beam test is an experiment in which particle detectors, or other experimental devices, are probed with known and controlled particle beams at an accelerator facility in order to study performances under a variety of realistic conditions. The beam test is one of the most important benchmarks in the development of a new detector, which usually follows more basic tests such as those based on light sources, radioactive sources, or cosmic rays.

We use 5 GeV electrons produced by the photon-factory advanced-ring at KEK, the High Energy Accelerator Research Organisation in Tsukuba, Japan, between 9:00 a.m. on Monday, 17th March 2025, and 9:00 a.m. on Wednesday, 19th March, 2025.
The GasPM prototype we use is configured to operate as a Cherenkov detector.

We install the GasPM and other standard detectors, in a row, in front of the collimated electron source.
In addition to the GasPM, the setup includes an RPC, a charged-particle gaseous detector that shares the same structure of the GasPM, except for the photocathode; a microchannel-plate photomultiplier; a multi-pixel photon-counter; and plastic scintillators.
Comparison with the RPC is used for determining the rate of uninteresting ionisation signals in GasPM data. The microchannel plate photomultiplier is used as reference to measure the GasPM time-of-flight resolution. 
The coincidence of two scintillation counters, located at the upstream and downstream ends of the setup, provides a trigger that signals the passage of a beam particle, which is therefore expected to have traversed the apparatus, and enables readout.

The ultimate goal is to measure the time resolution of the GasPM prototype and compare it with previous results.
When the beam is activated, electrons begin travelling through the apparatus.
For each triggered event, we record the arrival time of Cherenkov photons, generated by the electrons passing through the radiators in front of both the reference and the test detectors, and we compute the time difference $\Delta t = t_{\textrm{test}} -  t_{\textrm{ref}}$.

The few-centimetres distance between the two detectors results in a $\Delta t$ distribution not centred in zero, because of the finite speed of electrons. However, we are only interested in the time resolution, which is the spread $\sigma_{\textrm{TOF}}$ of the $\Delta t$ distribution, dominated by the timing uncertainties of the detectors themselves. Since we know the resolution of the reference detector $\sigma_{\textrm{ref}}$, we retrieve the resolution of the test detector $\sigma_{\textrm{test}} = \sqrt{\sigma^2_{\textrm{TOF}} - \sigma^2_{\textrm{ref}}}$.

While the conceptual scheme is straightforward, several experimental challenges make the experiment non trivial.
A variety of undesired effects might occur in the test detector, that need to be understood and subtracted for a reliable measurement. 
An important nuisance is photon feedback.
Resolving secondary peaks originating from photon feedback is important and calls for high-frequency signal sampling.
Adequate time and voltage calibration of the digitisation device must therefore be achieved, to reduce the associated time uncertainty. 
Finally, only events consisting of a single electron that impinges on the GasPM should be considered to avoid confusing results due to event overlap. This requires discrimination against multiple-electron events.

\section{Beam properties}
The advanced-ring test beam line is based at the KEK-PF/AR~\cite{Honda:2021pgx}, which is an electron storage ring that produces electron–positron pairs by inserting a wire target into the beam halo, where a small fraction of circulating electrons interact with the target material. These interactions produce electron–positron pairs via bremsstrahlung photons and subsequent pair production.
The resulting secondary electrons are extracted and transported to form the test beam.
The beam has a selectable energy of up to 6.5 GeV. We set the beam momentum to 5.0 GeV and measure a trigger rate of 23 Hz. The beam transverse size is approximately 8 mm in both the horizontal and vertical directions.

\section{Instrumentation} 
The technical specifics of the detectors used follow.

\begin{description}
       \item[Trigger counters]
       two scintillation counters, each made of a 25$\times$25 mm squared plastic scintillator connected to a photomultiplier tube by a light guide provide the trigger signal.
       \item[MPPC] a multi-pixel photon counter, also known as silicon photomultiplier, is used to select single-electron events. Produced by Hamamatsu, it is a matrix of 8$\times$8 channels, of size 3$\times$3 mm$^2$ each, for a total of 5.76 cm$^2$ active surface. It samples Cherenkov rings produced by charged particles passing through a 3-mm-thick acrylic resin radiator placed upstream in front of it. 
       \item [GasPM] the prototype detector under study, detailed in Sec. \ref{chap:GasPM}. In this test it is operated as a Cherenkov detector with a 5-mm-thick MgF$_2$ radiator, a CsI photocathode, and a 2.8 kV voltage over a 150 $\mu$m gap filled with 90\% R134a and 10\% SF$_6$.
       \item [MCP-PMT] two microchannel-plate photomultipliers~\cite{matsuoka2019performance} provide a reliable reference for time-of-flight resolution. They are produced by Hamamatsu, with a quartz radiator mounted upstream in front of them and have 35 ps single-photon time resolution.
       \item [RPC] a resistive-plate chamber is used to compare hit-rates to exclude ionisation signals from the total GasPM hit-rate. It is produced in house. Charged particles directly pass through the gas and ionise it generating a signal, the same way photoelectrons do in the GasPM.
    \end{description}
   These are read out by the following electronic boards: 
     \begin{description}
       \item [DRS4]a digitiser developed and produced by the Paul Scherrer Institute. It is a domino-ring sampler version-4 chip, which is a switched-capacitor array designed for waveform digitisation. The board allows for sampling rates up to 5 GSPS and features 4 input channels, each with 1024 sampling cells and time resolution of 14 ps. We use it as the default acquisition system as in the previous beam test.
       \item [DSA-C10-8+] development version of a new fast-sampling digitiser prototype, based on the ARRDVARC v3 chips, with 10 GSPS sampling frequency (Chap.~\ref{chap:NALU}). We newly use it in this beam test for specific studies on photon feedback.
       \item[EASIROC] an analogue integrated readout chip, developed by OMEGA microelectronics for reading out silicon photomultipliers~\cite{CALLIER20121569}.
\end{description}

\section{Improvements upon the 2023 beam test}
The previous beam test was conducted in 2023 using a GasPM prototype with different photocathode, radiator, and other configuration details. The first objective of my beam test is to match the performance achieved in 2023, and then to further improve upon it. 
The commonalities of the beam test reported in this work with the previous beam test are (i) beam; (ii) reference measurement, provided by two precise and reliable MCP-PMTs; (iii) RPC, used for result comparison; (iv) trigger; (v) DRS4 as the main data acquisition device.
The upgrades we implement are all aimed at suppressing probable causes of time-resolution degradation (Sec. \ref{sec:GasPM_issues}) and include

\begin{itemize}
    \item a reduction of the gas gap from 200 \textmu m to 150 \textmu m, which increases the electric field to \(187 \ \mathrm{kV/cm}\); this is an important improvement for time resolution as it increases the gain and the photoelectron drift velocity.
    We shorten the gas gap at unchanged bias instead of increasing bias voltage to reduce the risk of unwanted, and potentially damaging, discharges. The downside is that UV photons have a shorter path to reach the photocathode. The resulting photon-feedback signals have a larger overlap with the primary signals and are therefore more difficult to discriminate;
    \item an improved radiator that replaces quartz with MgF$_2$, which is more transparent to UV photons, and is made thicker, from 2.4 mm to 5.0 mm, to increment the number of generated photons;
    \item a high-sampling-frequency digitiser, used in part of the data, to better resolve closely spaced photon-feedback signals and primary signals; 
    \item a multi-pixel photon counter to select preferentially single-electron events, in order to avoid overlapping signals from multiple incident electrons.

\end{itemize}

 \begin{figure}[tb]
  \begin{center}
    \includegraphics[width=0.8\textwidth]{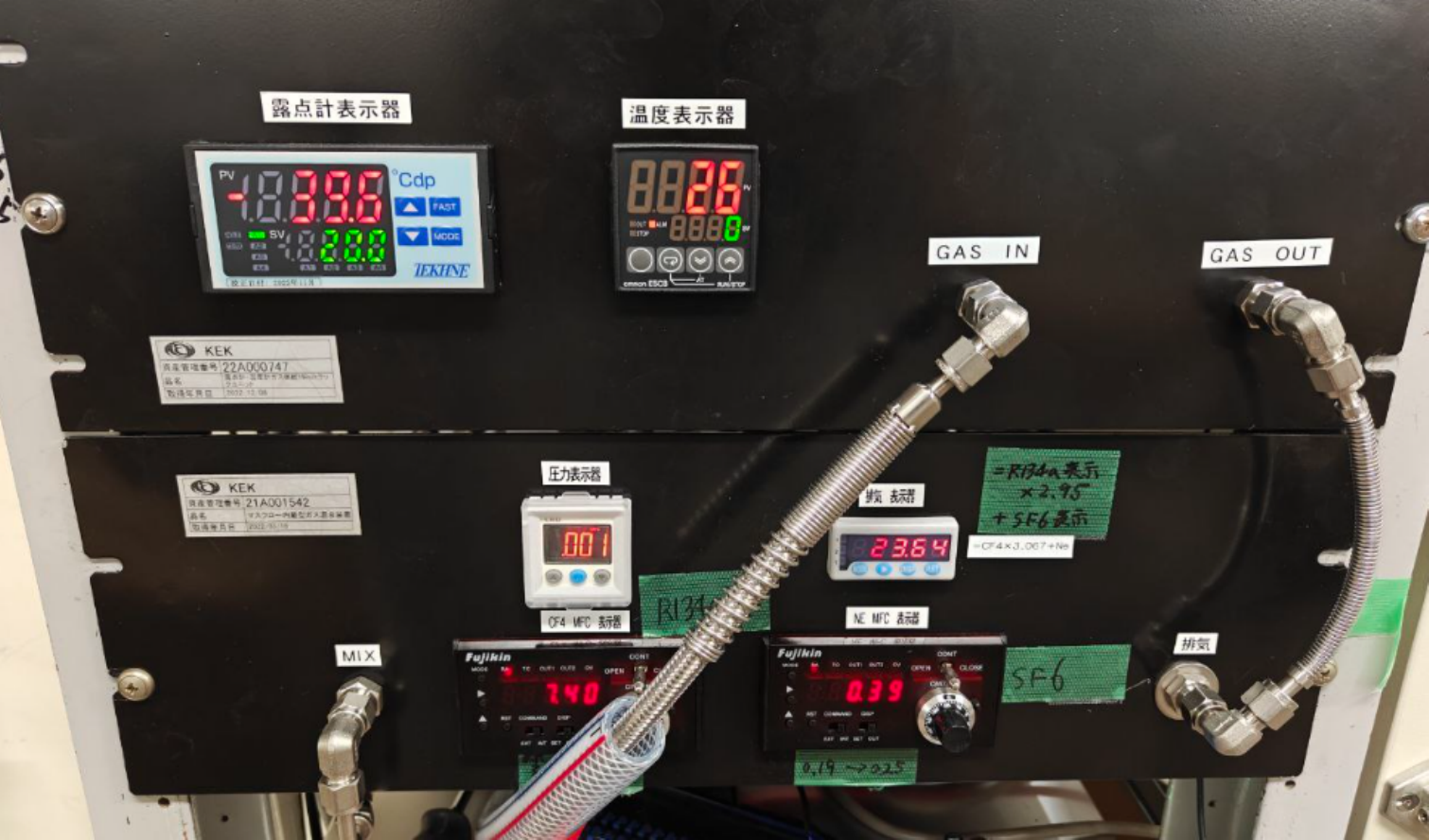}
  \end{center}
  \caption{Command panel of the gas mixing system. The two symmetrical displays at the bottom are the input thermal mass flow-meters for R134a and SF$_6$ respectively. The proportion are selected by turning the knobs under the displays. The display in the top-right corner provides the output flow-rate. Other quantities are provided including temperature, pressure and dew point.}
\label{fig:gas_system}
\end{figure}

\section{Preparation}
After assembling the aluminium structure that houses the photodetectors, we align the experimental set-up using precise laser-line generators collinear with the ideal beam line. 
We then connect to the gas bottles the circulation system that provides the proper gas mixture to GasPM and RPC and switch it on to begin to fill the detectors and purge any air inside them. Two thermal-mass flow-meters are used to select the desired proportion of each gas (Fig.~\ref{fig:gas_system}). The initial gas conditions are 4.5 cm$^3$/min flow-rate for R134a and 0.5 cm$^3$/min for SF$_6$, dew point of $-10.3^{\circ}\,\mathrm{C}$ and downstream flow-rate of 22.41 display units. The displayed flow-rates differ from the actual values because the system was originally calibrated for different gases.
Then, I set up two PCs for data collection and a third PC for online analysis, and I check the proper operation of the acquisition boards and respective software.
I also prepare the electronic modules dedicated to supply power (HV), trigger, and data collection.
We connect all computers and the HV module to the same HUB via LAN cables, allowing remote control of the full system from a single terminal. This is convenient as 5 GeV electrons are ionizing particles that damage biological tissues and render the area near the beam line inaccessible during beam operation, or following beam induced issues, due to radiation hazard. 
Remote control allows to ramp the detector HV down, which is important to avoid damage to the CsI photocathode.

\begin{figure}[tb]
  \begin{center}
    \includegraphics[width=\textwidth]{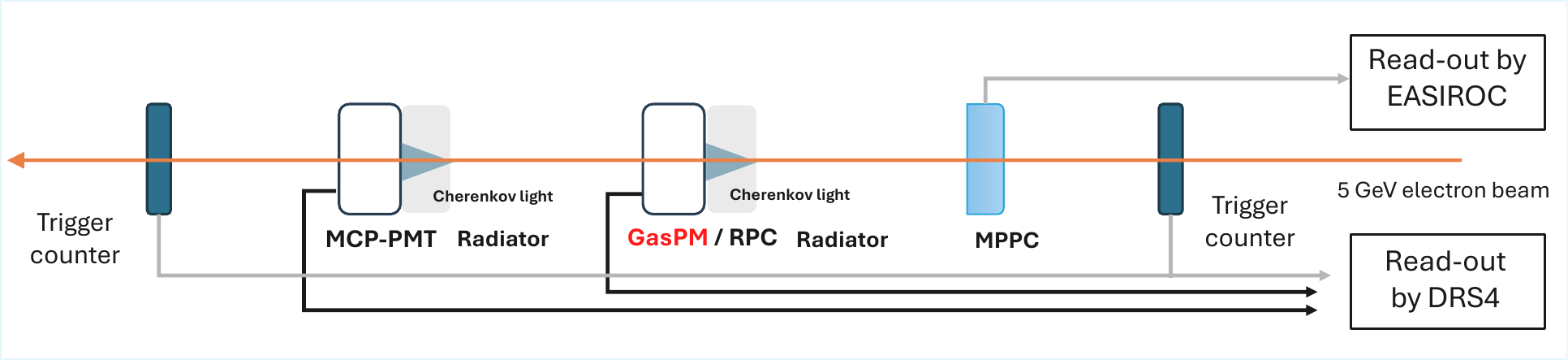}
  \end{center}
  \caption{Sketch of the detector configuration set-up (top view). The arrow indicates the direction of the beam. For clarity, only one MCP–PMT is shown althouth two are used.}
\label{detector_sketch}
\end{figure}

\subsection{Layout}
\begin{figure}[tb]
  \begin{center}
    \includegraphics[width=0.7\textwidth]{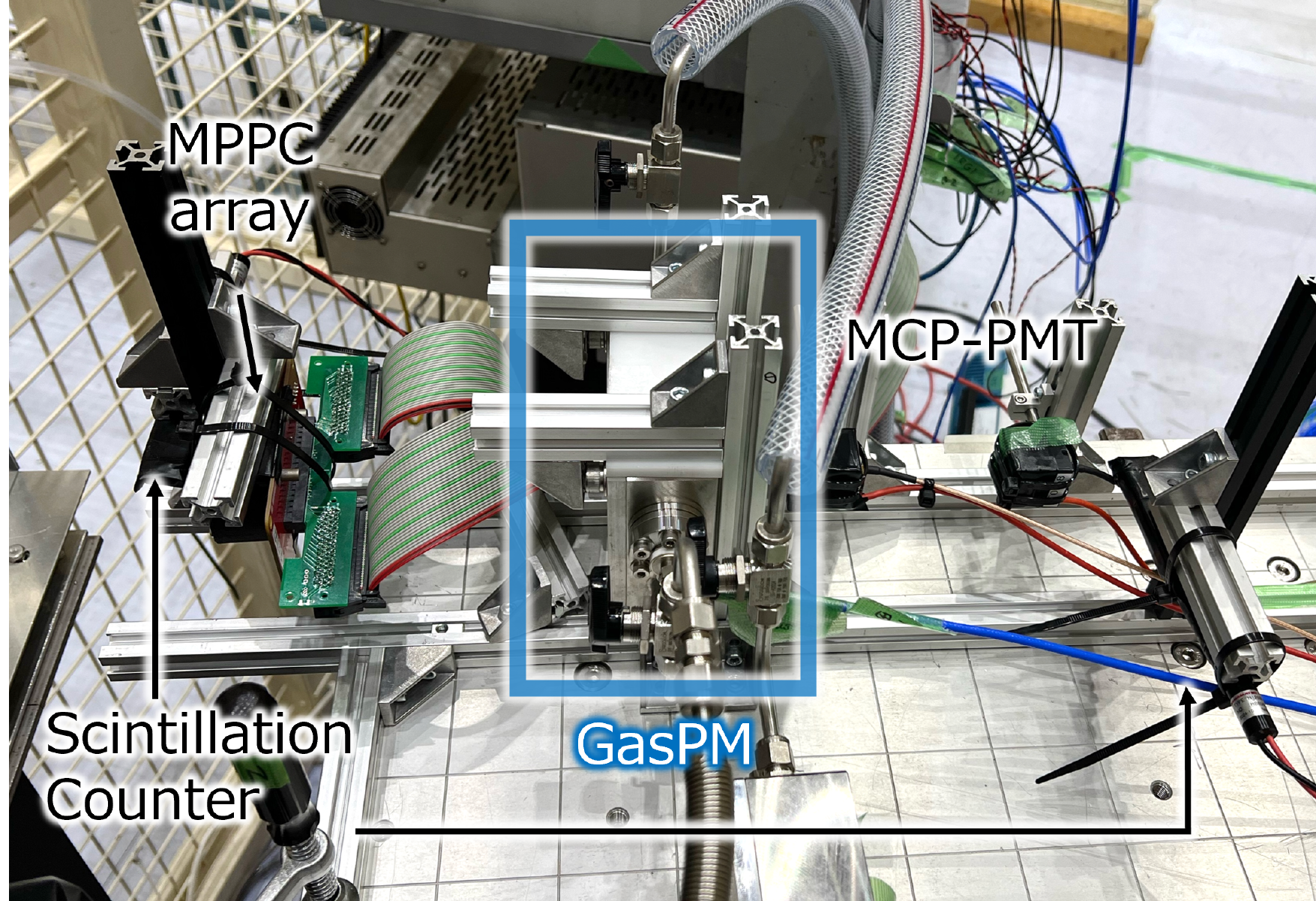}
  \end{center}
  \caption{Photo of the detector set-up. Electrons come from the left-most end.}
\label{detector_photo}
\end{figure}

A sketch of the experimental set-up is shown in Fig.~\ref{detector_sketch} and a picture is in Fig.~\ref{detector_photo}. The first detector downstream of the electron source is a scintillation counter; in connection with its counterpart in the downstream-most position, it provides a trigger through a coincidence circuit. At about 3 cm downstream, the multi-pixel photon counter distinguishes single electron hits from multiple electrons. About another 4 cm downstream, the GasPM is installed. A pair of MCP-PMTs follows, displaced by roughly 3 cm. These detectors are used together, or in coincidence with others, for a reference time-of-flight measurement.
The GasPM is occasionally swapped with the RPC to compare hit-rate and exclude GasPM events from ionisation.
All detectors are connected to the HV and the read-out systems (DRS4/NALU and EASIROC) through a LV amplifier to preserve the read-out device from possible discharges.
A low-pass filter between the high–voltage supply and the GasPM suppresses HV noise and ripple from the bias line, ensuring stable operating voltage and thereby maintaining a uniform electric field within the chamber.

\begin{figure}[tb]
  \begin{center}
    \includegraphics[width=0.7\textwidth]{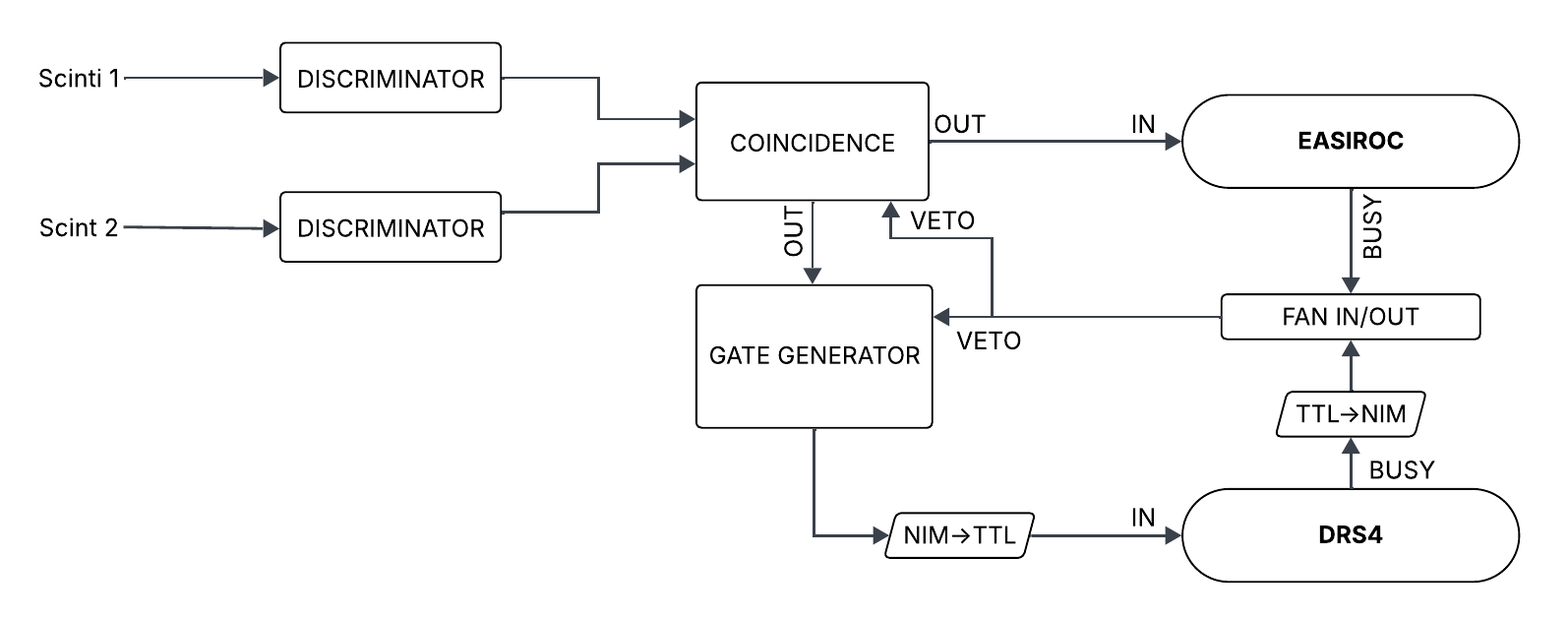}
  \end{center}
  \caption{Scheme of the trigger circuit that synchronises the two readout systems.}
\label{fig:trig_circuit}
\end{figure}

\subsection{DAQ setup}
Since data are acquired using independent devices, synchronization is necessary to enable analysis of combined datasets.
The trigger logic, schematised in Fig.~\ref{fig:trig_circuit}, generates a synchronised trigger for the digitisers DRS4 or DSA-C10-8+, dedicated to the main detectors, and the EASIROC used for the MPPC. The two scintillator outputs are fed each to a discriminator, to reduce accidental triggers, which outputs a logic pulse of 10~ns width when the input exceeds 10~mV. The discriminator outputs are sent to a coincidence module configured with a 200~ns gate. When both scintillators fire within the gate, the coincidence module generates an output signal, indicating a valid event. The coincidence output is then split and sent both to the EASIROC module directly and to the digitiser, passing through a gate generator that outputs a 80~$\mu$s gate that enables readout. Before entering the digitiser, the signal needs to be converted from NIM to TTL by a level adapter for compatibility. A FAN IN/OUT accepts the busy signals of the two readout devices as input and outputs a veto into the coincidence and gate generator. 
In this way the silicon photomultiplier is not read until the end of the digitiser acquisition. This prevents the triggers to the two systems from losing synchronization. 
The parameters chosen for the modules are based on previous beam tests and empirically modified when needed.

\section{Data acquisition}  \label{sec:daq}
Once everything is online and working, data acquisition starts. The first acquisition, with a preliminary determination of time resolution as goal, is carried out using the DRS4 digitiser.
We first measure time-of-flight using the two MCP-PMT reference detectors to make sure that the configuration is adequate. 
The observed time-of-flight resolution $\sigma_{\textrm{TOF}} = 23.5 \pm 0.07$ ps does not include the digitiser resolution, and confirms that the experimental setup is working properly.

\begin{figure}[tb]
  \begin{center}
    \includegraphics[width=0.4\textwidth]{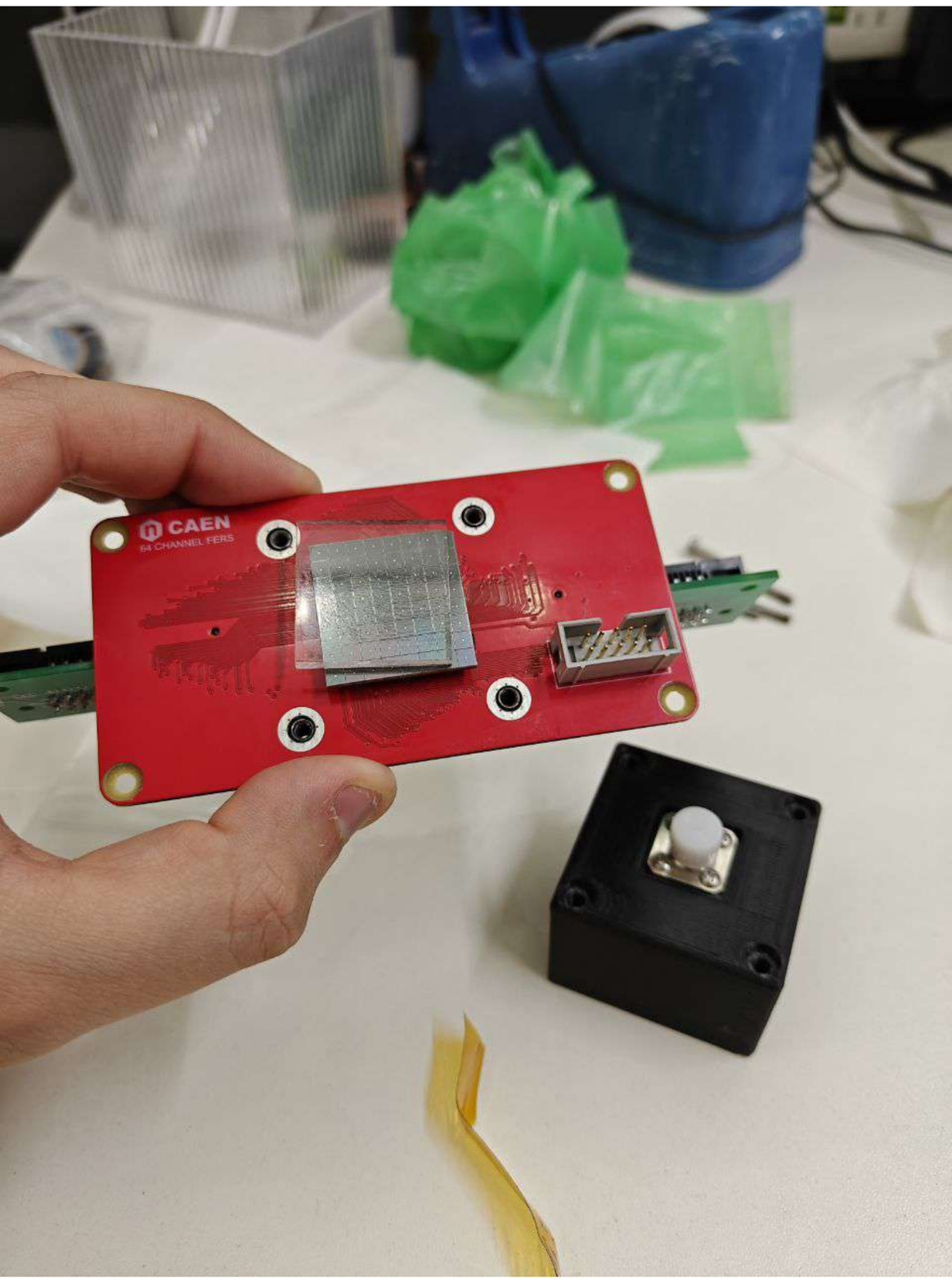}
  \end{center}
  \caption{MPPC opened during radiator replacement.}
\label{fig:MPPC_radiator}
\end{figure}

During this phase, we also acquire MPPC data, to quickly study them on-line and determine the final configuration in preparation for multiple-vs-single electron discrimination studies. We vary the MPPC radiator thickness from its 6-mm default to optimise the dimensions of the Cherenkov ring.
The Cherenkov-cone opening angle only depends on the refractive index of the radiator and of the electron velocity $\beta$; the resulting ring radius depends on that angle and on the propagation distance across the radiator. The default thickness yields a radius of 7 mm, which is comparable with the dimensions of the MPPC matrix.
Distinguishing geometrically between single- and multiple-electron hits is therefore difficult because rings overlap. 
A 3-mm radiator would make the Cherenkov ring smaller and may render selecting single-electron events geometrically easier. 
The downside is that the associated lower photon yield degrades the energetic discrimination between signal and noise.
We therefore use noise events acquired with the default radiator to form a noise model, which is independent of the radiator, and can be statistically subtracted from the data taken with the thin radiator.
Figure~\ref{fig:MPPC_radiator} shows the MPPC during the radiator replacement.

We then proceed by powering the RPC, which temporarily replaces the GasPM, to verify the suitability of the setup for gas detectors. Measuring the RPC hit-rate also serves to later exclude ionisation signals from the GasPM dataset. 
We increase electromagnetic shielding and shorten the cables to reduce initially observed electromagnetic environmental noise.

\begin{figure}[tb]
    \centering
    \includegraphics[width=0.495\textwidth]{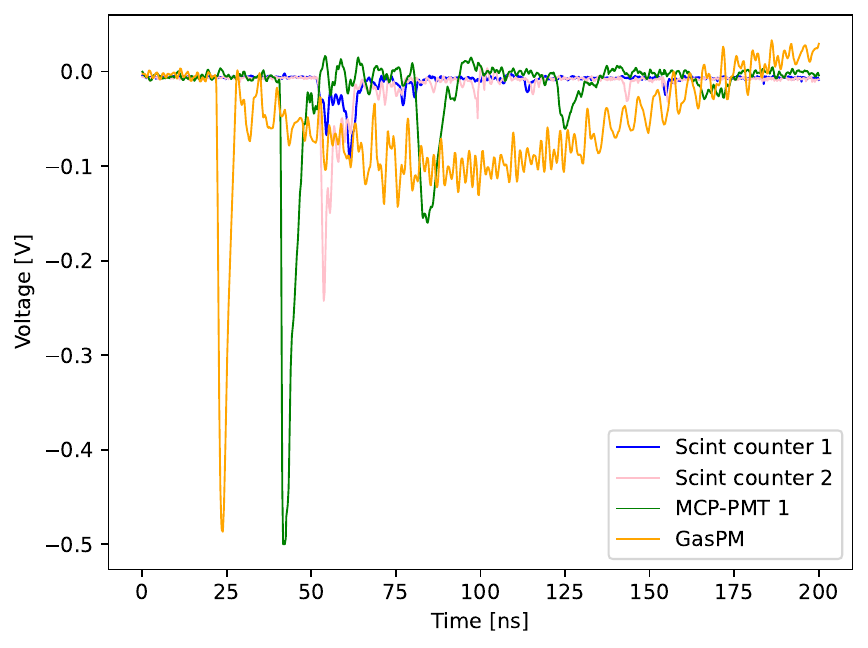}
    \hfill
    \includegraphics[width=0.495\textwidth]{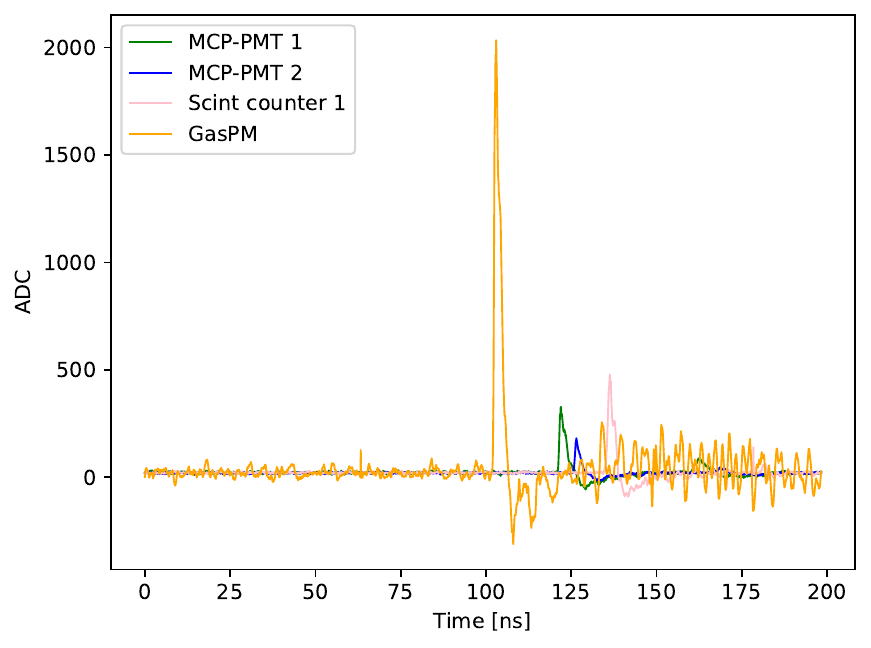}
    \captionsetup{width=0.9\linewidth} 
    \caption{Example of an oscilloscope display of a beam-test event acquired with the DRS4 (left) and NALU digitisers (right).}
    \label{fig:ev_displays}
\end{figure}

The RPC is then replaced with the GasPM, and we acquire roughly 35000 events. Figure~\ref{fig:ev_displays}(left) shows an example event. The orange curve is the GasPM signal, whereas the green curve is from the MCP-PMT. The other two curves are the trigger scintillator signals. The extended secondary deformation observed in the GasPM signal amplitude is attributed to ion feedback. Its extended duration arises from the significantly lower drift velocity of ions compared to that of electrons.

During acquisition the gas condition inside the GasPM is constantly monitored to ensure that purity levels are as expected.
We then replace the DRS4 digitiser with the more performant NALU DSA-C10-8+. The channel configuration mirrors the DRS4 one, with one channel connected to the GasPM output, one to each MCP-PMT, and one to each trigger scintillator. The GasPM output is processed through a $-20$ dB attenuator followed by a 43 dB amplifier in series, whereas the MCP–PMT signals are processed only through a $-70$ dB attenuator, keeping the same DRS4 amplification configuration.

A few short acquisitions are used to check the proper NALU operation. Here, the GasPM is powered off to preserve the photocathode and the digitiser parameters are tuned to observe signal from the MCP-PMT reference detectors. The acquisition window is set to 2048 samples, corresponding to 204.8 ns. Saturation is observed: the signal amplitude exceeds the dynamic range of the digitiser so that if the input reaches the boundaries, any increase causes no appreciable change in the output.
We therefore modify the signal attenuation to $-90$ dB for the MCP-PMTs and we add a $-6$ dB attenuator for the GasPM. 

The GasPM is then powered on and 5000 events are acquired. An example event is shown in Fig.~\ref{fig:ev_displays}(right). The orange signal is the GasPM, the green and blue curves are the two MCP-PMT signals, simultaneously connected in this acquisition. The pink curve is the first scintillator signal, the second scintillator is not shown. The amplitudes of the signals are expressed in ADC counts since this is an event display from the beam test and the calibration was only applied a posteriori. Due to the DSA-C10-8+ input stage configuration, which includes an inverting amplifier, the signal has positive polarity. We then study the dependence of amplification on the electric field inside the GasPM. We increase the voltage applied to the detector from 2.8 kV to 3.0 kV and proceed to acquire roughly 5000 events.

\begin{figure}[tb]
  \begin{center}
    \includegraphics[width=0.7\textwidth]{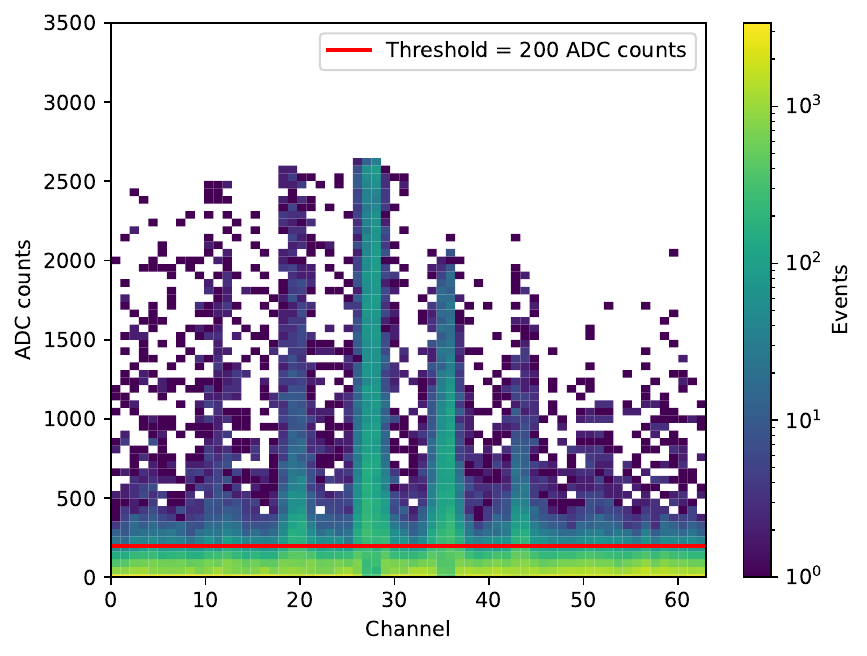}
  \end{center}
  \caption{Distribution of hit multiplicity in data taken with a 3-mm radiator. The 200 ADC threshold for a photon hit is shown.}
\label{fig:ADC_COUNTS}
\end{figure}

\begin{figure}[tb]
  \begin{center}
    \includegraphics[width=0.6\textwidth]{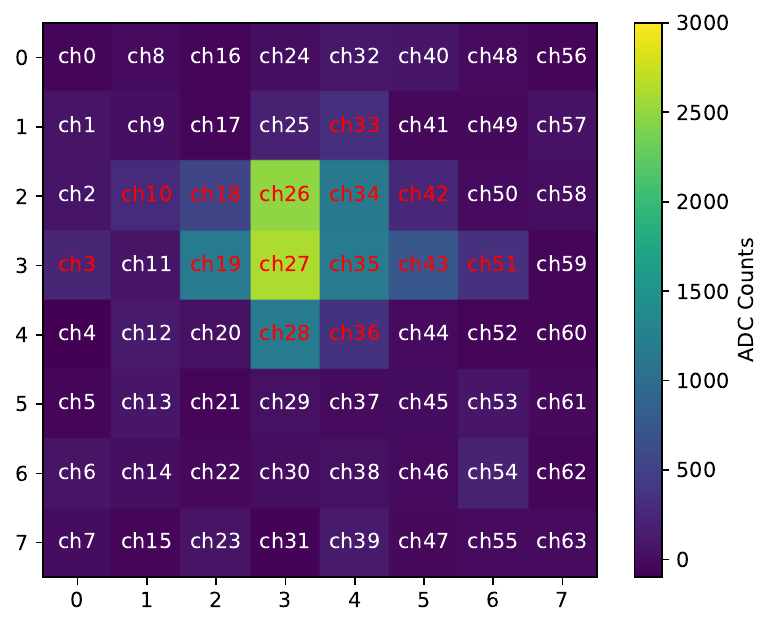}
  \end{center}
  \caption{Signal, in colour-coded ADC counts, detected in the MPPC channels for a beam-test event. Red labels channels that detect a signal exceeding the 200-ADC-counts photon-hit threshold.}
\label{fig:MPPC_map}
\end{figure}

\begin{figure}[tb]
    \centering
    \includegraphics[width=0.45\textwidth]{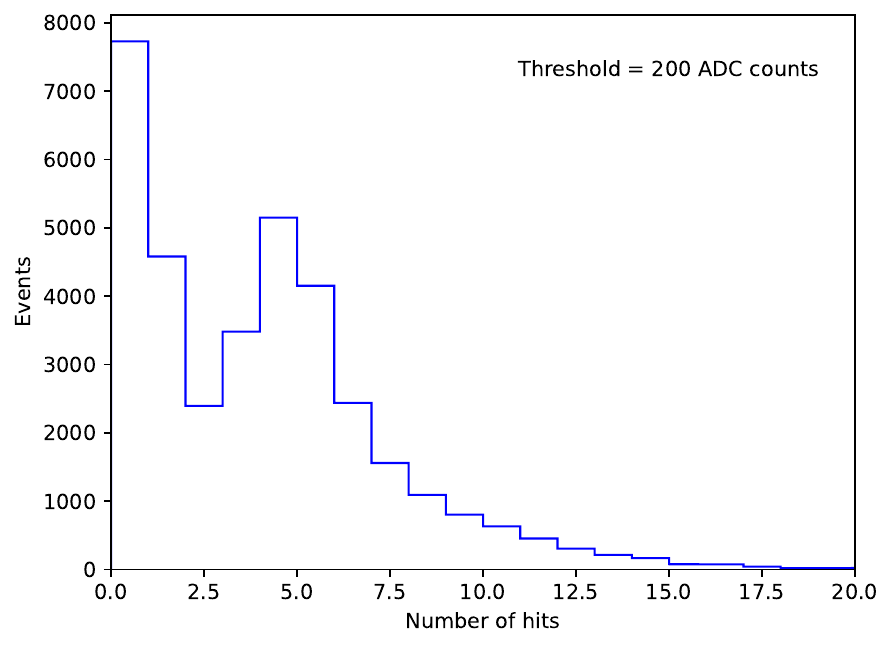}
    \hfill
    \includegraphics[width=0.45\textwidth]{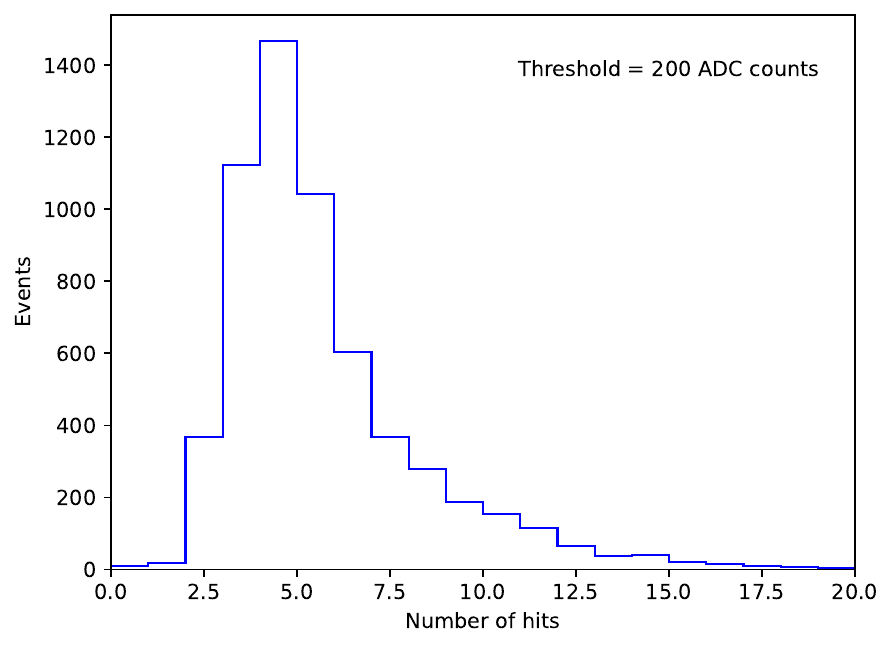}
    \captionsetup{width=0.9\linewidth} 
    \caption{Distribution of hit multiplicity in beam-test data taken with a 3-mm radiator before (left) and after (right) baseline selection.  Photon-hit threshold is 200 ADC counts.}
    \label{fig:n_hits_200}
\end{figure}

\begin{figure}[tb]
  \begin{center}
    \includegraphics[width=\textwidth]{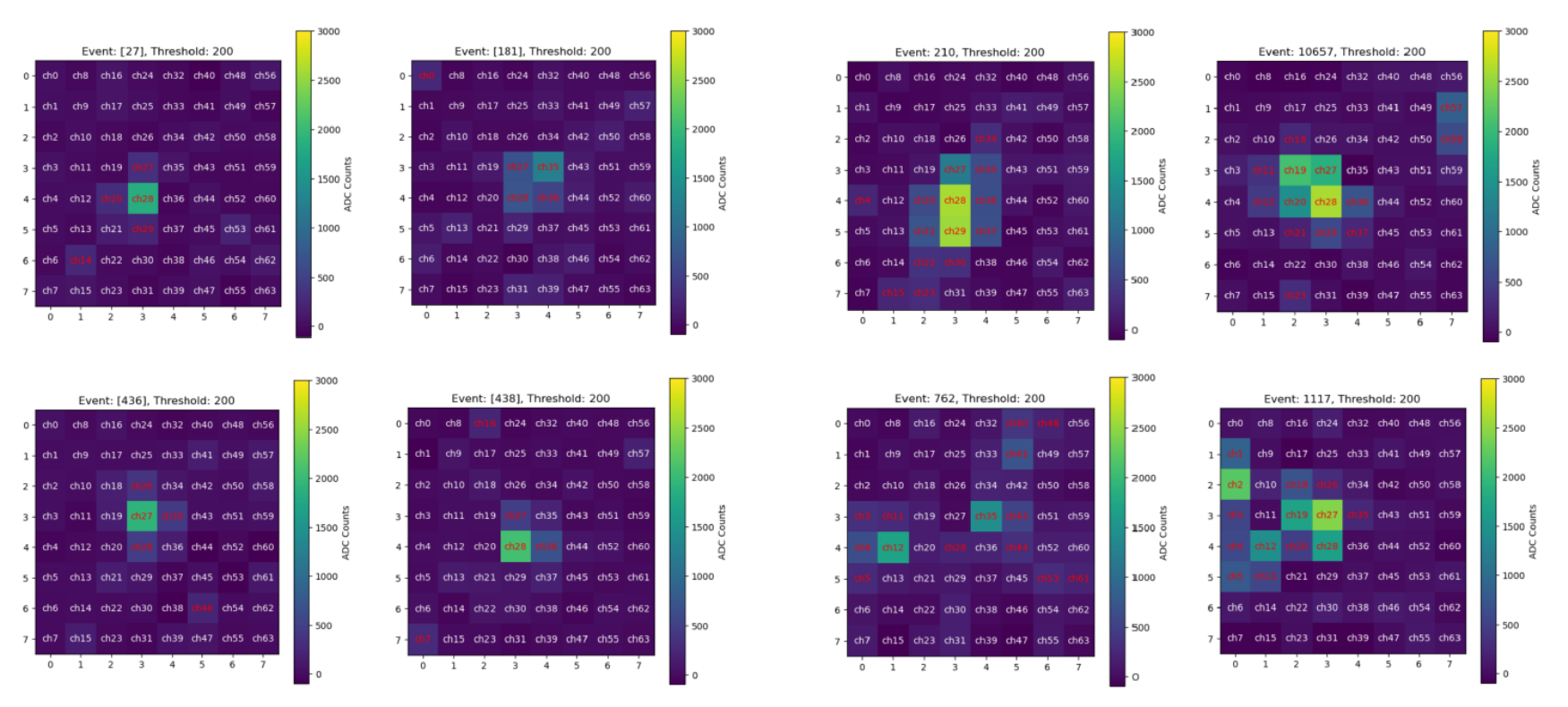}
  \end{center}
  \caption{Signal, in colour-coded ADC counts, detected in the MPPC channels for 8 beam-test events. In the 4 events displayed in the left (right) panels, hits are detected in 5 (14) channels. }
\label{fig:MPPC_hit_events}
\end{figure}

\section{Single-electron event selection}
Multiple electrons can reach the GasPM due to upstream $\delta$-rays, which are secondary electrons produced when energetic charged particles ionise the detector material.
Events with multiple electrons simultaneously impinging on the GasPM bias the time-resolution measurement as the resulting number of collected photoelectrons grows, and the correspondingly larger gain enhances statistical precision.
We use the newly added MPPC detector to suppress events from multiple electrons. This is an improvement over the 2023 beam test.
An inconvenient feature of the MPPC is the presence of dark counts, spurious signals recorded in the absence of actual photon hits. The most common cause is thermionic emission, where thermal energy causes spontaneous electron emission from the sensor active area. Leakage current in the readout system too can produce dark counts. These effects are independent of the MPPC position relative to the beam and are therefore expected to appear uniformly across channels.
Figure~\ref{fig:ADC_COUNTS} shows the ADC count for each channel. The dark counts populate the yellow region at low ADC counts. The asymmetry of the distribution shows residual alignment imperfections.
I empirically define a 200-ADC-counts threshold to determine if a channel records a sufficient number of photons to distinguish signal from dark noise.  We assign the same threshold to all channels after verifying that they all have similar gain.  

The event display in Fig.~\ref{fig:MPPC_map} shows the channel response in ADC counts for a typical event. Red numbers label channels that exceed the 200-ADC-counts threshold, hence indicating a photon hit, as per our choice. 
The hit-multiplicity distribution is shown in Fig.~\ref{fig:n_hits_200}(left panel). The first peak at zero hits is due to dark counts that are not read as photon hits, whereas the second, centred at five hits, represents one electron passing through the radiator.

I then apply a baseline filtering of the events suited for GasPM analysis, based on the choices made during the 2023 beam test. Occasionally, the trigger fires on Cherenkov interactions of beam-halo electrons with the light guide rather than on genuine scintillation light from its scintillators. To enrich the sample in events whose electron traversed the active volume of all detectors, we impose a threshold on scintillation-pulse height and charge, which are significantly lower for non-scintillation signals. In addition, a selection on the GasPM pulse heights reduces saturated signals.
This ensures that events generated by electrons traversing the entire detector setup are preferentially used.

The resulting distribution of hits is shown in Fig.~\ref{fig:n_hits_200}(right panel). As expected, the peak at zero hits disappears, owing to the baseline selection that suppresses triggered events with no electron, thus no photons, hitting the MPPC. In this distribution, we would ideally expect to observe distinct peaks corresponding to events with one, two, or more electrons. However, the possible presence of these distinct peaks is difficult to discern, as the distribution grows rapidly until its mode at about five hits, and then degrades smoothly. This is likely due to the reduced photon yield caused by the thinner, 3 mm, radiator discussed in Sec.~\ref{sec:daq}. 
I therefore inspect event displays to isolate structures possibly useful for single- vs multiple-electron discrimination.
Event displays show noticeable differences among events with different number of hits. Figure~\ref{fig:MPPC_hit_events} compares events with 5 hits, corresponding to the mode of the distribution, and events with 14 hits. Trying to visually identify the presence of different Cherenkov rings is challenging. 

\begin{figure}[tb]
    \centering
    \begin{minipage}[t]{0.45\textwidth}
        \centering
        \includegraphics[width=\textwidth]{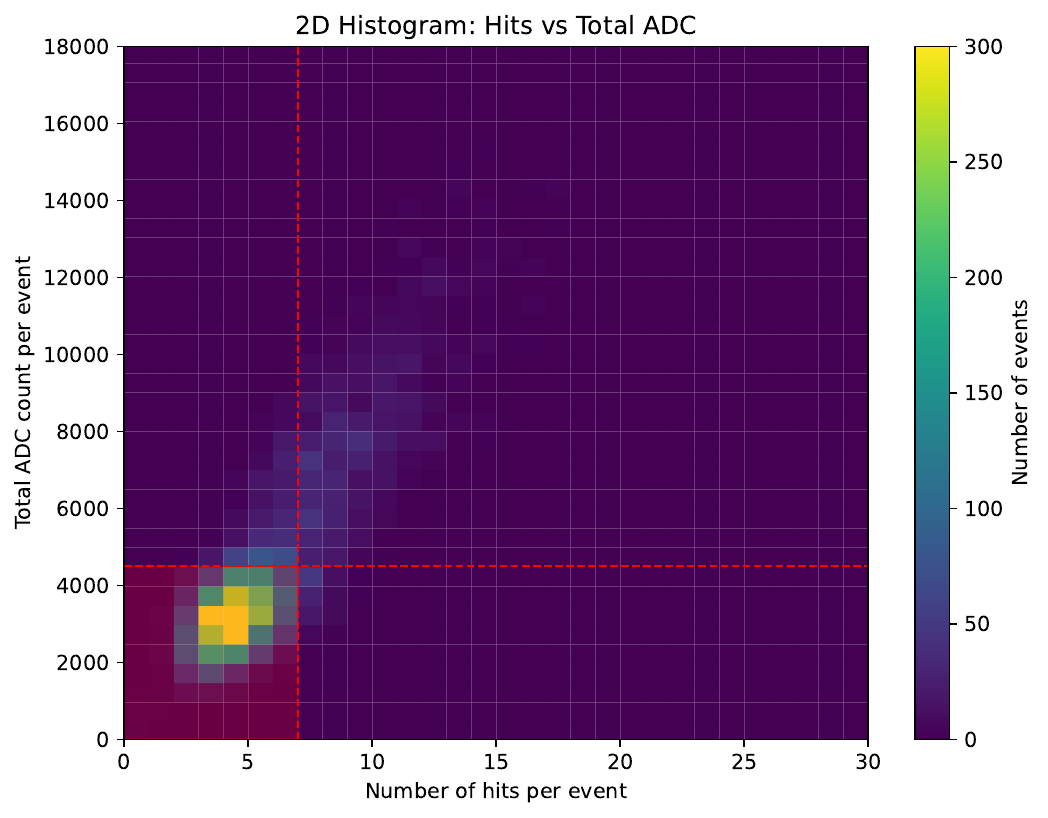}
        \captionsetup{width=\linewidth} 
        \caption{Two-dimensional histogram of the number of photon hits per event versus the total collected ADC counts. The red hatched region indicates the accepted region for single-electron events.}
        \label{fig:MPPC_cuts}
    \end{minipage}\hfill
    \begin{minipage}[t]{0.45\textwidth}
        \centering
        \includegraphics[width=\textwidth]{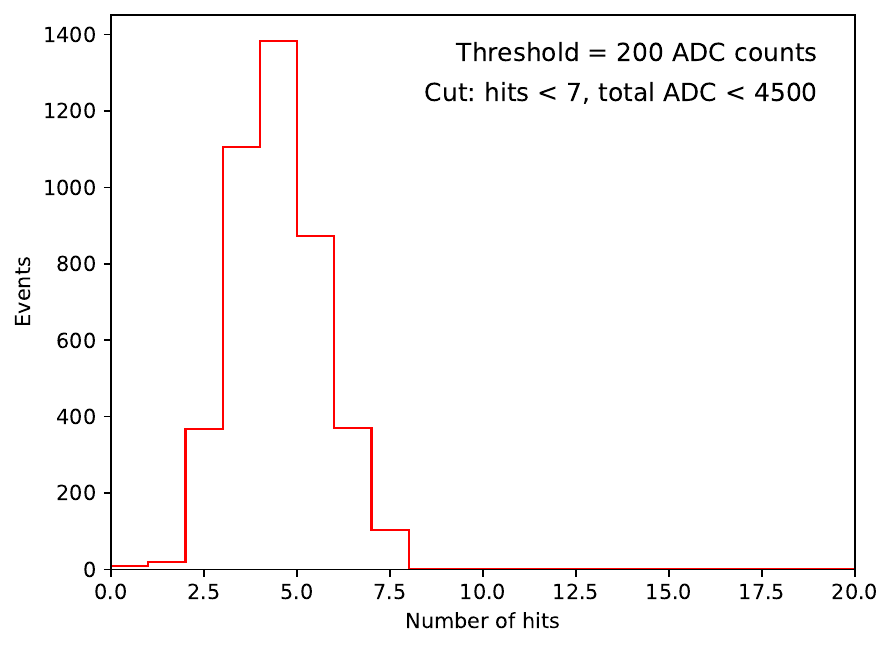}
        \captionsetup{width=\linewidth} 
        \caption{Distribution of the hit multiplicity after the single-electron selection. The threshold for a photon hit is set at 200 ADC counts.}
        \label{fig:n_hits_after}
    \end{minipage}
\end{figure}

I therefore explore a higher-dimensional discriminating observable. I define single-electron events based on the correlation of two observables: the total collected ADC counts, which is sensible to presence of multiple photons on the same MPPC channel, and the number of hits. These two quantities are good discriminants for identifying single-electron events as shown in Fig.~\ref{fig:MPPC_cuts}. A distinctive structure suggests the presence of single-electron events, which typically show relatively small multiplicity and a total ADC-count sum inferior to approximately 4500, thus constituting a visible cluster. In contrast, multiple-electron events result in broader spatial distributions of hits and significantly higher total number of photons. Hence, I empirically require fewer than 7 hit channels and 4500 total ADC counts to select single electron events, as shown by the red lines in Fig. \ref{fig:MPPC_cuts}. The resulting hit-multiplicity distribution is in Fig.~\ref{fig:n_hits_after}. This criterion tags as multiple-electron events (28.1 $\pm$ 0.6)\% of the total events. 

This analysis, although empirical and preliminary, confirms that we are able to suppress events originating from multiple electrons hitting the detector simultaneously by employing an MPPC. 
This represents a meaningful improvement over the 2023 beam test and will be adopted in the next beam test.

\section{Photon feedback study}
This section addresses an important topic of this work, the study and suppression of photon feedback, one of the main limitations observed to date in GasPM operations. Mitigating its impact on the time-resolution measurement is important to optimise the GasPM toward the intended goals.
Photon feedback originates from UV photons, emitted during the excitation and de-excitation of gas molecules, travelling backwards, and impinging onto the photocathode. The ensuing avalanche, which is only partially mitigated by the quenching gas,  appears as a secondary signal in the detector output, spoiling time resolution.

Identification of the events affected by photon feedback is made more challenging by the reduced gas gap that shortens the path to the photocathode, thus reducing the delay of the photon-feedback avalanche.
We therefore develop a procedure to discriminate against events with photon feedback using the upgraded 10 GHz digitisation system.

\subsection{Signal shape survey}
Since beam time is limited, the available dataset consists of only 4365 events. I therefore do not apply the full set of baseline selection criteria, but just a simple requirement on the delay between the signal and the trigger: I consider only signals exceeding 20~mV that occur between 33~ns and 35~ns before the scintillation signals, which is a known delay due to the cable length. The remaining sample contains 1227 events.

\begin{figure}[h!]
  \begin{center}
    \includegraphics[width=\textwidth]{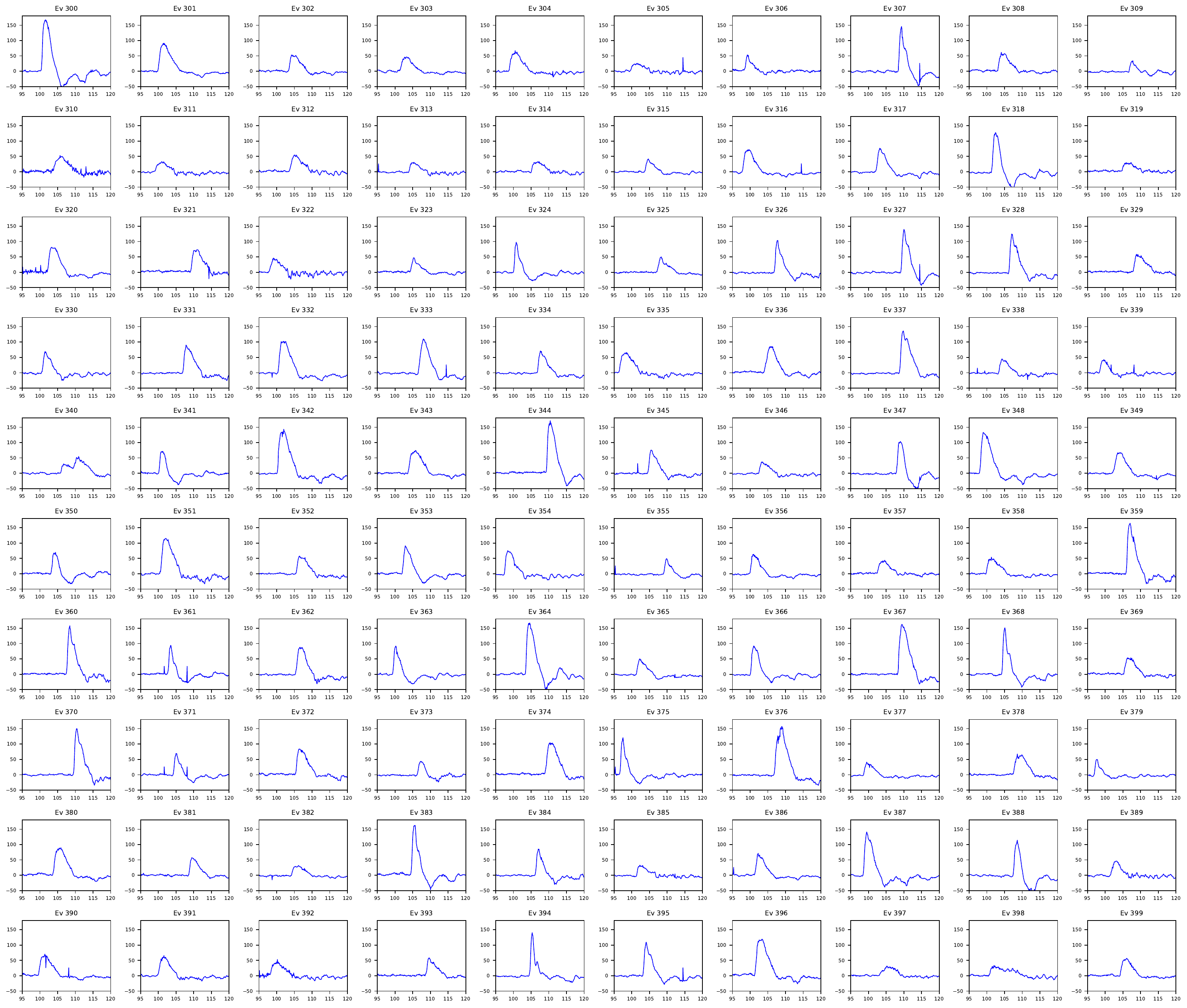}
  \end{center}
  \caption{A random sample of 100 signal shapes from the NALU beam test acquisition. }
\label{fig:shapes}
\end{figure}

Figure \ref{fig:shapes} shows a random subset of 100 signals. 
Classifying systematically these signals is challenging, as they exhibit markedly different shapes and amplitudes. Even identifying distinctive features is non-trivial, since multiple, simultaneously concurring, effects determine the signal shape. The sample contains events with single- or multiple-electron signals (the MPPC was not used during this acquisition), events with or without photon feedback, events involving different Cherenkov-photon yield, and voltage reflections introduced by the low-pass filter placed between the GasPM and the HV generator.
The average number of photons impinging on the photocathode that generate a photoelectron, and are therefore detected by the GasPM, is obtained by the Poisson distribution considering the probability of observing no hit $p = 1- \textrm{GasPM efficiency}$. When no signal is observed,  $n$ equals zero, and the average number of detected photons is $\lambda = -\ln(1-p)=1.71\pm0.08$, using the (63 $\pm$ 2)\% observed efficiency.

\begin{figure}[tb]
    \centering
    \includegraphics[width=0.48\textwidth]{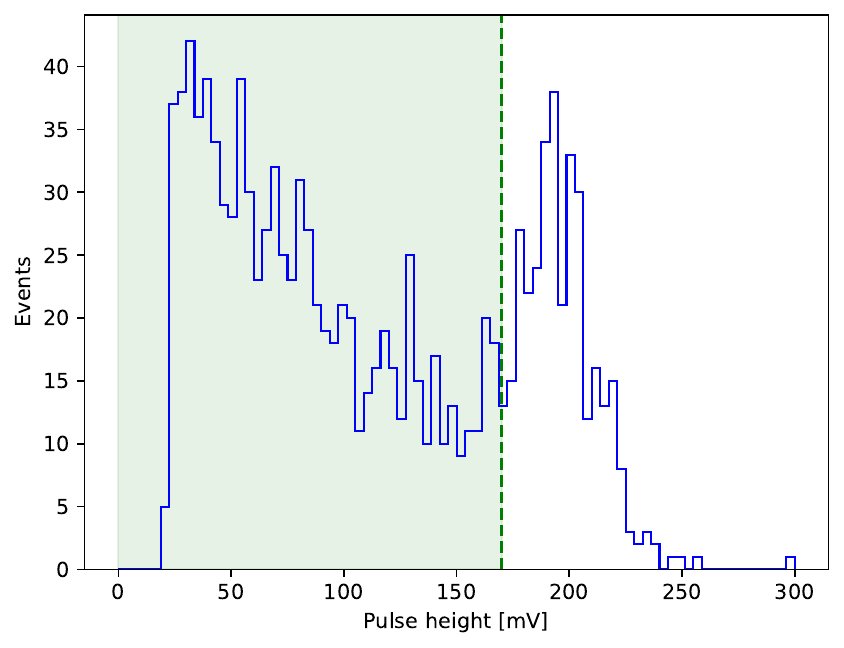}
    \hfill
    \includegraphics[width=0.5\textwidth]{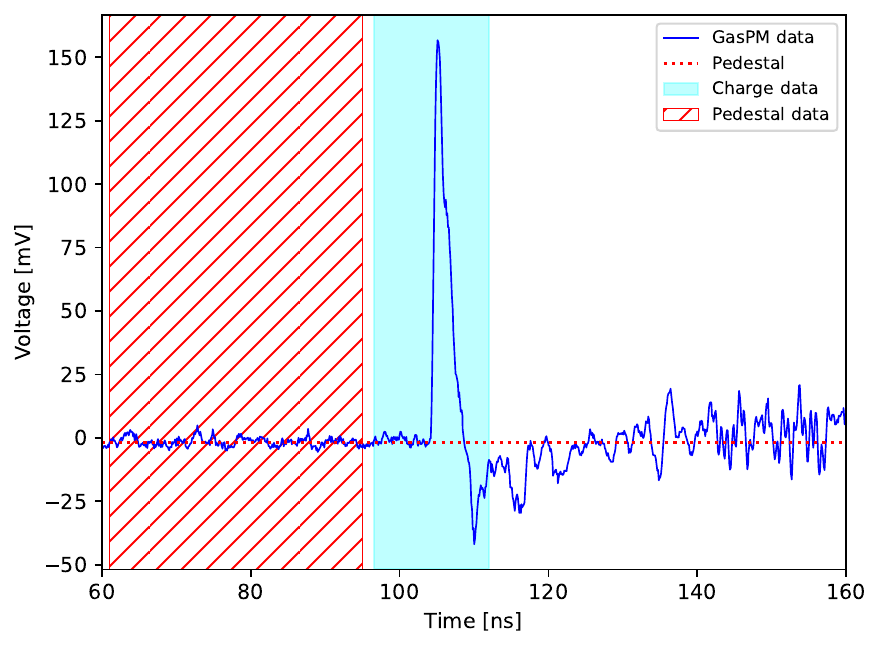}
    \captionsetup{width=0.9\linewidth} 
    \caption{(Left panel) Pulse-height distribution from beam test data. The green region indicates the events retained for the analysis. (Right panel) Typical signal of the GasPM prototype recorded by the NALU digitiser with 27 dB external amplification. The red dotted line is the pedestal, calculated from the sampling points in the red hatched region.}
    \label{fig:heights}
\end{figure}

The pulse heights span a wide spectrum, from weak signals, around 30~mV, up to large ones exceeding 220~mV. Our hypothesis is that the observed broad spectrum arises from distortions in the drift field in events with multiple detected photons: the first photoelectron initiates an avalanche; other photoelectrons, generated by additional Cherenkov photons from the same primary electron, would encounter a locally reduced electric field due to the space charge generated by the first avalanche, leading to a lower gain. The gain degradation would depend on the relative distance between the photoelectrons. In addition, ionisation processes further modify the electric field, affecting the signal shape as well.
Figure~\ref{fig:heights}(left) shows the pulse-height distribution after amplification. The distribution rises to its mode at roughly 40~mV, then it decreases smoothly up to about 170~mV, where it starts to increase again. We attribute the second peak to saturation due to input amplitudes exceeding the dynamic range of the NALU digitiser. We therefore discard all events with pulse above 170~mV, reducing the event yield to 895.

The charge of the pulses varies too. Signals waveforms appear more complicated than those obtained in the earlier laser test performed with a different photocathode~\cite{Okubo-GasPM}. This is ascribed mainly to differences in source and environment, and obfuscate any potential correlation between charge, rise time, and pulse height that could indicate photon feedback as used in previous laser tests.

A typical signal, shown in Fig.~\ref{fig:heights}(right), rises steeply within approximately 1~ns and then decays over a longer timescale of typically 4--5~ns. Decay is mostly determined by the capacitance between the GasPM electrodes. The red hatched region is used to compute the pedestal and thus the pulse height; the cyan region is used for charge computation. Since our goal is the study of the photon feedback, which mostly affects the rise-time, we focus on the rising edge of the waveform.

\subsection{Classification}

\begin{figure}[htb]
  \begin{center}
    \includegraphics[width=0.7\textwidth]{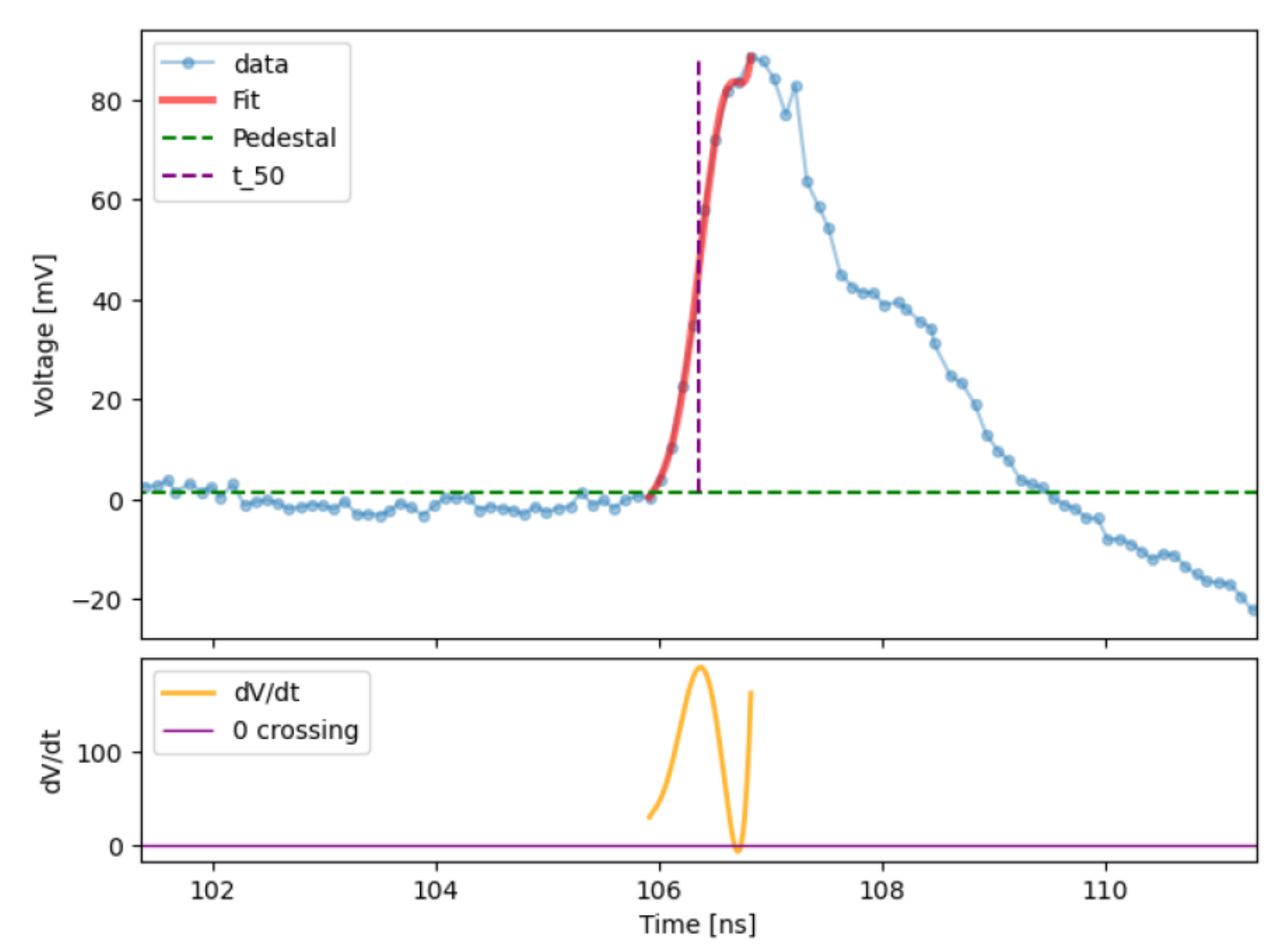}
  \end{center}
  \caption{Event classified as affected by photon feedback using the first-derivative zero-crossing method. The first derivative is shown in the bottom panel. The red curve represents a 20th-order polynomial fit. The pedestal is indicated by the green dashed line, and the signal timing is indicated by the vertical dashed line at the 50\% of the pulse height.}
\label{fig:loc_max}
\end{figure}

I develop criteria to classify events as affected by photon feedback focussing on the rising edge.
I first fit the rising edge of the signal with an 8th-order polynomial from the baseline to the peak. I compute the derivative of the fit function, and search for zero-crossings, which directly correspond to local maxima. I tag the events with a zero-crossing as those likely affected by photon feedback. Figure \ref{fig:loc_max} shows an example. The red curve represents the fit of the rising edge, while the black dashed line marks the signal time, defined as the time corresponding to the 50\% of the pulse height. The pedestal is the green horizontal dashed line. The lower panel shows the first derivative of the fit function, with the zero-crossing highlighted.

This method tags 35.6\% of the events as affected by photon feedback. By inspecting events individually though, I notice that only events with a visibly clear secondary bump are getting selected. Even though these events are likely affected by photon feedback, they are so evident that also the lower-frequency DRS4 digitiser could recognise them. In other words, the method fails to classify the events in which the two peaks are strongly overlapping.

\begin{figure}[h!]
    \centering
    
    \begin{subfigure}[b]{0.495\textwidth}
        \centering
        \includegraphics[width=\textwidth]{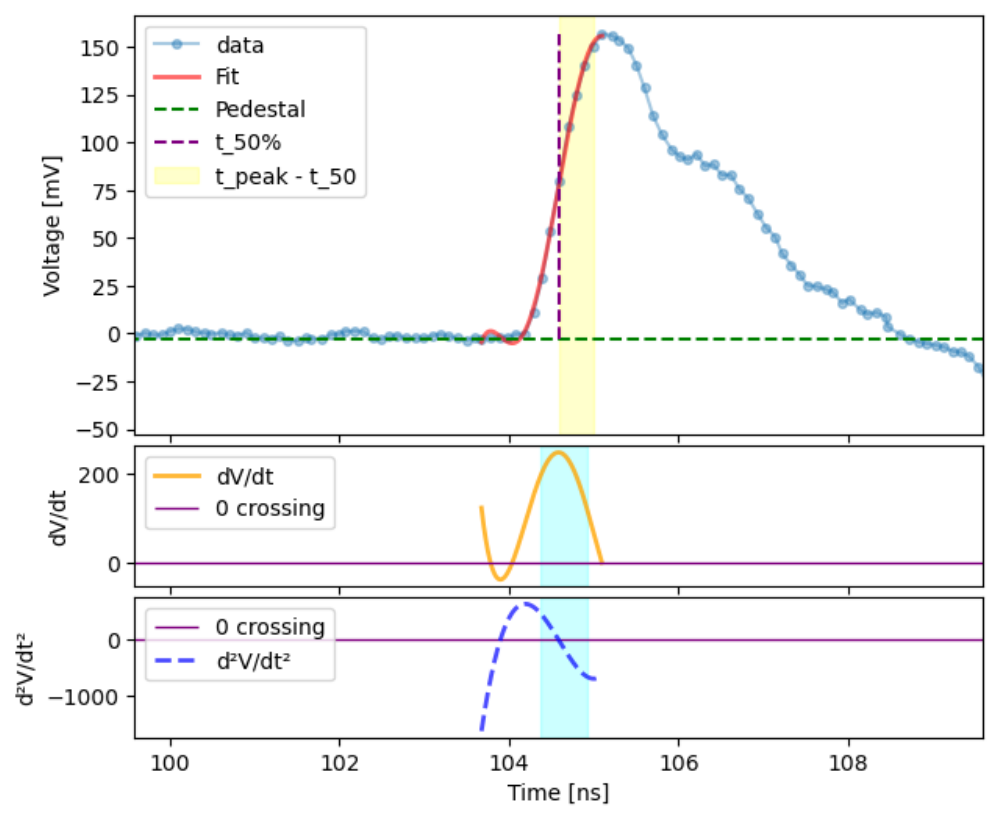}
        
        \includegraphics[width=\textwidth]{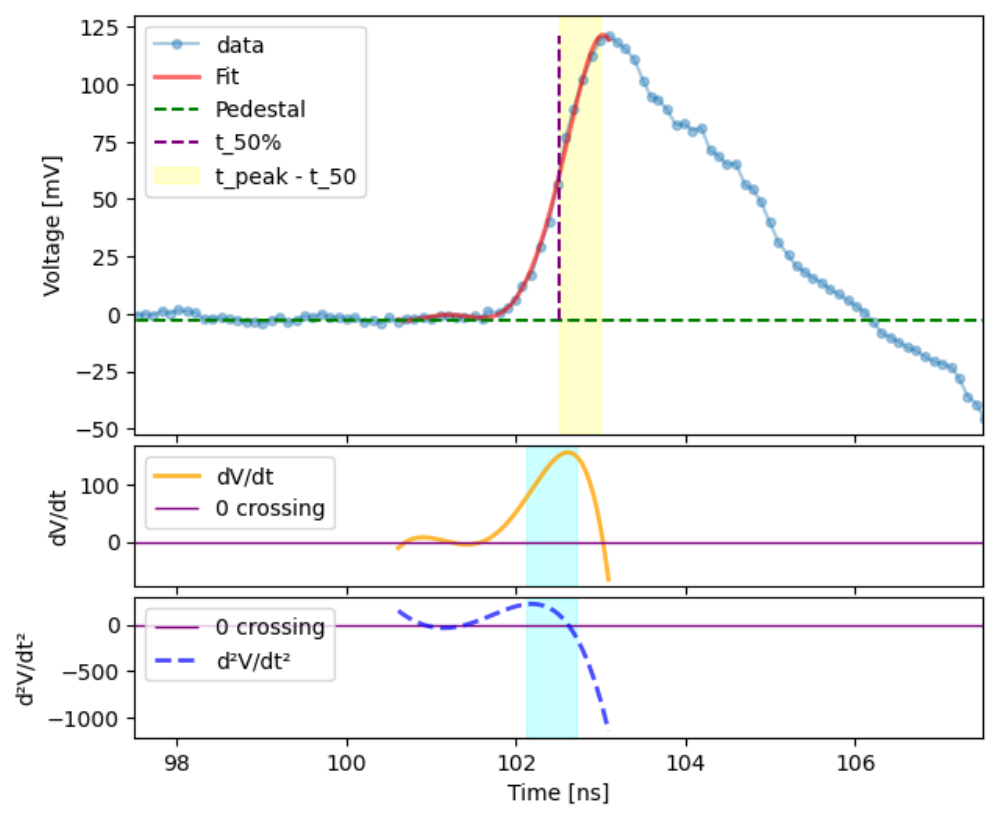}
        
        \caption{Single avalanche events.}
        \label{fig:col1}
    \end{subfigure}
    \hfill
    \begin{subfigure}[b]{0.483\textwidth}
        \centering
        \includegraphics[width=\textwidth]{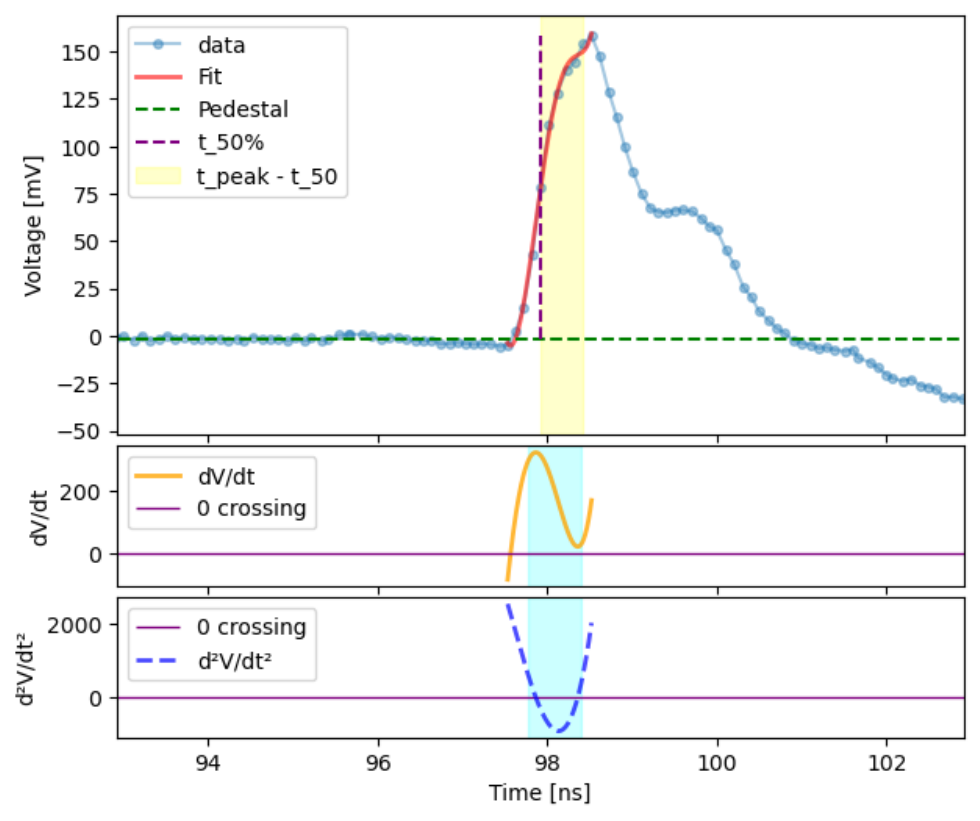}
        
        \includegraphics[width=\textwidth]{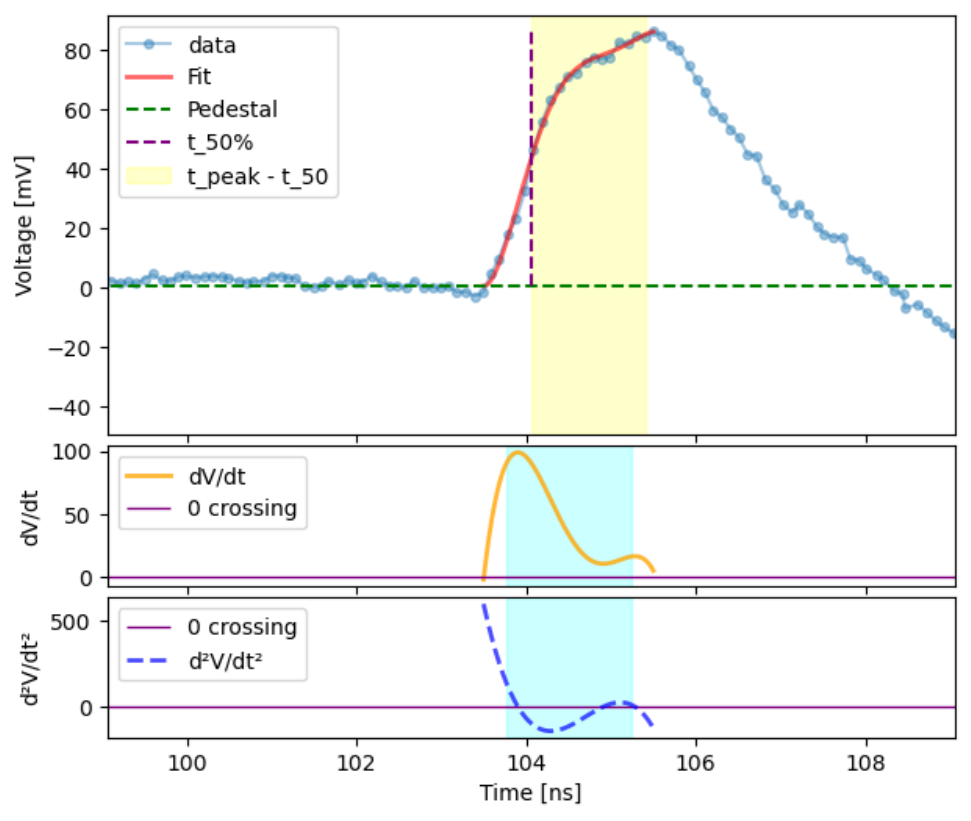}
        
        \caption{Photon feedback events.}
        \label{fig:col2}
    \end{subfigure}
    
    \caption{Signal waveforms observed during the beam test and  analysed using the second derivative zero-crossing method for events tagged as without (left panel) and with (right panels) photon feedback. First and second derivatives are shown in the bottom subpanels. The red curve represents the 8th-order polynomial fit. Signal timing is defined at 50\% of the pulse height, while the yellow hatched areas indicate the rise time between 50\% of the pulse and the peak. In the derivative panels, the cyan hatched region marks the zero-crossing check interval.}
    \label{fig:flex_point}
\end{figure}

I therefore develop a more precise and selective method.
I search for local maxima in the first derivative, which correspond to zeros of the second derivative. These inflection points, mark the location of the steepest slope in the rising edge. The events with multiple such points, have a kink in the rising edge and are selected.
Figure~\ref{fig:flex_point} shows four events selected with this approach, where both the first and second derivatives are displayed. Events in Fig.~\ref{fig:col1} are presumably photon-feedback-free, whereas in Fig~\ref{fig:col2}, events are tagged as affected by photon feedback. In the lower panels, the cyan region indicates the search interval for zero-crossing of the second derivative. A fixed time interval is unsuitable for this search due to the variance in signal shapes, so I empirically define the interval as the fraction of the signal rise-time that spans from 10\% to 92\% of the rising edge. The search interval stops short of the peak to avoid bias from random fluctuations. The yellow region corresponds to 50\%–-100\% of the rise time.

This method classifies $(53.2 \pm 2.3)$\% of the events as affected by photon feedback, which includes all events previously identified using the simpler approach. This fraction is significantly larger than the approximately 30\% estimated during the laser test~\cite{Okubo-GasPM}. An increase is expected since the extremely low quantum efficiency in the laser test resulted in signals most likely produced by single photons, whereas here several photons are likely to contribute, leading to a higher rate. 
In addition, our CsI photocathode features higher quantum efficiency with respect to the previous one, and this also increases the probability of emitting a photoelectron from the UV photons responsible for photon feedback.

\clearpage

\begin{figure}[tb]
  \begin{center}
    \includegraphics[width=0.7\textwidth]{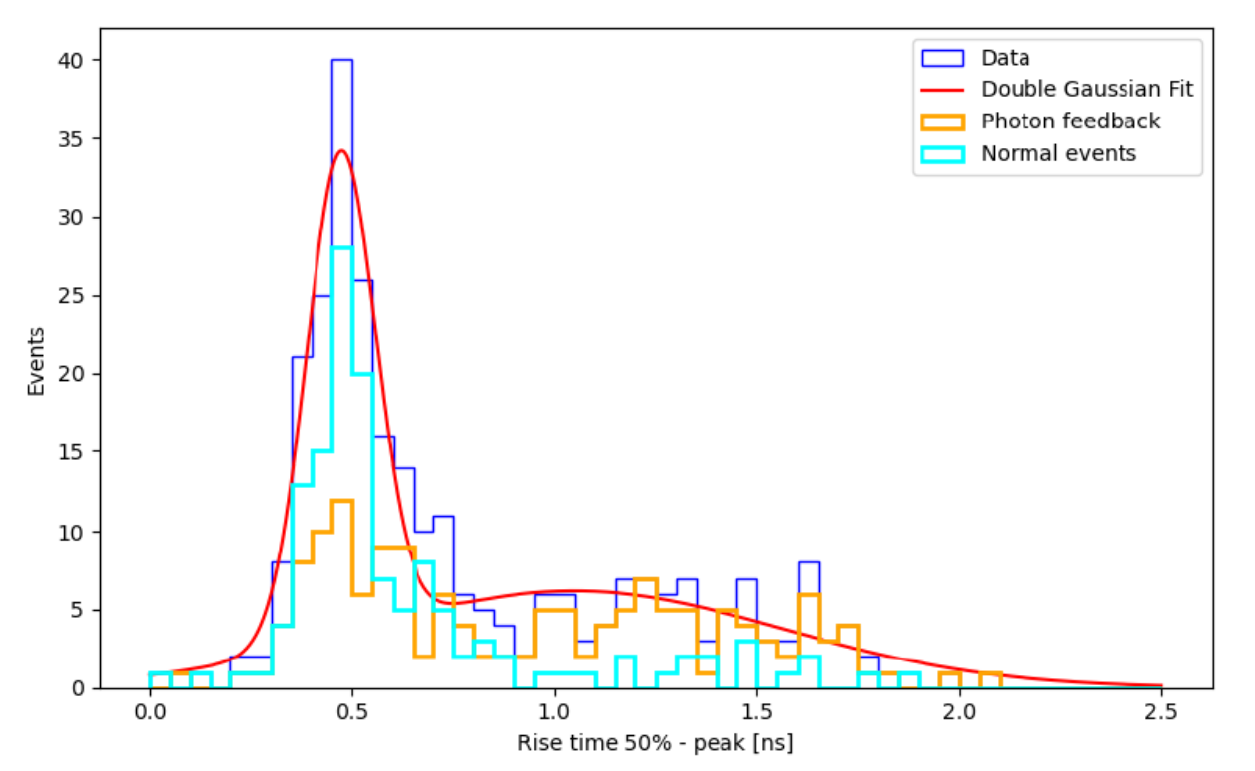}
  \end{center}
  \caption{Distribution of the pulse rise time from 50\%  to the peak. The blue histogram shows the full dataset with a double-Gaussian fit overlaid. The orange and cyan histograms represent the events tagged as affected by photon feedback and single-avalanche events, respectively, as selected using the second-derivative method.}
\label{fig:t50-peak}
\end{figure}

Unfortunately, a control sample featuring an independent assessment of photon-feedback contamination is not available for a closure test; hence, the efficiency of the selection criterion cannot be reliably determined.
In an attempt to move in this direction though, I focus on the 50\%-to-peak rise-time because it is expected to increase with closely spaced peaks. This quantity has a bimodal distribution, indicating a possible signature of photon feedback. I therefore use it for a quasi-independent consistency check. 
Applying the second-derivative zero-crossing selection and checking the correlation with rise time of the resulting classification may offer useful insights.
Figure \ref{fig:t50-peak} shows the total distribution of rise times in blue, with a double-Gaussian fit overlaid in red, and subsets of events classified as affected or unaffected by photon feedback in orange and cyan, respectively. The high rise-time region is dominated by photon-feedback events, which is encouraging. However, a large fraction of the low rise-time events is also tagged as photon feedback. Individual inspection of waveforms of a subset of events shows that many of these low rise-time signals still show secondary peaks, so the classification is consistent. While this feature alone does not provide a rigorous validation of the selection, it at least does not contradict its preliminary consistency.

Other definitions of rise-time, including the 50\% pulse-height constant fraction, and the 10\%--90\% rise interval do not offer any useful insights, as their distributions in our data are unimodal and offer no obvious evidence of distinct event classes.

A possible approach to better understand photon feedback in future and to establish a pure control sample of photon-feedback-free events is to use an RPC read out by the same digitiser. Comparing GasPM signals with RPC signals, which cannot exhibit photon feedback since RPCs have no photocathode, is likely to provide insights on useful signal-shape features. The shape of the RPC signals is expected to mirror that of GasPM signals not affected by photon feedback, as the underlying physical processes are identical, offering an independent benchmark for a pure single-avalanche signal.
In combination with the higher sampling frequency of the NALU digitiser, this could help devising effective classification procedures, as either a stand-alone method to identify photon-feedback events or in combination with other observables to improve the classification performance.

In summary, I report the development of a preliminary method to classify events affected by photon feedback, taking advantage of the higher sampling frequency of the digitiser. This procedure, opportunely refined, is ready to be applied in a future dedicated beam test.
\chapter{LaB$_6$ photocathode characterisation} \label{chap:cosmic}
This chapter describes preparatory work toward a possible future beam test with a GasPM equipped with a LaB$_6$ photocathode, known to be tolerant to ion feedback. To this end, I execute a cosmic ray test as a preliminary probe of quantum efficiency.

\section{Concept}

One of the challenges affecting the GasPM is ion feedback, which progressively damages the photocathode and reduces its efficiency. During the 2023 beam test, the CsI photocathode demonstrated poor resistance against ion feedback, showing damage and degraded efficiency. 
We therefore investigate alternative photocathode materials expected to have greater resistance to damage and potentially enhanced photoemission properties. 
We revert to lanthanum hexaboride (LaB$_6$), which was employed in an early laser test~\cite{Okubo-GasPM}.
This material is known to have higher resistance to ion feedback and exposure to air~\cite{osti_5693510}.
The drawback is low quantum efficiency for 375 nm photons~\cite{Okubo-GasPM}, which could represent an issue for a future beam test. 
The quantum efficiency   
\begin{equation}
    \mathrm{QE} = \frac{\text{number of photoelectrons emitted}}{\text{number of incident photons}},
\end{equation}
is wavelength-dependent. Cherenkov-photon wavelengths range from ultraviolet to visible. The LaB$_6$ quantum efficiency in the UV spectrum, relevant for Cherenkov photons, is unknown and must be tested prior to using LaB$_6$ in a future beam test. To this end, we conduct a cosmic-ray test of the GasPM equipped with the LaB$_6$ photocatode, and repeat it with an RPC to compare hit-rate results, thus isolating the rate of Cherenkov-photon signals from the contamination by ionization signals in the GasPM.

\section{Cosmic ray test} 

\begin{figure}[tb]
  \begin{center}
    \includegraphics[width=0.35\textwidth]{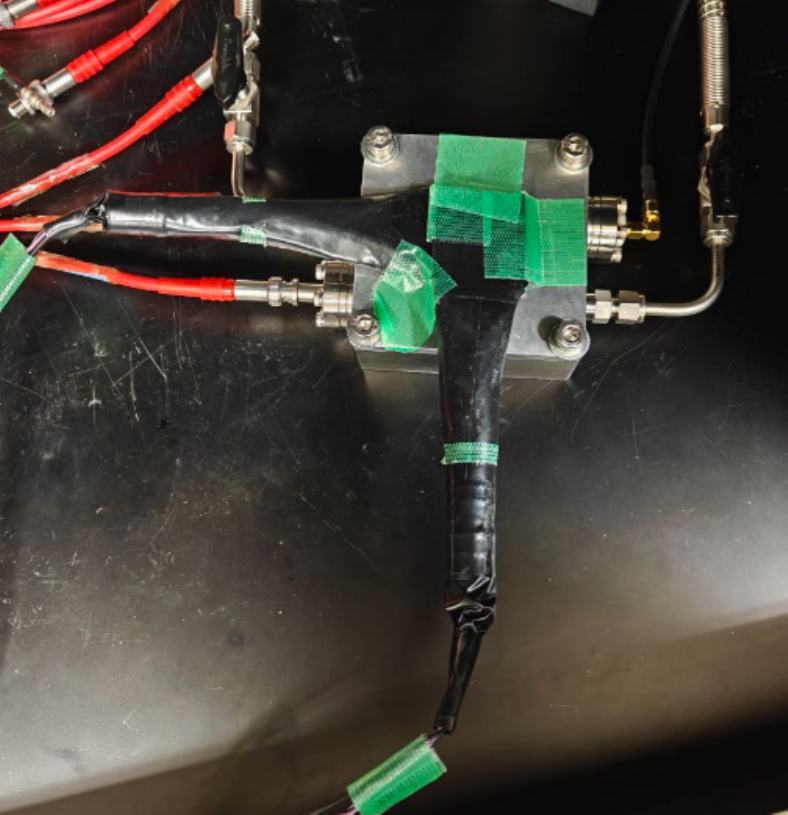}
  \end{center}
  \caption{Cosmic-ray test configuration.}
\label{fig:cosmic_config}
\end{figure}

\subsection{Layout}
The experimental configuration for the cosmic-ray test has analogies and differences with respect to the beam test (Fig.~\ref{fig:cosmic_config}). The 150~$\mu$m gas gap is unchanged. The trigger is provided by two $4\times4$~cm$^2$ plastic scintillators connected to PMTs by light guides. The scintillation counters do not sandwich the GasPM as in the beam test, but they are both placed above it. This setup facilitates installation, but it leads to a reduction in acceptance. 
We use the DRS4 digitiser for readout.

\subsection{Data acquisition}
I first prepare the coincidence logic for the trigger scintillation counters. I am interested only in events where both counters detect a signal, ensuring that a cosmic muon passed through. To monitor the trigger signals, each scintillator output is split and sent both to a channel of the digitiser and to a discriminator with a $-30$~mV threshold and a 40~ns output width. The two outputs from the discriminators are in coincidence, which is sent to the digitiser for trigger. The same output is also fed into a scaler to count the triggers. 

An amplifier, installed at the end of the GasPM signal cable right before the DRS4 input, protects the digitiser from discharges. A 20 dB attenuation balances this voltage amplification.
I then apply 1.1 kV to the  trigger counter PMTs. The trigger-rate is approximately $11 \,\mu$ min$^{-1}$, which is close to the expected value based on the sea-level rate of approximately 1 muon per cm$^{-2}$~min$^{-1}$ and the 16~cm$^2$ scintillator active area. 
We then prepare the gas mixing system as in the beam test and replace the photocathode immediately afterwards, so that the GasPM can be connected to the gas line without delay. This minimises any exposure of the photocathode to air.

I then gradually apply voltage to the GasPM, while I monitor its output on an oscilloscope to rule out unwanted discharges. Once achieved the target 2.64~kV, corresponding to an electric field of 176 kV/cm over the 150~$\mu$m gas gap, we connect the GasPM to the DRS4 and the trigger and we monitor the output.

\subsection{Studies and results}

Approximately 200 000 triggered events are collected after two weeks.
We expect a GasPM signal only in a small fraction of these, due to the photocathode quantum efficiency and the limited acceptance inherent to the experimental layout.  Cosmic rays arrive from all directions, while both scintillators are placed on top of the detector and the GasPM active area is few centimetres beneath them. A semiquantitative estimate based on geometric considerations suggests an expected acceptance relative to the scintillation counters of about 25\%, corresponding to about 0.05 Hz.  The quantum efficiency is expected to strongly reduce this value. 

\begin{figure}[tb]
  \begin{center}
    \includegraphics[width=0.7\textwidth]{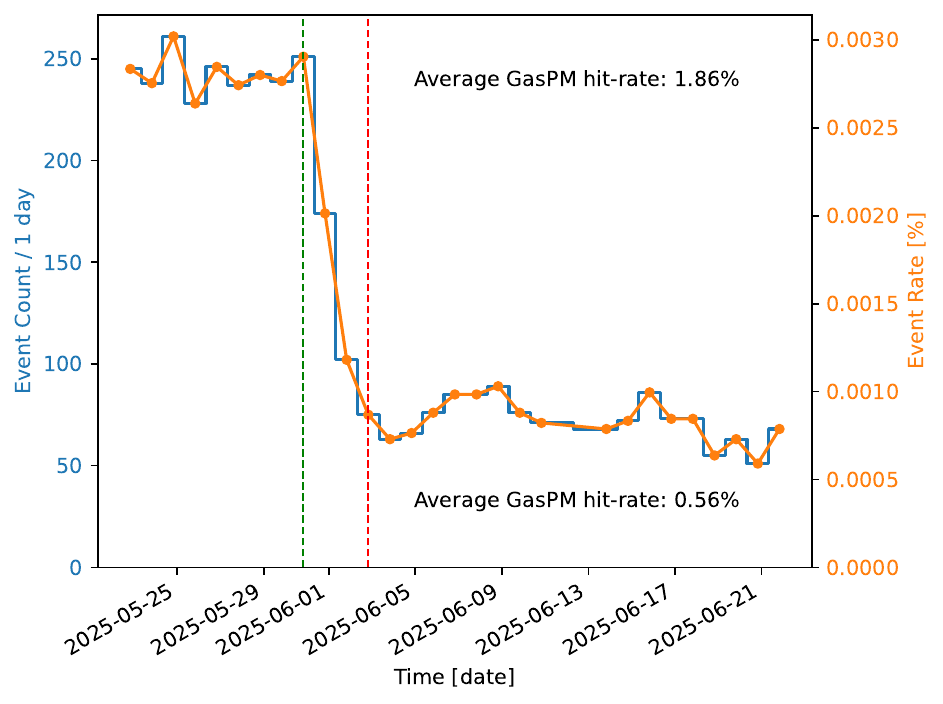}
  \end{center}
  \caption{GasPM hit-rate during the first 30 days of the cosmic-ray acquisition as a function of time.}
\label{fig:hit_rate}
\end{figure}

The observed GasPM hit-rate is extremely low and shows large, nonstatistical variance with time (Fig.~\ref{fig:hit_rate}), such as the large observed drop after approximately one week of data. 
This behaviour is unexpected and we study  signal shapes to obtain further information about the possible cause of this effect.

\begin{figure}[tb]
  \begin{center}
    \includegraphics[width=0.7\textwidth]{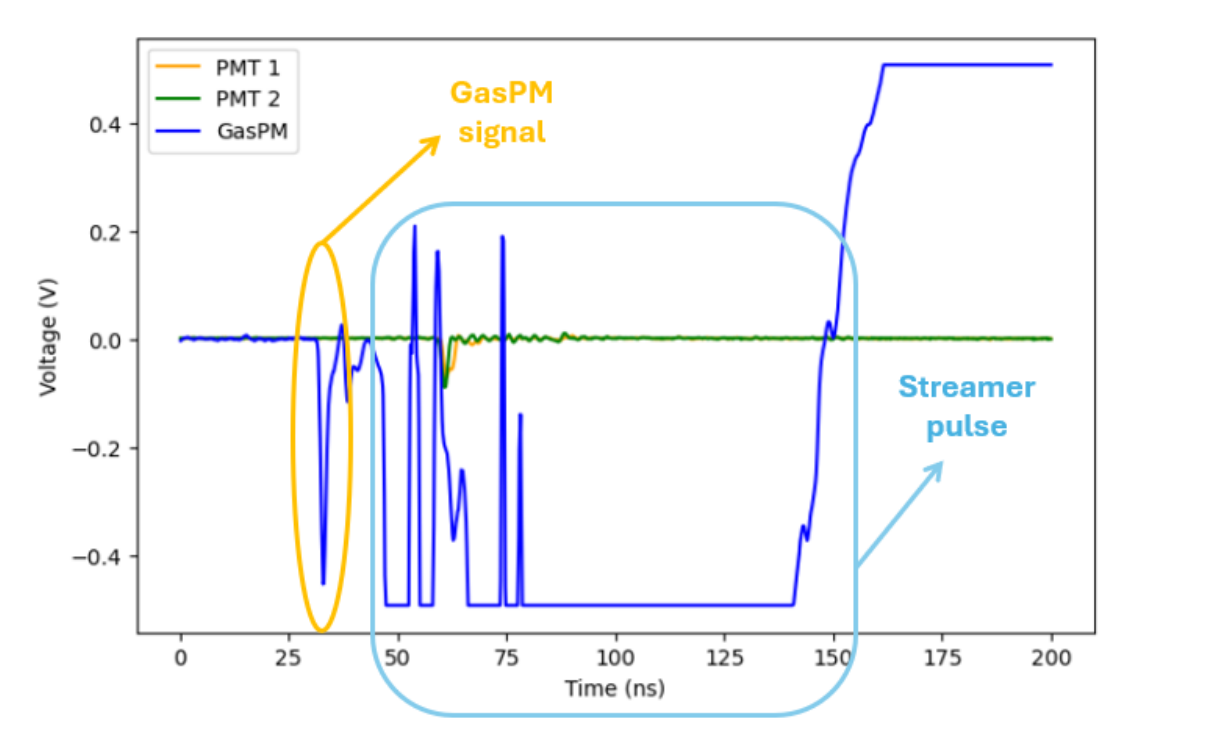}
  \end{center}
  \caption{Example of an oscilloscope display of an event acquired with the DRS4 digitiser.}
\label{fig:cosmic_event}
\end{figure}

Event displays, like the one shown in Fig. \ref{fig:cosmic_event}, where the blue line is the GasPM signal whereas the orange and green represent the two scintillation counters, show two distinct signal types. A large and saturated pulse follows a smaller one. The small pulse occurs regularly at $27.5$ ns from the PMT signals. The relative timing of the large signal varies.

The expected relative timing between the GasPM and PMT signals can be estimated from the cable lengths. This estimate is reliable because the GasPM signal is connected directly to the digitiser, passing only through attenuators and amplifiers, which introduce negligible delays. Likewise, the PMT signal passes through a splitter, which also introduces negligible delay. Assuming the typical cable propagation speed of $5$ ns/m, the expected delay for the GasPM signal is approximately $25$ ns. Accounting for a 2--3 ns difference in the detector response times, this matches well with the observed $27.5$ ns delay.
This indicates that the small signal is the genuine GasPM signal.

I then investigate the origin of the large pulse. 
To make sure to study genuine GasPM signals, for each event I extract the GasPM signal amplitude observed at 27.5 ns from the trigger signal and approximately estimate the gain, assuming a triangular signal shape. 
The resulting gain is around $10^6$ electrons, consistent with expectations for proportional operation.
A plausible hypothesis is then that the subsequent large signals correspond to streamer pulses (Sec.~\ref{sec:GasPM_issues}). Since the boundary between proportional and streamer regime is fuzzy, streamer has been observed to occasionally happen in proportional regime too. In the GasPM, the high resistivity of the resistive plate is intended to limit the current and, in combination with the quenching gas, limit discharges thus preventing a transition to a full spark. However, in this case these mitigation measures did not work. Although the GasPM is designed to operate in proportional mode, we suspect that high electric fields may drive it into limited-proportionality or streamer regime. The frequent (90\%) occurrence of large pulses is a concern, as it was not observed in previous measurements. 

\begin{figure}[tb]
    \centering
    \begin{minipage}[t]{0.35\textwidth}
        \centering
        \includegraphics[width=\textwidth]{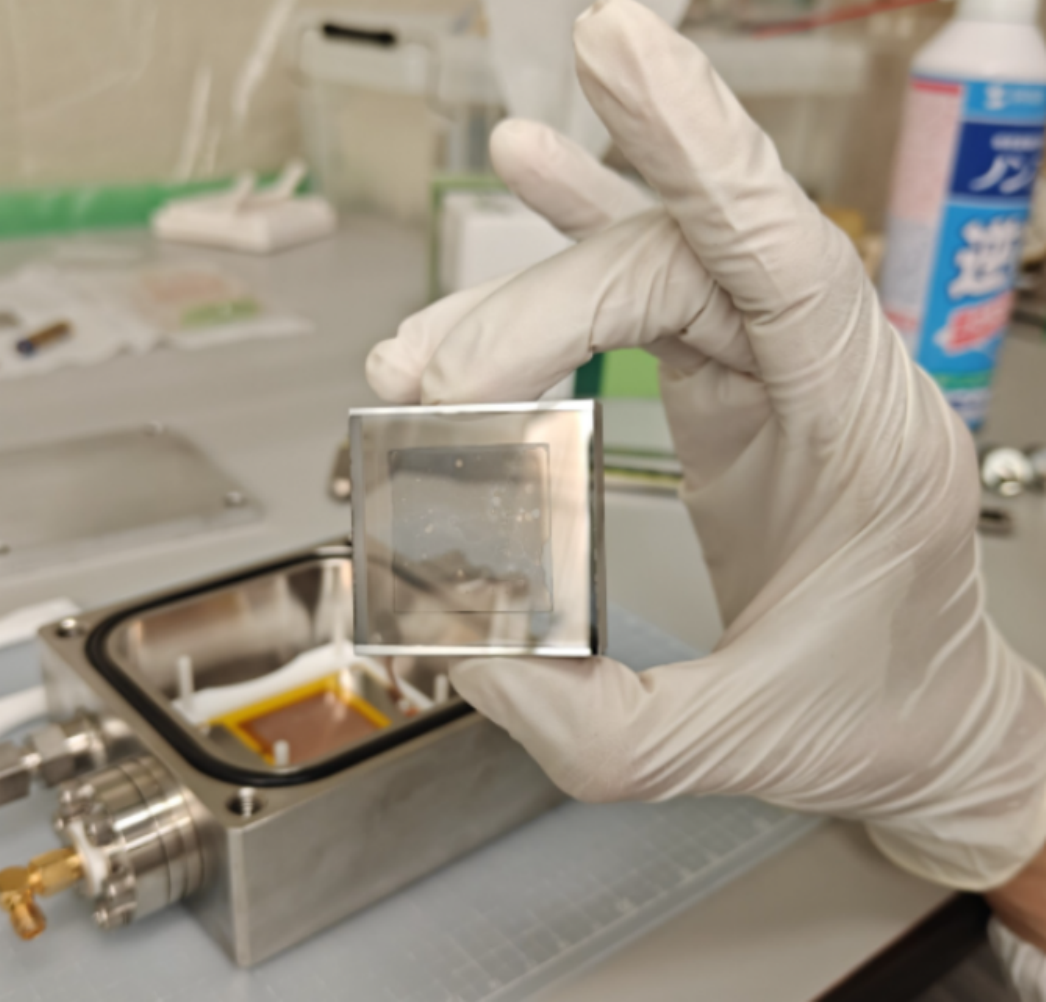}
        \captionsetup{width=\linewidth}
        \caption{LaB$_6$ photocathode damage after first weeks of data acquisition.}
        \label{fig:photocathode_damage}
    \end{minipage}%
    \hspace{0.05\textwidth}  
    \begin{minipage}[t]{0.35\textwidth}
        \centering
        \includegraphics[width=\textwidth]{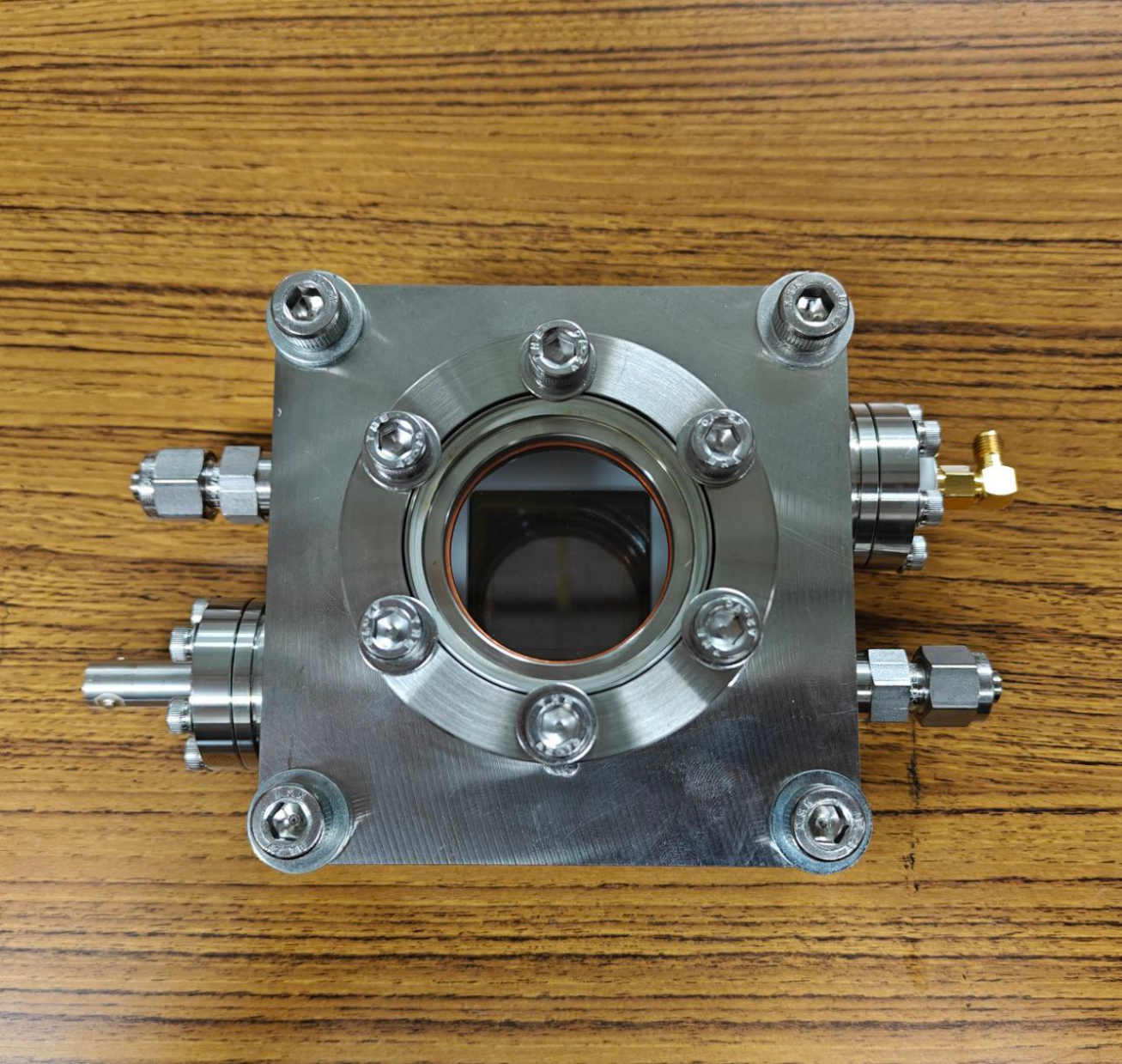}
        \captionsetup{width=\linewidth}
        \caption{GasPM with MgF$_2$ window.}
        \label{fig:window}
    \end{minipage}
\end{figure}

Since the origin of the observed efficiency drop and streamer phenomena, is unclear, we disassemble the GasPM and inspect its internal components.
Visible and significant damage affects the photocathode, in the form of a change of colour and the creation of small bubbles on its surface (Fig.~\ref{fig:photocathode_damage}). This damage might explain the drop in detection efficiency;   
a likely cause are too-frequent occurrences of streamer discharges.

We therefore revert to a previously tested configuration, known to be stable, by increasing the gas gap from $150$ to $200~\mu$m and maintaining the same electric field by adjusting the high voltage to $3.52$~kV.
Nothing changes. We then consider alternative options. The field itself is unlikely to be the cause since we have previously operated smoothly the GasPM at this electric field. Another plausible hypothesis is a gas-mixing-system malfunction that fails to supply quenching gas in correct concentration.
The quenching gas absorbs UV photons emitted by excited gas molecules, thereby preventing secondary avalanches and the onset of streamer discharges. If the SF$_6$ fraction is lower than the standard 1:9, large secondary pulses are expected. The flow-rates used are $0.50$~cm$^3$/min for the SF$_6$ and $4.50$~cm$^3$/min for the R134a.
Upon checking the system, I notice that the SF$_6$ mass-flow controller is set to its lower operational limit. Reducing it slightly would cause the flow to stop entirely. It is therefore possible that the SF$_6$ flow had stopped during the test, causing the GasPM to operate without quenching gas.
We therefore increase both gas flows proportionally to maintain the correct ratio and repeat the measurement.
The situation does not improve, and streamer pulses continue. We then attempt to further increase the SF$_6$ content, reaching a $50{:}50$ proportion. However, this change too proves ineffective.

\begin{figure}[tb]
  \begin{center}
    \includegraphics[width=0.35\textwidth]{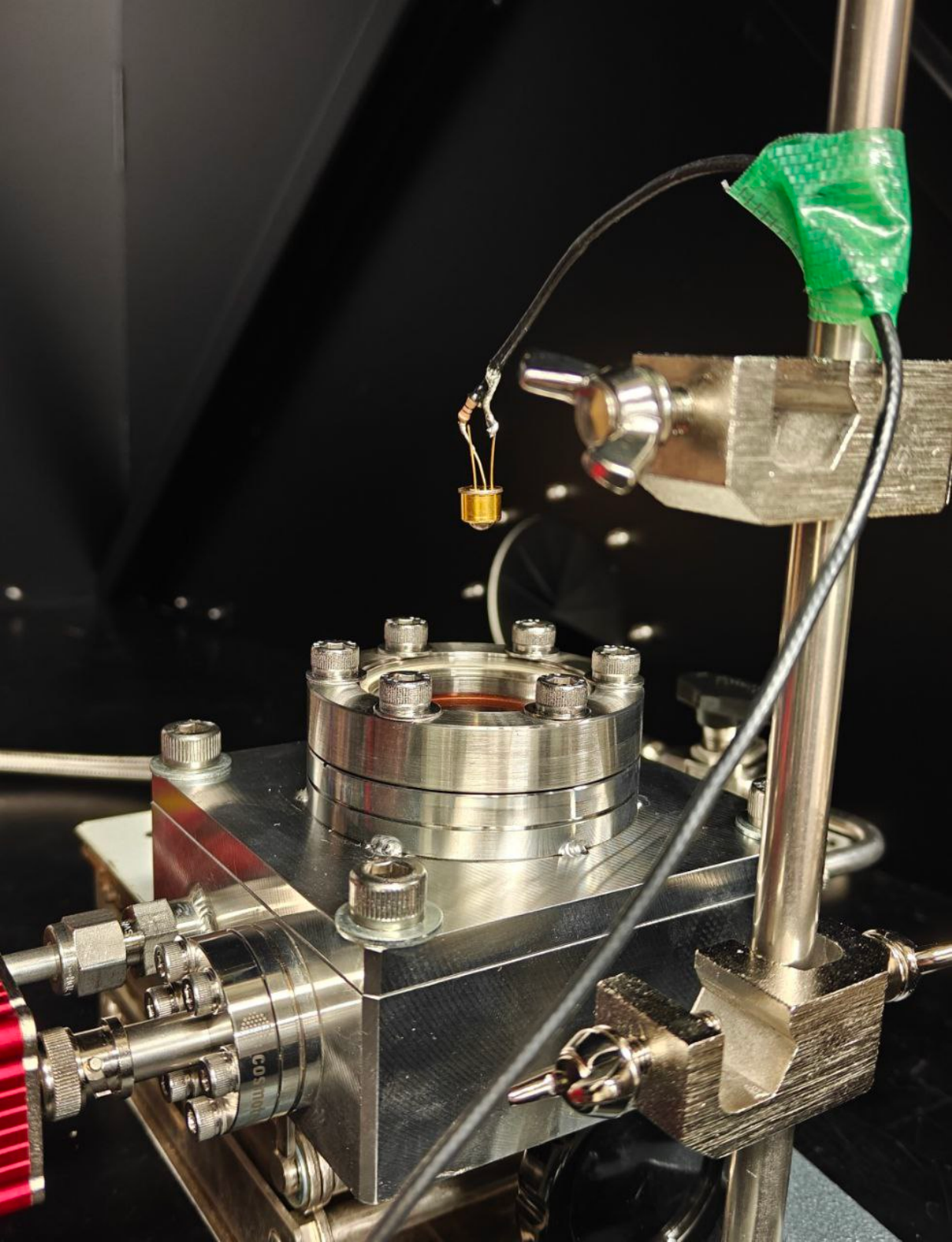}
  \end{center}
  \caption{LED test setup.}
\label{fig:LED}
\end{figure}

We then replace the photocathode with a standard metallic electrode, effectively converting the GasPM into a standard RPC. This allows us to determine whether the issue is specifically related to the photocathode, while also providing a way to assess the proportion between Cherenkov and ionization GasPM signal rates.
By comparing the hit rates recorded with the GasPM and the RPC configuration, I estimate the fraction of signal due to ionisation and that due to Cherenkov photons. This test also provides a suitable and safe configuration with no streamer, so that we can safely replace the damaged photocathode with a new one.

The first test with the RPC, under the same 176 kV/cm electric field, shows an enhanced efficiency, with a hit-rate around 7\%  over the total triggered events, with a streamer ratio of 49\%. The streamer ratio is the total number of GasPM events divided by the ones showing saturated signals. This represents an improvement, but our goal is to completely suppress the streamer phenomenon to protect the photocathode. We therefore begin gradually reducing the electric field and slightly modifying the gas mixture by increasing the quenching-gas fraction.
We finally find a safe and stable configuration at 2.5 kV with 20\% SF$_6$.
After completing data acquisition with the RPC, we return to the GasPM, installing a new photocathode and we collect more data.

The high streamer rate, unobserved in previous LED, laser, or beam tests at equal or higher electric fields, is unexplained. Since the objective of this study is to probe if the photocatode quantum efficiency at the target UV wavelength warrants a beam test, a test that does not require high gain, we proceed with data taking with this electric field.

The result of this test shows a GasPM hit-rate of $7.19 \pm 0.49$. This is consistent with the $7.66 \pm 0.18$ hit-rate measured by the RPC. 
The agreement in hit-rate between the two devices strongly suggests that no significant fraction of  Cherenkov photons are being detected.  
This may indicate that the quantum efficiency of the employed photocathode is too low for our purposes.  

To determine the quantum efficiency of the photocathode, I prepare a LED-light test (Fig.~\ref{fig:LED}). The LED is a Marktech Ultraviolet emitter MTE2350D-UV, with a central wavelength of 235 nm. An MgF$_2$ window replaces the upper metallic surface of the GasPM, as shown in Fig.~\ref{fig:window}, to allow for photons to reach the photocathode directly. A square wave of different amplitudes within the LED working range (5-8V) and a 1\% duty cycle provided by a waveform generator pilots the LED.
This test could not be completed in time for this thesis. It is the first step of our upcoming measurement campaign in autumn 2025.

\chapter{Summary and conclusions}   \label{chap:conclusion}

This thesis reports the latest developments on the gaseous photomultiplier (GasPM), a novel gaseous photodetector under study for a potential upgrade of the Belle~II experiment. The principal motivation is to enhance suppression of beam-induced background photons, which limit the performance of the Belle~II electromagnetic calorimeter. The GasPM associates a photocatode with the RPC avalanche-multiplication technology. High efficiency and timing resolution, along with scalability and cost-effectiveness, make it ideally suited to suppress out-of-collision-time beam-background photons, thus extending the Belle II physics reach. When I joined the project, an excellent $25$~ps single-photon time resolution had been achieved with a picosecond pulse laser impinging on a LaB$_6$ photocathode. However, a later beam test, featuring a CsI photocathode, showed considerably degraded performance. 

My work focusses on addressing a major cause of the degradation, secondary signals from UV photon emission during excitation and de-excitation of gas molecules (photon feedback).
I design and execute an improved beam test that, along with several configuration changes, newly introduces single-vs-multiple electron discrimination and higher frequency signal readout.
In addition, I conduct a cosmic-ray test to start probing  the quantum efficiency of the LaB$_6$-photocathode, known to be tolerant to damage from ions in the gas gap drifting backwards. The principal results of my work are
\begin{enumerate}
    \item development of an algorithm for the efficient discrimination of photon feedback;
    \item achievement of discrimination between single- and multiple-electron events;
    \item development of a preliminary calibration procedure for a newly deployed 10-GSPS prototype digitiser;
    \item preliminary exploration of the LaB$_6$ photocathode quantum efficiency  for usage in a future beam test.
\end{enumerate}

These results are being prepared for showing at the 7th International Workshop on New Photon-Detectors, upcoming in Bologna this December 2025.

In addition to the above results, my work informs future developments of the GasPM. It demonstrates the need for an LED-based test to measure the quantum efficiency of the LaB$_6$ photocathode; the need for an optimisation of signal attenuation to match optimally the GasPM output with the dynamic range of the digitiser; and it ultimately provides important information for conducting a beam test that incorporates the improvements devised in this thesis.

\bibliographystyle{unsrt}
\bibliography{references}

@article{TDR,
      author         = "Abe, T. and others",
      title          = "{Belle II Technical Design Report}",
      collaboration  = "Belle II Collaboration",
      year           = "2010",
      eprint         = "1011.0352",
      archivePrefix  = "arXiv"
}

@article{km,
     author    = {Makoto Kobayashi and Toshihide Maskawa},
     title     = {C{P}-{V}iolation in the {R}enormalizable {T}heory of {W}eak {I}nteraction},
     journal   = {Prog. Theor. Phys.},
     volume    = {49},
     year      = {1973},
     pages     = {652},
     doi       = {10.1143/PTP.49.652}
}

@article{cabibbo,
     author    = {Nicola Cabibbo},
     title     = {Unitary {S}ymmetry and {L}eptonic {D}ecays},
     journal   = {Phys. Rev. Lett.},
     volume    = {10},
     year      = {1963},
     pages     = {531},
     doi       = {10.1103/PhysRevLett.10.531}
}

@article{acc_design,
  title={Accelerator Design at {S}uper{KEKB}},
  author={Yukiyoshi Ohnishi and others},
  journal={Prog. Theo. Exp. Phys.},
  year={2013},
  pages={03A011},
  doi = {10.1093/ptep/pts083},
  publisher={Oxford University Press}
}

@article{nanobeam,
  title={Beam-Beam Issues for Colliding Schemes with Large {P}iwinski Angle and Crabbed Waist},
  author={Paolo Raimondi and others},
  eprint = "physics/0702033",
  archivePrefix  = "arXiv",
  year={2007}
}

@article{Salam:1968rm,
    author = "Salam, Abdus",
    title = "{Weak and Electromagnetic Interactions}",
    doi = "10.1142/9789812795915_0034",
    journal = "Conf. Proc. C",
    volume = "680519",
    pages = "367",
    year = "1968"
}

@inproceedings{betatronFunctionParameterization,
  author    = "D. R. Douglas and J. Kewisch and R. C. York",
  title     = "{Betatron Function Parameterization of Beam Optics Including Acceleration}",
  booktitle = "Proceedings of the 1988 Linear Accelerator Conference",
  year      = "(1988)"
}

@techreport{verduAndresLuminosity,
  author      = "S. Verdu-Andres",
  title       = "{Luminosity Geometric Reduction Factor from Colliding Bunches with Different Lengths}",
  type        = "BNL Report",
  number      = "BNL-114409-2017-IR",
  year        = "(2017)",
}

@article{Higgs:1964pj,
    author = "Higgs, Peter W.",
    editor = "Taylor, J.C.",
    title = "{Broken Symmetries and the Masses of Gauge Bosons}",
    doi = "10.1103/PhysRevLett.13.508",
    journal = "Phys. Rev. Lett.",
    volume = "13",
    pages = "508",
    year = "1964"
}

@article{Englert:1964et,
    author = "Englert, F. and Brout, R.",
    editor = "Taylor, J.C.",
    title = "{Broken Symmetry and the Mass of Gauge Vector Mesons}",
    doi = "10.1103/PhysRevLett.13.321",
    journal = "Phys. Rev. Lett.",
    volume = "13",
    pages = "321",
    year = "1964"
}

@article{1957AnPhy...2....1L,
       author = {{L{\"u}ders}, Gerhart},
        title = "{Proof of the TCP Theorem}",
      journal = {Annals of Physics},
         year = 1957,
       volume = {2},
        pages = {1},
          doi = {10.1016/0003-4916(57)90032-5},
       adsurl = {https://ui.adsabs.harvard.edu/abs/1957AnPhy...2....1L}
}

@article{Jarlskog:1985ht,
    author = "Jarlskog, C.",
    title = "{Commutator of the Quark Mass Matrices in the Standard Electroweak Model and a Measure of Maximal CP Violation}",
    reportNumber = "USIP-85-14",
    doi = "10.1103/PhysRevLett.55.1039",
    journal = "Phys. Rev. Lett.",
    volume = "55",
    pages = "1039",
    year = "1985"
}

@article{Jarlskog:1985cw,
    author = "Jarlskog, C.",
    title = "{A Basis Independent Formulation of the Connection Between Quark Mass Matrices, CP Violation and Experiment}",
    reportNumber = "CERN-TH-4242/85",
    doi = "10.1007/BF01565198",
    journal = "Z. Phys. C",
    volume = "29",
    pages = "491",
    year = "1985"
}

@article{1957AnPhy...2....1L2,
       author = {Pauli, Wolfgang},
        title = "{Exclusion Principle, Lorentz Group, and Reversal of Space-time and Charge}",
      journal = {Niels Bohr and the Development of Physics},
      pages = {1},
         year = 1955
}

@article{milesi,
	author = {{Milesi, Marco} and {Tan, Justin} and {Urquijo, Phillip}},
	title = {Lepton identification in Belle II using observables from the electromagnetic calorimeter and precision trackers},
	DOI= "10.1051/epjconf/202024506023",
	url= "https://doi.org/10.1051/epjconf/202024506023",
	journal = {EPJ Web Conf.},
	year = 2020,
	volume = 245,
	pages = "06023",
}

@article{Aulchenko:2015,
  author    = {V. Aulchenko and others},
  title     = {Electromagnetic calorimeter for Belle II},
  journal   = {J. Phys. Conf. Ser.},
  volume    = {587},
  number    = {1},
  pages     = {012045},
  year      = {2015},
  doi       = {10.1088/1742-6596/587/1/012045}
}

@misc{adams2023axion,
      title={Axion Dark Matter}, 
      author={C. B. Adams and others},
      year={2023},
      eprint={2203.14923},
      archivePrefix={arXiv},
      primaryClass={hep-ex}
}

@article{Oddone,
      author         = "Oddone, P.",
      title          = "{Detector Considerations}",
      journal        = "in D. H. Stork, proceedings of Workshop on Conceptual Design of a Test Linear Collider: Possibilities for a BB Factory",
      year           = "1987",
      pages          = "423"
}

@Article{GellMann-Levy,
author="Gell-Mann, M.
and L{\'e}vy, M.",
title="The Axial Vector Current in Beta Decay",
journal="Il Nuovo Cimento",
year="1960",
day="01",
volume="16",
pages="705",
issn="1827-6121",
doi="10.1007/BF02859738"
}

@article{GIM,
  title = {Weak Interactions with Lepton-Hadron Symmetry},
  author = {Glashow, S. L. and Iliopoulos, J. and Maiani, L.},
  journal = {Phys. Rev. D},
  volume = {2},
  issue = {7},
  pages = {1285},
  numpages = {0},
  year = {1970},
  publisher = {American Physical Society},
  doi = {10.1103/PhysRevD.2.1285}
}

@article{SM-glashow,
title = "Partial-Symmetries of Weak Interactions",
journal = "Nuclear Physics",
volume = "22",
pages = "579",
year = "1961",
issn = "0029-5582",
doi = "https://doi.org/10.1016/0029-5582(61)90469-2",
author = "Sheldon L. Glashow",
}

@article{SM-weinberg,
  title = {A Model of Leptons},
  author = {Weinberg, Steven},
  journal = {Phys. Rev. Lett.},
  volume = {19},
  issue = {21},
  pages = {1264},
  numpages = {0},
  year = {1967},
  publisher = {American Physical Society},
  doi = {10.1103/PhysRevLett.19.1264}
}

@article{wolfenstein,
  title = {Parametrization of the Kobayashi-Maskawa Matrix},
  author = {Wolfenstein, Lincoln},
  journal = {Phys. Rev. Lett.},
  volume = {51},
  issue = {21},
  pages = {1945},
  numpages = {0},
  year = {1983},
  publisher = {American Physical Society},
  doi = {10.1103/PhysRevLett.51.1945}
}

@phdthesis{Manfredi,
  author = {Riccardo Manfredi},
  title = {Measurement of the properties of $B^+ \to \rho^+ \rho^0$ decays at Belle II},
  school = {University of Trieste},
  year = {2022},
  note = {\url{https://arts.units.it/handle/11368/3060398}}
}

@article{PDG,
    author = "Workman, R. L. and Others",
    collaboration = "Particle Data Group",
    title = "{Review of Particle Physics}",
    doi = "10.1093/ptep/ptac097",
    journal = "PTEP",
    volume = "2022",
    pages = "083C01",
    year = "2022"
}

@article{adachi2018detectors,
  title={Detectors for extreme luminosity: Belle II},
  author={Adachi, I and Browder, TE and Kri{\v{z}}an, P and Tanaka, S and Ushiroda, Y and Belle II Collaboration and others},
  journal={Nuclear Instruments and Methods in Physics Research Section A: Accelerators, Spectrometers, Detectors and Associated Equipment},
  volume={907},
  pages={46--59},
  year={2018},
  publisher={Elsevier}
}

@article{Okubo-GasPM,
    author = "Okubo, R. and others",
    title = "{Development of a picosecond-timing Cherenkov detector using gaseous photomultiplification}",
    eprint = "2502.17025",
    archivePrefix = "arXiv",
    primaryClass = "physics.ins-det",
    doi = "10.1088/1748-0221/20/08/C08017",
    journal = "JINST",
    volume = "20",
    number = "08",
    pages = "C08017",
    year = "2025"
}

@article{MATSUOKA2023168378,
title = {Demonstration of a 25-picosecond single-photon time resolution with gaseous photomultiplication},
journal = {Nuclear Instruments and Methods in Physics Research Section A: Accelerators, Spectrometers, Detectors and Associated Equipment},
volume = {1053},
pages = {168378},
year = {2023},
issn = {0168-9002},
doi = {https://doi.org/10.1016/j.nima.2023.168378},
url = {https://www.sciencedirect.com/science/article/pii/S0168900223003686},
author = {K. Matsuoka and R. Okubo and Y. Adachi}
}

@inproceedings{Honda:2021pgx,
    author = "Honda, Tohru and Kobayashi, Yukinori and Mitsuda, Chikaori and Nagahashi, Shinya and Takai, Ryota and Takaki, Hiroyuki",
    title = "{Operational Status of Photon Factory Light Sources}",
    booktitle = "{12th International Particle Accelerator Conference~}",
    doi = "10.18429/JACoW-IPAC2021-MOPAB093",
    month = "8",
    year = "2021"
}

@unknown{Eidelman,
author = {Eidelman, S. and Nefediev, A. and Pakhlov, P. and Zhukova, V.},
year = {2020},
month = {12},
pages = {},
title = {Superfactory of bottomed hadrons Belle II},
doi = {10.48550/arXiv.2012.05147}
}

@article{SRitt,
    author = "Stricker-Shaver, D. A. and Ritt, S. and Pichler, B. J.",
    title = "{Novel Calibration Method for Switched Capacitor Arrays Enables Time Measurements with Sub-Picosecond Resolution}",
    eprint = "1405.4975",
    archivePrefix = "arXiv",
    primaryClass = "physics.ins-det",
    doi = "10.1109/TNS.2014.2366071",
    journal = "IEEE Trans. Nucl. Sci.",
    volume = "61",
    number = "6",
    pages = "3607--3617",
    year = "2014"
}

@article{RIEGLER2003144,
    title = {Detector physics and simulation of resistive plate chambers},
    journal = {Nuclear Instruments and Methods in Physics Research Section A: Accelerators, Spectrometers, Detectors and Associated Equipment},
    volume = {500},
    number = {1},
    pages = {144-162},
    year = {2003},
    note = {NIMA Vol 500},
    issn = {0168-9002},
    doi = {https://doi.org/10.1016/S0168-9002(03)00337-1},
    url = {https://www.sciencedirect.com/science/article/pii/S0168900203003371},
    author = {Werner Riegler and Christian Lippmann and Rob Veenhof}
}

@article{Sandilya:2016rpm,
    author = "Sandilya, S.",
    collaboration = "Belle-II",
    title = "{Particle Identification at Belle II}",
    eprint = "1610.00264",
    archivePrefix = "arXiv",
    primaryClass = "physics.ins-det",
    reportNumber = "BELLE2-CONF-PROC-2016-006, UCHEP-16-05, UNIVERSITY-OF-CINCINNATI-PREPRINT --UCHEP-16-05",
    doi = "10.1088/1742-6596/770/1/012045",
    journal = "J. Phys. Conf. Ser.",
    volume = "770",
    number = "1",
    pages = "012045",
    year = "2016"
}

@article{Liptak:2021tog,
    author = "Liptak, Zachary J. and others",
    title = "{Measurements of beam backgrounds in SuperKEKB Phase 2}",
    eprint = "2112.14537",
    archivePrefix = "arXiv",
    primaryClass = "physics.ins-det",
    doi = "10.1016/j.nima.2022.167168",
    journal = "Nucl. Instrum. Meth. A",
    volume = "1040",
    pages = "167168",
    year = "2022"
}

@misc{tektronix,
  author       = "{Tektronix}",
  title        = "{PSPL2600C Datasheet}",
  year         = {2017},
  note         = {\url{https://www.tek.com/en/datasheet/pspl2600c-pulse-generator-datasheet}}
}

@misc{Keysight,
  author       = "{KEYSIGHT}",
  title        = "{81150A Pulse Function Arbitrary Noise Generator}",
  year         = {2018},
  note          = {\url{https://shorturl.at/bHAXQ}},
}

@INPROCEEDINGS{NALU,
  author={Macchiarulo, Luca and Mostafanezhad, Isar and Varner, Gary and Rotter, Benjamin and Uehara, Dean and Liu, Gang and Chock, Christopher},
  booktitle={2020 IEEE Nuclear Science Symposium and Medical Imaging Conference (NSS/MIC)}, 
  title={Design Improvements and First results for the Revision 3 of AARDVARC Waveform Sampling System On Chip}, 
  year={2020},
  volume={},
  number={},
  pages={1-2},
  doi={10.1109/NSS/MIC42677.2020.9507950}}

@article{peskov2009research,
  title={Research on discharges in micropattern and small gap gaseous detectors},
  author={Peskov, V and Fonte, P},
  journal={arXiv preprint arXiv:0911.0463},
  year={2009}
}

@misc{Hearty,
  author       = {Christopher Hearty},
  title        = {The ECL: an overview},
  howpublished = {Presentation slides},
  note         = {Belle II Physics Week, KEK, Japan},
  year         = {2023}
}

@Article{instruments2040019,
    AUTHOR = {Nolet, Frédéric and Parent, Samuel and Roy, Nicolas and Mercier, Marc-Olivier and Charlebois, Serge A. and Fontaine, Réjean and Pratte, Jean-Francois},
    TITLE = {Quenching Circuit and SPAD Integrated in CMOS 65 nm with 7.8 ps FWHM Single Photon Timing Resolution},
    JOURNAL = {Instruments},
    VOLUME = {2},
    YEAR = {2018},
    NUMBER = {4},
    ARTICLE-NUMBER = {19},
    URL = {https://www.mdpi.com/2410-390X/2/4/19},
    ISSN = {2410-390X},
    DOI = {10.3390/instruments2040019}
}

@inproceedings{matsuoka2019performance,
  title={Performance of the MCP-PMTs of the TOP Counter in the First Beam Operation of the Belle II Experiment},
  author={Matsuoka, Kodai},
  booktitle={Proceedings of the 5th International Workshop on New Photon-Detectors (PD18)},
  pages={011020},
  year={2019}
}

@article{Conneely_2015,
  doi = {10.1088/1748-0221/10/05/C05003},
  url = {https://dx.doi.org/10.1088/1748-0221/10/05/C05003},
  year = {2015},
  month = may,
  volume = {10},
  number = {05},
  pages = {C05003},
  author = {Conneely, T.~M. and {van Dijk}, M.~W.~U. and D'Ambrosio, C. and Brook, N. and García, L.~Castillo and Cowie, E.~N. and Cussans, D. and Forty, R. and Frei, C. and Gao, R. and Gys, T. and Harnew, N. and Howorth, J. and Lapington, J. and Milnes, J. and Piedigrossi, D. and Slatter},
  title = {The TORCH PMT: a close packing, multi-anode, long life MCP-PMT for Cherenkov applications},
  journal = {Journal of Instrumentation}
}

@article{10.1063/1.1388868,
    author = {Gol’tsman, G. N. and Okunev, O. and Chulkova, G. and Lipatov, A. and Semenov, A. and Smirnov, K. and Voronov, B. and Dzardanov, A. and Williams, C. and Sobolewski, Roman},
    title = {Picosecond superconducting single-photon optical detector},
    journal = {Applied Physics Letters},
    volume = {79},
    number = {6},
    pages = {705-707},
    year = {2001},
    month = {08},
    issn = {0003-6951},
    doi = {10.1063/1.1388868},
    url = {https://doi.org/10.1063/1.1388868},
    eprint = {https://pubs.aip.org/aip/apl/article-pdf/79/6/705/18559493/705_1_online.pdf},
}

@article{Korzh:2018oqv,
    author = "Korzh, Boris and others",
    title = "{Demonstration of sub-3 ps temporal resolution with a superconducting nanowire single-photon detector}",
    eprint = "1804.06839",
    archivePrefix = "arXiv",
    primaryClass = "physics.ins-det",
    reportNumber = "FERMILAB-PUB-20-149-V",
    doi = "10.1038/s41566-020-0589-x",
    journal = "Nature Photon.",
    volume = "14",
    number = "4",
    pages = "250--255",
    year = "2020"
}

@article{Sohl_2020,
    doi = {10.1088/1748-0221/15/04/C04053},
    url = {https://dx.doi.org/10.1088/1748-0221/15/04/C04053},
    year = {2020},
    month = {apr},
    publisher = {},
    volume = {15},
    number = {04},
    pages = {C04053},
    author = {Sohl, L. and Aune, S. and Bortfeldt, J. and Brunbauer, F. and David, C. and Desforge, D. and Fanourakis, G. and Gallinaro, M. and García, F. and Giomataris, I. and Gustavsson, T. and Guyot, C. and Iguaz, F.J. and Kebbiri, M. and Kordas, K. and Legou, P. and Liu, J. and Lupberger, M. and Manthos, I. and Müller, H. and Niaouris, V. and Oliveri, E. and Papaevangelou, T. and Paraschou, K. and Pomorski, M. and Resnati, F. and Ropelewski, L. and Sampsonidis, D. and Schneider, T. and Schwemling, P. and Scorsone, E. and Stenis, M. van and Thuiner, P. and Tsipolitis, Y. and Tzamarias, S.E. and Veenhof, R. and Wang, X. and White, S. and Zhang, Z. and Zhou, Y.},
    title = {Single photoelectron time resolution studies of the PICOSEC-Micromegas detector},
    journal = {Journal of Instrumentation}
}

@article{Santonico1981,
  title     = {Development of resistive plate counters},
  author    = {Santonico, R. and Cardarelli, R.},
  journal   = {Nuclear Instruments and Methods in Physics Research},
  volume    = {187},
  number    = {2},
  pages     = {377--380},
  year      = {1981},
  month     = may,
  doi       = {10.1016/0029-554X(81)90363-3},
  url       = {https://www.sciencedirect.com/science/article/pii/0029554X81903633}
}

@article{FRANCKE20041,
    title = {Photosensitive gaseous detectors and their applications},
    journal = {Nuclear Instruments and Methods in Physics Research Section A: Accelerators, Spectrometers, Detectors and Associated Equipment},
    volume = {525},
    number = {1},
    pages = {1-5},
    year = {2004},
    note = {Proceedings of the International Conference on Imaging Techniques in Subatomic Physics, Astrophysics, Medicine, Biology and Industry},
    issn = {0168-9002},
    doi = {https://doi.org/10.1016/j.nima.2004.03.017},
    url = {https://www.sciencedirect.com/science/article/pii/S0168900204003377},
    author = {T. Francke and V. Peskov}
}

@article{osti_5693510,
  author       = {Mroczkowski, S J},
  title        = {Electron emission characteristics of sputtered lanthanum hexaboride},
  doi          = {10.1116/1.577369},
  url          = {https://www.osti.gov/biblio/5693510},
  journal      = {Journal of Vacuum Science and Technology, A (Vacuum, Surfaces and Films); (USA)},
  issn         = {ISSN 0734-2101},
  volume       = {9:3},
  place        = {United States},
  year         = {1991},
  month        = {05}}

@article{CALLIER20121569,
    title = {EASIROC, an Easy \& Versatile ReadOut Device for SiPM},
    journal = {Physics Procedia},
    volume = {37},
    pages = {1569-1576},
    year = {2012},
    note = {Proceedings of the 2nd International Conference on Technology and Instrumentation in Particle Physics (TIPP 2011)},
    issn = {1875-3892},
    doi = {https://doi.org/10.1016/j.phpro.2012.02.486},
    url = {https://www.sciencedirect.com/science/article/pii/S1875389212018688},
    author = {Stéphane Callier and Christophe Dela Taille and Gisèle Martin-Chassard and Ludovic Raux}
}

@article{ParticleDataGroup:2024cfk,
    author = "Navas, S. and others",
    collaboration = "Particle Data Group",
    title = "{Review of particle physics}",
    doi = "10.1103/PhysRevD.110.030001",
    journal = "Phys. Rev. D",
    volume = "110",
    number = "3",
    pages = "030001",
    year = "2024"
}

@article{10.1093/ptep/pts083,
    author = {Ohnishi, Yukiyoshi and Abe, Tetsuo and Adachi, Toshikazu and Akai, Kazunori and Arimoto, Yasushi and Ebihara, Kiyokazu and Egawa, Kazumi and Flanagan, John and Fukuma, Hitoshi and Funakoshi, Yoshihiro and Furukawa, Kazuro and Furuya, Takaaki and Iida, Naoko and Iinuma, Hiromi and Ikeda, Hoitomi and Ishibashi, Takuya and Iwasaki, Masako and Kageyama, Tatsuya and Kamada, Susumu and Kamitani, Takuya and Kanazawa, Ken-ichi and Kikuchi, Mitsuo and Koiso, Haruyo and Masuzawa, Mika and Mimashi, Toshihiro and Miura, Takako and Mori, Takashi and Morita, Akio and Nakamura, Tatsuro and Nakanishi, Kota and Nakayama, Hiroyuki and Nishiwaki, Michiru and Ogawa, Yujiro and Ohmi, Kazuhito and Ohuchi, Norihito and Oide, Katsunobu and Oki, Toshiyuki and Ono, Masaaki and Satoh, Masanori and Shibata, Kyo and Suetake, Masaaki and Suetsugu, Yusuke and Sugahara, Ryuhei and Sugimoto, Hiroshi and Suwada, Tsuyoshi and Tawada, Masafumi and Tobiyama, Makoto and Tokuda, Noboru and Tsuchiya, Kiyosumi and Yamaoka, Hiroshi and Yano, Yoshiharu and Yoshida, Mitsuhiro and Yoshimoto, Shin-ichi and Zhou, Demin and Zong, Zhanguo},
    title = {Accelerator design at SuperKEKB},
    journal = {Progress of Theoretical and Experimental Physics},
    volume = {2013},
    number = {3},
    pages = {03A011},
    year = {2013},
    month = {03},
    issn = {2050-3911},
    doi = {10.1093/ptep/pts083},
    url = {https://doi.org/10.1093/ptep/pts083},
    eprint = {https://academic.oup.com/ptep/article-pdf/2013/3/03A011/4439973/pts083.pdf},
}

@article{Shin:2022ybc,
    author = "Shin, Shawn and others",
    title = "{Advances in the Large Area Picosecond Photo-Detector (LAPPD\textsuperscript{TM}): 8'' {\texttimes} 8'' MCP-PMT with Capacitively Coupled Readout}",
    eprint = "2212.03208",
    archivePrefix = "arXiv",
    primaryClass = "physics.ins-det",
    doi = "10.1088/1748-0221/19/06/P06040",
    journal = "JINST",
    volume = "19",
    number = "06",
    pages = "P06040",
    year = "2024"
}


\clearpage

\chapter*{Acknowledgements}
I acknowledge KEK and Nagoya University for hosting me during my stay in Japan, the Belle II group at INFN Trieste for its financial support and hospitality, and the IMAPP program for making this possible.

I am grateful to Matsuoka-san for accepting me in his group, for his guidance during my period at KEK and comments on the manuscript, and generally for his help, time, and patience. I thank Diego for throwing me at this project and for his guidance when I was not at KEK and during thesis writing. I thank Okubo-san for his valuable advice, for comments on the manuscript,  and for teaching me many things about this project and photodetectors in general. I thank Ueda-san for helping me getting familiar with the project and sharing the DRS4 software code and the baseline selection requirements. 

My coworkers and I acknowledge the key support of KEK in our beam test at the PF-AR test beamline.


\end{document}